\documentclass[12pt, draftclsnofoot, onecolumn]{IEEEtran}
\IEEEoverridecommandlockouts

\usepackage{cite}
\usepackage{algorithm}
\usepackage{algpseudocode}
\usepackage{graphicx}
\usepackage{textcomp}
\usepackage{xcolor}
\usepackage{bbm}
\def\BibTeX{{\rm B\kern-.05em{\sc i\kern-.025em b}\kern-.08em
		T\kern-.1667em\lower.7ex\hbox{E}\kern-.125emX}}
\usepackage{subfigure}
\usepackage{array}
\usepackage{epsf}
\usepackage{times}
\usepackage{epsfig}
\usepackage{epstopdf}
\usepackage{hyperref}
\usepackage{mathtools}
\usepackage{amsthm}
\usepackage{amssymb}
\usepackage{comment}
\usepackage{siunitx}
\usepackage{lipsum}
\usepackage{kantlipsum}
\usepackage{dblfloatfix}
\usepackage{adjustbox}
\usepackage[font = small]{caption}
\usepackage{subcaption}
\usepackage{amsmath}
\usepackage{amsfonts}
\usepackage{mathtools}
\usepackage{amscd}
\usepackage{bm}
\usepackage{tikz}
\usepackage{pgfplotstable}
\usepackage{rotate}

\definecolor{rubblue}{cmyk}{1,0.5,0,0.6}
\definecolor{rubgreen}{cmyk}{0.5,0,1,0}
\definecolor{rubgray}{cmyk}{0.03,0.03,0.03,0.1}

\usepgfplotslibrary{units}

\usetikzlibrary{%
patterns,%
calc,%
fit,%
arrows,%
plotmarks,%
shadows,%
chains,%
shapes%
}

\tikzset{>=latex'} % requires arrows-library
\tikzstyle{every picture}+=[remember picture] % overlays
\pgfdeclarelayer{background}
\pgfdeclarelayer{foreground}
\pgfsetlayers{background,main,foreground}

\tikzstyle{blueblock}=[draw=rubblue, rectangle, thick, drop shadow, minimum width=20mm, minimum height=8mm,fill=rubblue!20, text width=20mm, text centered]
\tikzstyle{bluebox}=[draw=rubblue, rectangle, thick, drop shadow, minimum width=8mm, minimum height=8mm,fill=rubblue!20, text width=8mm, text centered]%, rounded corners=3pt]
\tikzstyle{greenblock}=[draw=rubgreen, rectangle, thick, drop shadow, minimum width=20mm, minimum height=8mm,fill=rubgreen!20, text width=20mm, text centered]
\tikzstyle{dot} = [draw, circle, minimum size=0.2pt,scale=0.3,fill=black,black]
\tikzstyle{smalldot} = [draw, circle, minimum size=0.1pt,scale=0.2,fill=black,black]
\tikzstyle{reddot}  =[draw,circle,minimum size=0.2pt,scale=0.8,fill=red,thin]
\tikzstyle{greendot}  =[draw,circle,minimum size=0.2pt,scale=0.8,fill=Green,thin]
\tikzstyle{bluedot}  =[draw,circle,minimum size=0.2pt,scale=0.8,fill=blue,thin]
\tikzstyle{whitedot}=[draw,circle,minimum size=0.2pt,scale=0.8,fill=white,thin]
\tikzstyle{blackdot} = [draw, circle, minimum size=0.2pt,scale=0.7,fill=black,black]
\tikzstyle{sum} = [drop shadow, draw=rubblue, thick, fill=rubblue!20, circle]
\tikzstyle{relay} = [blueblock, minimum width=5mm, minimum height=20mm, text width=5mm, rounded corners=2pt]
\tikzstyle{relay2} = [blueblock, minimum width=5mm, minimum height=15mm, text width=5mm, rounded corners=2pt]
\tikzstyle{relay3} = [blueblock, minimum width=5mm, minimum height=25mm, text width=5mm, rounded corners=2pt]
\tikzstyle{relay4} = [blueblock, minimum width=5mm, minimum height=10mm, text width=5mm, rounded corners=2pt]
\tikzstyle{relay5} = [blueblock, minimum width=5mm, minimum height=50mm, text width=5mm, rounded corners=2pt]
\tikzstyle{relay6} = [blueblock, minimum width=5mm, minimum height=5mm, text width=5mm, rounded corners=2pt]
\tikzstyle{circgreen} = [draw, circle, inner sep=2pt, fill=rubgreen, drop shadow, thick]
\tikzstyle{circwhite} = [draw, circle, inner sep=2pt, fill=white, drop shadow, thick]
\tikzstyle{circdashed} = [draw, dashed, circle, inner sep=2pt, fill=rubgray, drop shadow, thick]
\tikzstyle{vertbox} = [rectangle, draw=rubblue, thick, rotate=90, text centered, minimum width=16.5mm, minimum height=8mm, text width=16.5mm, inner sep=0pt, fill=rubblue!20, drop shadow]
\tikzstyle{vertboxb} = [rectangle, draw=rubblue, thick, rotate=90, text centered, minimum width=16.5mm, minimum height=8mm, text width=16.5mm, fill=rubblue!20, drop shadow]
\tikzstyle{vertboxshort} = [rectangle, draw=rubblue, thick, rotate=90, text centered, minimum width=10mm, minimum height=8mm, text width=10mm, inner sep=0pt, fill=rubblue!20, drop shadow]
\tikzstyle{smalldotgreen} = [draw=rubgreen, circle, minimum size=0.2pt,scale=0.8,fill=rubgreen!20]
\tikzstyle{antenna} = [regular polygon, regular polygon sides=3, draw, shape border rotate=180, minimum size=0.2pt, scale=0.3]

\tikzstyle{poly} = [regular polygon, regular polygon sides=6, shape aspect=0.5, minimum width=1.5cm, minimum height=0.35cm, draw, dashed]

\definecolor{cff9e00}{RGB}{255,158,0}
\definecolor{c4fff00}{RGB}{79,255,0}
\definecolor{cff0012}{RGB}{255,0,18}
\definecolor{c00c5ff}{RGB}{0,197,255}
\definecolor{c046f00}{RGB}{4,111,0}
\definecolor{c004b9d}{RGB}{0,75,157}
\newlength{\mylen}
\usetikzlibrary{petri}
\usetikzlibrary{shapes}
\usetikzlibrary{positioning,arrows,patterns}
\usepackage{scalefnt}
\usetikzlibrary{calc,decorations.markings}
\pgfplotsset{compat=1.10}
\usetikzlibrary{arrows.meta}
\usepackage[columnwise,switch,mathlines]{lineno} % this adds line numbers for proof reading

\usepackage{pgfplots}
\pgfplotsset{compat=newest}
\usepgfplotslibrary{groupplots}

\IEEEoverridecommandlockouts
\newtheorem{theorem}{Theorem}

\newtheorem{defn}{\textbf{Definition}}
\newtheorem{lemma}{\textbf{Lemma}}
\newtheorem{assmon}{\textbf{Assumption}}

\DeclareMathOperator*{\argmin}{arg\,min}
\newcommand{\norm}[1]{\left\lVert#1\right\rVert}

\begin{document}
	%\title{Personalized Federated Learning for Cellular VR: Online Learning and Dynamic Adaptation}
	
	\title{Personalized Federated Learning for Cellular VR: Online Learning and Dynamic Caching}
	
	\author{Krishnendu S. Tharakan, \IEEEmembership{Member, IEEE}, Hayssam Dahrouj, \IEEEmembership{Senior Member, IEEE}, Nour Kouzayha, \IEEEmembership{Member, IEEE}, Hesham ElSawy, \IEEEmembership{Senior Member, IEEE}, and Tareq Y. Al-Naffouri, \IEEEmembership{Senior Member, IEEE}
		\thanks{K. S. Tharakan and H. ElSawy are with the School of Computing, Queen’s University, Kingston, Canada. E-mail:\{k.tharakan, hesham.elsawy\}@queensu.ca.}
		\thanks{H. Dahrouj is with the Department of Electrical Engineering, University of Sharjah, Sharjah, United Arab Emirates. E-mail: hayssam.dahrouj@gmail.com.}
		\thanks{N. Kouzayha, and T. Y. Al-Naffouri are with the Division of Computer, Electrical and Mathematical Sciences and Engineering, King Abdullah University of Science and Technology, Thuwal 23955-6900, Saudi Arabia. E-mail: \{nour.kouzayha, tareq.alnaffouri\}@kaust.edu.sa.}
		\thanks{This paper has been accepted for publication in the IEEE Transactions on Communications.}
		
	}
	
	\maketitle
	
	\begin{abstract}
		Delivering an immersive experience to virtual reality (VR) users through wireless connectivity offers the freedom to engage from anywhere at any time. Nevertheless, it is challenging to ensure seamless wireless connectivity that delivers real-time and high-quality videos to the VR users. This paper proposes a field of view (FoV) aware caching for mobile edge computing (MEC)-enabled wireless VR network. In particular, the FoV of each VR user is cached/prefetched at the base stations (BSs) based on the caching strategies tailored to each BS. Specifically,  decentralized and personalized federated learning (DP-FL) based caching strategies with guarantees are presented. Considering VR systems composed of multiple VR devices and BSs, a DP-FL caching algorithm is implemented at each BS to personalize content delivery for VR users. The utilized DP-FL algorithm guarantees a probably approximately correct (PAC) bound on the conditional average cache hit. Further, to reduce the cost of communicating gradients, one-bit quantization of the stochastic gradient descent (OBSGD) is proposed, and a convergence guarantee of $\mathcal{O}(1/\sqrt{T})$ is obtained for the proposed algorithm, where $T$ is the number of iterations. Additionally, to better account for the wireless channel dynamics, the FoVs are grouped into multicast or unicast groups based on the number of requesting VR users. The performance of the proposed DP-FL algorithm is validated through realistic VR head-tracking dataset, and the proposed algorithm is shown to have better performance in terms of average delay and cache hit as compared to baseline algorithms.
	\end{abstract}
	\IEEEpeerreviewmaketitle

	\section{Introduction}
	\subsection{Overview}
	Virtual reality (VR) promises revolutionized interactions between users and their surroundings. VR has the potential to unite users from around the world within immersive virtual environments, transcending geographical limitations. The global VR market is expected to grow at a compound annual growth rate of $27.5$\% from $2023$ to $2030$ to reach $\$435.36$ billion by $2030$~\cite{grandview23}. The wireless-connected VR devices present a promising solution for delivering ubiquitous user experiences anytime, anywhere, and have the potential to unlock numerous innovative applications. However, the challenges to provide seamless connectivity over unstable wireless channels for real-time VR applications are yet to be addressed~\cite{guo24, fenghe20}. In this context, this paper proposes a field of view (FoV) aware caching scheme to facilitate high quality VR services over mobile edge computing (MEC)-enabled wireless network. The proposed approach leverages personalized caching strategies to cache or prefetch each VR user's FoV at the serving base station (BS). A decentralized and personalized federated learning (DP-FL) based caching algorithm is then presented for optimizing the considered VR network.
	
	In pursuit of enhancing the quality of personal experience (QoPE), various architectures, such as fog computing, cloud computing, and MEC, have been suggested in the recent literature. Among such alternatives, MEC has emerged as a promising paradigm by placing edge servers at the periphery of wireless access networks, near mobile devices and users~\cite{tharakan19}. Due to the constrained processing capabilities of users' local devices in VR applications, local devices can only handle the generation of basic or 2D models. Consequently, the video frames captured by VR sensors may be forwarded to edge servers to provide data for computer vision tasks~\cite{liu21,fenghe20}. The VR systems can leverage the MEC paradigm to execute tasks like swift and precise 3D graphics rendering, interaction with multiple sensors, caching the VR files at the edge server, and displaying high-resolution content on mobile devices possessing constrained computational capabilities. This approach involves shifting computing tasks closer to the end-users, reducing latency, and improving real-time responsiveness. 
	
	Immersive videos that encompass a $360$-degree multi-perspective view play a pivotal role in enabling the utilization of VR applications~\cite{chen18}. There is a plethora of video coding solutions in the literature that adjust streaming based on users' attention by monitoring their visual area of interest~\cite{Zink2019,lungaro18}. The shared objective in~\cite{Zink2019,lungaro18} is to exclusively stream users' viewpoints, which refers to the portion of the sphere within a user's FoV. Streaming real-time tile-based FoV content to a network of VR users entails several time-consuming procedures. For instance, edge servers must obtain and analyze  data to determine the set of tiles within the FoV, and subsequently schedule their transmission. The overall delay in this process is significant and cannot be ignored. Thus, with the increasing number of users, operating within the delay budget becomes challenging. This realization necessitates innovative solutions across various network layers. For example, proactively caching the VR content can significantly improve the efficiency of content delivery~\cite{guo24}.
	
	Specifically, caching VR content at the edge BS offers significant benefits, including reduced latency and improved performance. Nevertheless, the dynamic nature of wireless-connected VR imposes several challenges such as low-delay constraints, constant evolution of new content, user mobility, limited cache sizes, and dynamic popularity distribution over time. Such unique VR aspects challenge the conventional edge-caching assumption of stationary file popularity~\cite{ozfatura22}. Consequently, we approach the caching problem from the perspective of online learning without relying on particular apriori statistical assumptions about the sequence of file requests~\cite{bharath18, paria21}. In  particular, we propose a DP-FL caching algorithm for VR systems so as to improve the user's QoPE subject to latency constraints.
	
	\subsection{Related Work}
	It can be noted that the majority of existing studies on VR content delivery concentrate on enhancing the throughput of wireless VR networks. For instance, the authors in~\cite{chen22} develop an iterative algorithm for optimizing wireless multiplayer interactive VR game transmission frameworks based on mobile edge computing. The algorithm employs a truncated first-order Taylor approximation of the objective function, which is iteratively refined. The work in~\cite{gupta23} proposes a two-step approach for VR user assignment to mmWave access points. The first step involves a graph-theoretic assignment, followed by optimization using geometric programming algorithms. In~\cite{dang23}, the authors propose a multi-dimensional resource allocation method for wireless mixed reality in dynamic time division duplex networks. The approach in~\cite{dang23} aims to optimize resource allocation based on the quality of user experience (QoE) while mitigating multi-cell interference. In~\cite{jakob23}, the authors present a millimeter-wave beamforming algorithm specifically designed for head-mounted display use in mobile VR applications. The approach in~\cite{zhang22} utilizes deep reinforcement learning to optimize edge server resource allocation for serving multiple VR headsets. In~\cite{li23}, the authors present a MEC-enabled VR streaming system that integrates viewport prediction and resource allocation for efficient delivery. Existing research on VR applications primarily addresses transmission delay reduction by employing wireless resource allocation techniques.
	
	The development of machine learning (ML) networks has made VR video streaming possible by multicasting single-view images instead of the entire three degrees of freedom VR content. Despite the potential of VR communication, its practical application poses a significant challenge. From a communication perspective, VR video streaming requires substantial bandwidth on both backhaul and wireless links to deliver high-resolution videos to VR headsets. Usually, the prevailing trend is toward centralized algorithms. However, transmitting raw data from wireless edge devices to a central processor for learning poses significant challenges, as achieving a well-trained model requires vast amounts of data~\cite{bakambekova2024interplay}. It can be highly demanding on energy and bandwidth resources, introduce substantial delays, and potentially compromise user privacy. A more promising and practical alternative involves shifting the learning process to the edge themselves, leveraging collaborative machine learning techniques, specifically federated learning (FL)~\cite{Zhagypar2023, amiri20, krishnendu22}. FL enables the utilization of local datasets and processing capabilities of edge BSs, necessitating minimal communication. Specifically, FL can efficiently manage distributed ML tasks across edge BSs. However, FL encounters its own set of challenges. In its fundamental form, such as FedAvg~\cite{pmlr-mcmahan17a}, FedProx~\cite{pmlr-karimireddy20a}, the classical FL involves learning a single, global statistical model from data stored on tens to potentially millions of remote devices, thus disregarding their individual contributions. When training multiple instances of a model on diverse edge BSs with heterogeneous data, each instance may conform to a different statistical relationship, potentially diminishing the accuracy of the global model's inferences. Inspired by this, in the current work, the statistical challenges in the federated setting are addressed by learning separate models for each BS. Since structure between the different models exists, it becomes natural to model them using online learning, which this paper aims to promote.

	Thus, developing efficient FL algorithms that address statistical heterogeneity while ensuring convergence guarantee is crucial. FL encounters two primary challenges: (i) the heterogeneous nature of data and (ii) the communication overhead during training. Statistical heterogeneity emerges when data-generating distributions differ among users. This data generation paradigm deviates from the assumptions of identically and independently distributed (iid) data in distributed optimization, leading to a higher risk of stragglers and potentially introducing complexity in terms of modeling, analysis, and evaluation. This is common in FL, where data on each device tends to be personalized~\cite{tian19,smith17}. For instance, consider sentence completion from text messages, where users of varying ages and backgrounds use different wording and structures, resulting in imbalanced datasets.

	The aforementioned considerations prompt us to pose the following two questions: (i) In the realm of VR systems, how can we design a personalized and decentralized federated algorithm for a BS by leveraging the caching strategies of its neighboring BSs to improve its own performance? (ii) How can we reduce the communication cost within a decentralized FL setting? To answer the first question, a weighted combination of caching strategies in spatial and temporal domains is proposed. Specifically, the paper formulates the caching objective as a weighted linear combination of all the neighboring BSs and users' caching strategies, as well as the past caching strategies. This leads to new performance guarantees for VR networks. Towards answering the second question, the paper adopts a one-bit quantization of the stochastic gradient descent (OBSGD) framework while communicating the gradients. Such an OBSGD framework particularly relies on exchanging the sign of the stochastic gradient between the BSs, as opposed to the stochastic gradient exchange adopted in the classical literature.
	
	To the best of our knowledge, the approach of using a personalized and decentralized federated model for VR content delivery, which takes into account the unique characteristics of spatial and temporal correlation of requested FoVs, has not yet been explored in the context of wireless communication. The current work, therefore, presents a comprehensive framework for theoretical guarantees. 
	
	%In particular, a Probably Approximately Correct (PAC) bound on the conditional average cache hit rate and convergence analysis of the proposed algorithm is derived. Further, the DP-FL algorithm leverages OBSGD to minimize communication costs. By quantizing gradients to a single bit, OBSGD significantly reduces the amount of data transmitted during training. Additionally, to account for the wireless channel's characteristics, we group FoVs for transmission. This grouping considers user demand, employing multicast when multiple VR users request the same FoV and unicast for individual requests. The modified problem now incorporates the delay constraint inherent in transmitting VR tiles, ensuring a more realistic representation of the network environment.
	
	\subsection{Contributions and Organization}
	%Hence, one of the key take-away points from this paper is that there exist computationally cheap caching strategies that perform excellently in an online setting even with non-iid request sequence.
	In this paper, a novel DP-FL caching algorithm for VR systems that improves the QoPE is proposed. The paper examines a VR system with multiple VR devices and BSs. The use of decentralized FL facilitates the computation and communication tasks occurring locally amongst BSs. The user demands are assumed to be non-stationary and correlated \footnote{for instance, VR users seated in a stadium watching a cricket match would have overlapping FoVs.}, thus enabling the design of caching strategies that leverage this structure in an online fashion. The main contributions of this article are summarized as follows:
	
	\begin{itemize}
		\item In order to enhance the QoPE of users, a  DP-FL based caching strategy is deployed to predict the FoV of each VR user. The aim is to learn separate models (i.e. the caching strategies) at each of the edge BSs. The caching strategy is assumed to be a weighted combination of the temporal and spatial caching strategies. The weights are updated in an online manner and a distributed learning algorithm is proposed. This leads to a Probably Approximately Correct (PAC) bound on the conditional average cache hit using the Martingale difference equation~\cite{lipsterMATH89}. 
		\item To lower the communication overhead of gradient transmission, the paper proposes using an OBSGD-based algorithm, which is shown to converge at a rate of $\mathcal{O}(1/\sqrt{T})$, where $T$ represents the number of iterations. Unlike previous FL algorithms, the proposed algorithm learns distinct but related caching models at each BS, resulting in a personalized approach.
		\item To take the wireless characteristics of the channel into account, the FoVs are grouped together into multicast or unicast groups depending on how many VR users request them. Additionally, the modified problem incorporates the delay constraint associated with transmitting VR tiles. Here, the convergence rate of the algorithm is shown to be in the order of $\mathcal{O}(1/{T})$.
		\item The performance of the proposed algorithm is compared to the state-of-the-art baseline algorithms. The results illustrate the effectiveness of the proposed algorithm under various settings in terms of average delay and cache hit. The simulation results further highlight that the performance is scalable for large network sizes.
	\end{itemize}

	An outline of the remainder paper is as follows. The system model and problem formulation are described in Section~\ref{sec:sys_model}. Section~\ref{sec:online_algo} discusses the distributed online FL algorithm. The theoretical guarantees and the algorithm are introduced in Section~\ref{sec:theo_guarant}. The simulation results and conclusions are described in Section~\ref{sec:sim_res} and Section~\ref{sec:concl}, respectively.
	
	{\color{red} }
	
	%\subsection{Notations}
	% Bold uppercase letter denotes matrices. $\mathbb{E}(\cdot)$ and $\mathbb{P}(\cdot)$ denotes the statistical expectation operator and probability respectively. $f(\cdot)$ represents the probability density function (PDF). Superscript $(\cdot)^{T}$ represents transposition. $||\cdot||_F$, $||\cdot||_{op}$ and $vec$ indicates the Frobenius norm, operator norm and vector respectively. $I_d$ represents the $d \times d$ identity matrix.

	\section{System Model and Problem Formulation} \label{sec:sys_model}
	\begin{figure}[t!]
		\centering
		\subfigure[]{\includegraphics[width=0.36\textwidth]{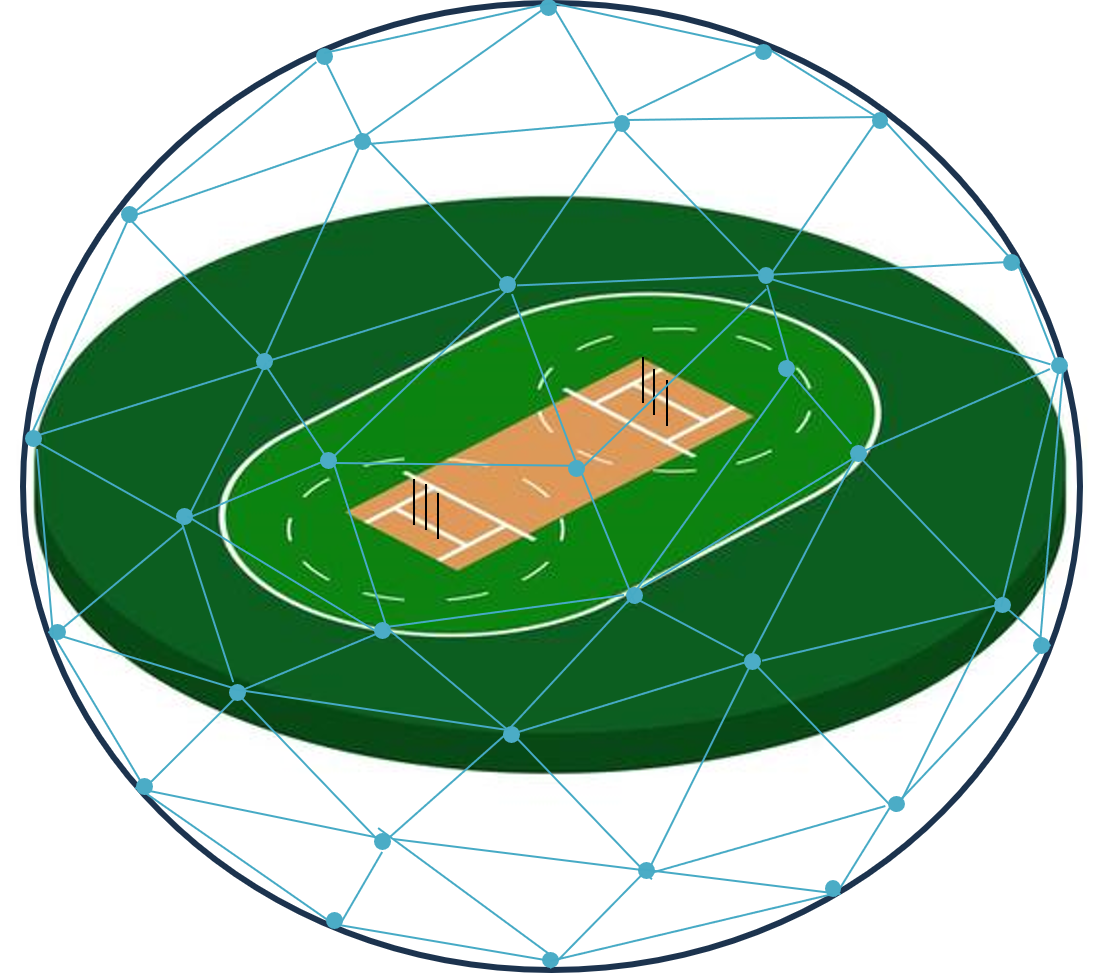}} 
		\subfigure[]{\includegraphics[width=0.43\textwidth]{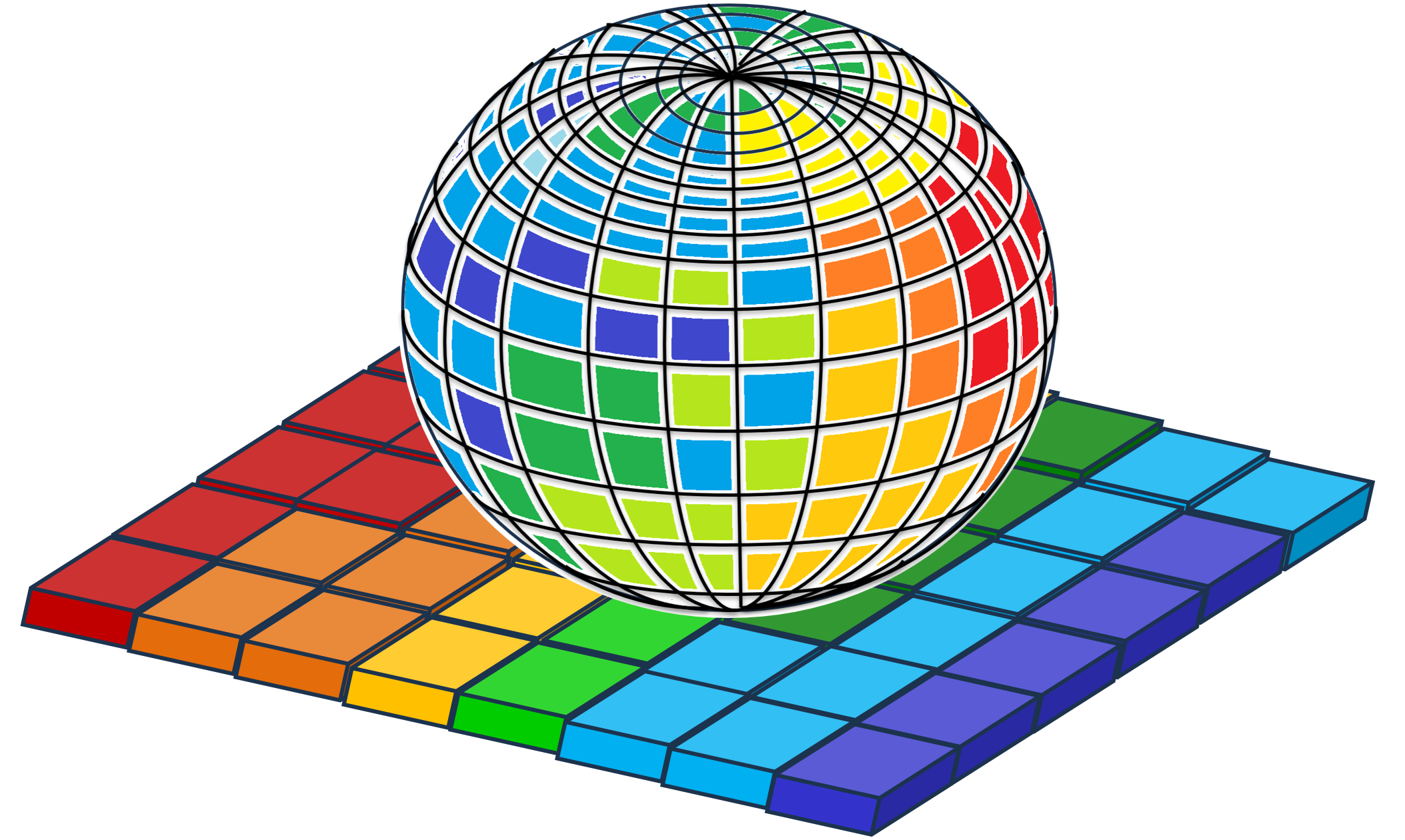}} 
		\subfigure[]{\includegraphics[width=0.43\textwidth]{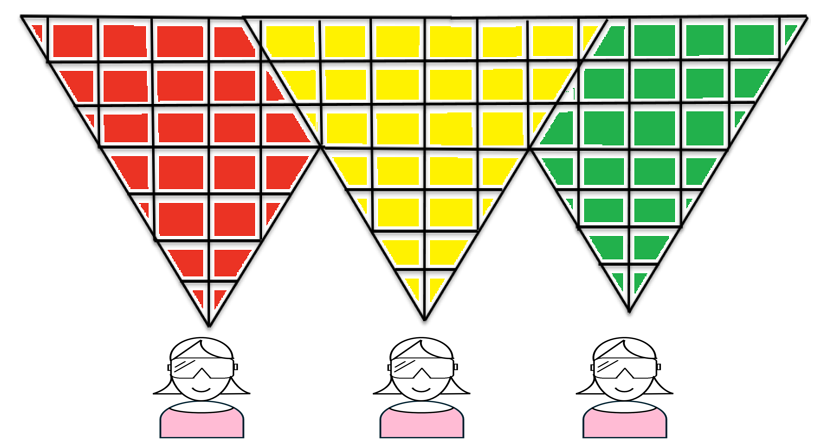}}
		\caption{(a) Event scenario with viewpoints captured  (b) FoV-based view model (c) User correlation in FoV}
		\label{fig:foobar}
	\end{figure}
	Fig.~\ref{fig:foobar} illustrates the system model. In Fig.~\ref{fig:foobar}(a), VR users interact in a sports event from multiple angles. In Fig.~\ref{fig:foobar}(b), the $360^{\circ}$ video is divided into tiles and then projected onto an equirectangular projection. Each user requests tiles based on their FoV. Overlapping FoVs among users leads to correlations in their tile requests, as shown in Fig.~\ref{fig:foobar}(c). The system model consists of $B$ BSs denoted by the set $\mathcal{B} = \{1,2,\ldots,b,\ldots,B\}$ and $U$ users denoted by the set $\mathcal{U} = \{1,2,\ldots,i,\ldots,U\}$. Since a decentralized setting is considered, it is assumed that the BSs can communicate with each other through limited capacity links and exchange the information through these links. Additionally, it can be assumed that user locations are static in many fully immersive VR applications \cite{rzhang24}. Further, each BS is assumed to have a limited computational and storage capacity of $C_b$ contents. 
	
	Based on their received VR requests, the users with overlapping FoVs are grouped together. BSs then multicast the required FoVs to users within the same group or unicast to individual users with unique FoVs. The user groups can be categorized as follows: (i) $M$ multicast groups, denoted by the set $\mathcal{G}^{m}_b = \{ \mathcal{G}_{b1}^{m}, \mathcal{G}_{b2}^{m},\ldots, \mathcal{G}^{m}_{bM} \}$, and (ii) $U$ unicast groups, denoted by the set $\mathcal{G}^{u}_b = \{\mathcal{G}^{u}_{b1}, \mathcal{G}^{u}_{b2}, \ldots, \mathcal{G}^{u}_{bU} \}$.
	For the $f$-th tile in the FoV of the $i$-th VR user in the $j$-the multicast group, the multicast signal between the $b$-th BS and the $i$-th VR user in the $j$-th multicast group at the $t$-th time slot can be expressed as:
	\begin{eqnarray}
		\bm{y}^{t}_{b,ij,f} = {(\bm{h}_{b,ij,f}^t)}^H\bm{v}_{b,j}^{t}x_{b,f}^{t} 
		+ \sum_{\mathcal{G}^{m}_{b^{'}s} \in \mathcal{G}^{m}_b/\mathcal{G}^{m}_{bs} } {(\bm{h}_{b^{'},ij,s}^t)}^H\bm{v}_{b^{'},s}^{t} x_{b^{'},s}^{t} + \bm{n}_{ij}^{t},
	\end{eqnarray}
	where $\bm{h}_{b,ij,f}^t$ is the independent Rayleigh fading channel vector, and $\sum_{\mathcal{G}^{m}_{b^{'}s} \in \mathcal{G}^{m}_{b}/\mathcal{G}^{m}_{bs} } {(\bm{h}_{b^{'},ij,s}^t)}^H\bm{v}_{b^{'},s}^{t} x_{b^{'},s}^{t}$ are the interference from the other BSs. $\bm{v}_{b,j}^{t}$ is the multicast vector from the $b$-th BS connected to the VR users in the $j$-th multicast group, and $x_{b,f}^t$ is the intended tile in the multicast group. In addition, $\bm{n}_{ij}^t$ is the additive white Gaussian noise in the $j$-th multicast group for the $i$-th VR user. Using similar notation, for the VR users in the unicast group, the unicast signal between the $b$-th BS and the $i$-th VR user in the $l$-th unicast group at the $t$-th time slot can be expressed as 
	\begin{eqnarray}
		\bm{y}^{t}_{b,il,f} = {(\bm{g}_{b,il,f}^t)}^H\bm{w}_{b,l}^{t}x_{b,f}^{t} + \sum_{\mathcal{G}^{u}_{b^{'}s} \in \mathcal{G}^{u}_b/\mathcal{G}^{u}_{bs} } {(\bm{g}_{b^{'},il,s}^t)}^h\bm{w}_{b^{'},s}^{t} x_{b^{'},s}^{t} + \bm{n}_{li}^t,
	\end{eqnarray}
	where $\bm{g}_{b,il,f}^t$ is the uncorrelated Rayleigh fading channel vector between the $b$-th BS and the $i$-th VR user for the $f$-th tile in the $l$-th unicast group, and $\sum_{\mathcal{G}^{u}_{b^{'}s} \in \mathcal{G}^{u}_b/\mathcal{G}^{u}_{bs} } {(\bm{g}_{b^{'},il,s}^t)}^H\bm{w}_{b^{'},s}^{t} x_{b^{'},s}^{t}$ are the interference from the other BSs. $\bm{v}_{b,l}^{t}$ is the unicast vector from the $b$-th BS connected to the VR users in the $l$-th unicast group. In addition, $\bm{n}_{li}^t$ is the additive white Gaussian noise in the $j$-th unicast group for the $i$-th VR user.\footnote{Note that the wireless characteristics of the channel are taken into account in the second optimization problem.} 
	
	It is further assumed that only a portion of the $360^{\circ}$ VR video is requested by a user, which is a widely utilized assumption~\cite{Mangiante2017V, lungaro18}. This observable portion is called FoV. Additionally, we assume that the pre-processing procedures which includes stitching, equirectangular projection, extraction and projection of the FoV are computed at the BS itself. Similar to the file request pattern of each user, the FoV also follows similar patterns for different users. Initially, a $360$-degree video is projected into a two-dimensional video plane. Then it is divided into $N\times P$ tiles, and the user requests a subset of such $F$ tiles, i.e., $ F = N\times P$. More precisely, the FoV allocation time is assumed to be slotted, where at each time slot, each VR user requests the FoVs from these $N\times P$ tiles. Let the set of users associated with the $b$-th BS in the $t$-th time slot be denoted by $\mathcal{U}_b^t$. The total demand for the $b$-th BS in the $t$-th time slot is given by $D_{b}^t = \sum_f\sum_{i \in \mathcal{U}_b^t} d_{b,i,f}^t$, where $d_{b,i,f}^t$ is the demand for the $f$-th FoV at the $b$-th BS by the $i$-th user in the $t$-th time slot. 
	
	Let $d_{b,i,f}^t$ be the initial FoV requested by the $i$-th user and $c_{b,i,f}^t$ be the $f$-th cached FoV at the $b$-th BS. The mean squared error (MSE)\footnote{MSE is one of the popular metrics used to evaluate VR video quality~\cite{liu21}.} for the $i$-th user at the $b$-th BS for the $f$-th FoV at $t$-th time is defined as:
	\begin{equation}
		{\rm{MSE}}_{b,i,f}^t = (c_{b,i,f}^t - d_{b,i,f}^t)^2.
	\end{equation}
	Cache placement of FoVs at each BS $b$ occurs periodically, specifically at the end of each time slot. In this work, DP-FL based caching strategy is considered, i.e., at the end of slot $t-1$, the overall caching strategy for the next $t$-th time slot is given by $\bm{\phi_b^t}$, where $\bm{\phi_{b}^t}: = \times_{i = 1}^U \times_{f = 1}^F \phi_{b,i,f}^t$, and $\times$ denotes the Cartesian product, i.e., the set that contains all possible ordered pairs. Note that $\bm{\phi_b^t}$ is a matrix of dimension $U \times F$, where each of matrix entries, i.e., $\phi_{b,i,f}^t$, denotes the the caching strategy employed at the $b$-th BS for the $i$-th user at the $t$-th time slot for the $f$-th FoV. A good way to measure the effectiveness of a caching scheme is by looking at how often users find the content they need already stored in the caches of their connected BSs. Thus, combining the caching strategies and QoPE measure we get the following cache hit metric as follows:
	\begin{equation}
		\mathcal{Q}(\bm{\phi_{b}^t}) = \sum_i\sum_f{\phi_{b,i,f}^t}\log_{10}(1/{\rm{MSE}}_{b,i,f}^{t}).
	\end{equation}
	\noindent Due to the inherent randomness of the hit rate, its average value with respect to global demands is employed to obtain a reliable measure of caching performance. To evaluate the performance of the proposed algorithm, we focus on the objective of maximizing the average cache hit rate. Since the $b$-th BS only has access to its local data at the $t$-th time, the appropriate performance metric is the conditional mean.
	Thus, the optimization problem considered in this paper can be written as follows:
	\begin{eqnarray} 
		\label{eq:the_problem}
		&&\max_{\bm{\phi}} {\sum_{b}\mathbb{E} \{\mathcal{Q}(\bm{\phi}_{b}^t)|D_{b}^{t-1} \}} \nonumber \\
		&&\text{s. t.} \sum_i \sum_f \phi_{b,i,f}^t  \leq C_b,
	\end{eqnarray}
	\noindent where $D_b^t = \sum_f\sum_{i \in \mathcal{U}_b^t} d_{b,i,f}^t$. While the conditional expectation in the objective of the problem in~\eqref{eq:the_problem} offers a theoretical basis for optimizing caching, the above formulation~\eqref{eq:the_problem} is complex to optimize, particularly in dynamic environments where the underlying data distributions may change over time. This makes solving the caching problem in~\eqref{eq:the_problem} using the conditional expectation complex in practice. One potential approach to tackle the problem is by first involving estimating the conditional expectation value and using it as a proxy for online updates as user demands arrive. This paper, however, adopts a different approach. That is, instead of using online estimate updates, a distributed online solution is proposed for the caching problem that leverages readily available ``local'' data. Furthermore, using conventional FL in a distributed setting would result in learning a single caching strategy across all BSs, which would fail to capture statistical and spatial heterogeneity, ultimately leading to poor performance. Inspired by multi-task learning, we address these statistical and spatial challenges by learning separate caching strategies for each BS and personalizing them accordingly~\cite{smith17}. 
	The following section details this approach for online distributed FL-based caching scenarios.
	
	%Let us first denote the $\mathcal{F}_b(\bm{\phi}_b^t) := {\mathbb{E} \{\mathcal{Q}(\bm{\phi}_{b}^t)|D_{b}^{t-1} \}}$. In the conventional FL setting, the objective is to solve the following optimization problem: 
	%\begin{eqnarray}
	%   \max_{\bm{\phi}_b^t} \sum_b \frac{D_b^t}{D^t}\mathcal{F}_b(\bm{\phi}_b^t), 
	%\end{eqnarray}
	%where the total dataset is given by $D^t := \sum_b D_b^t$. However, to account for the statistical he
	
	\section{Distributed Online FL Algorithm}
	\label{sec:online_algo}
	Building upon the concept of distributed online learning, we now introduce a few structural assumptions related to the caching strategy so as to lay the foundation for the proposed solution. For instance, we leverage the principles of  FL, which revolves around training a unified statistical model using data distributed across numerous remote devices. However, directly applying canonical FL to the caching problem presents unique challenges due to the non-identically distributed nature of data across BSs. In particular, naively optimizing an aggregate function could inadvertently favor or disadvantage certain devices. This bias might arise because the learned model could lean towards devices with more extensive datasets or, if devices are weighted equally, towards frequently encountered groups of devices. Thus, it is natural to learn separate models (i.e. caching strategies) for each of the BSs using the local data as depicted below in Fig.~\ref{fig:sys_model}. This leads to decentralized and personalized caching strategies.
	\begin{figure}[t!]
		\centering
		\begin{tikzpicture}
			\node (system) at (0,0)
			{\includegraphics[width=9cm]{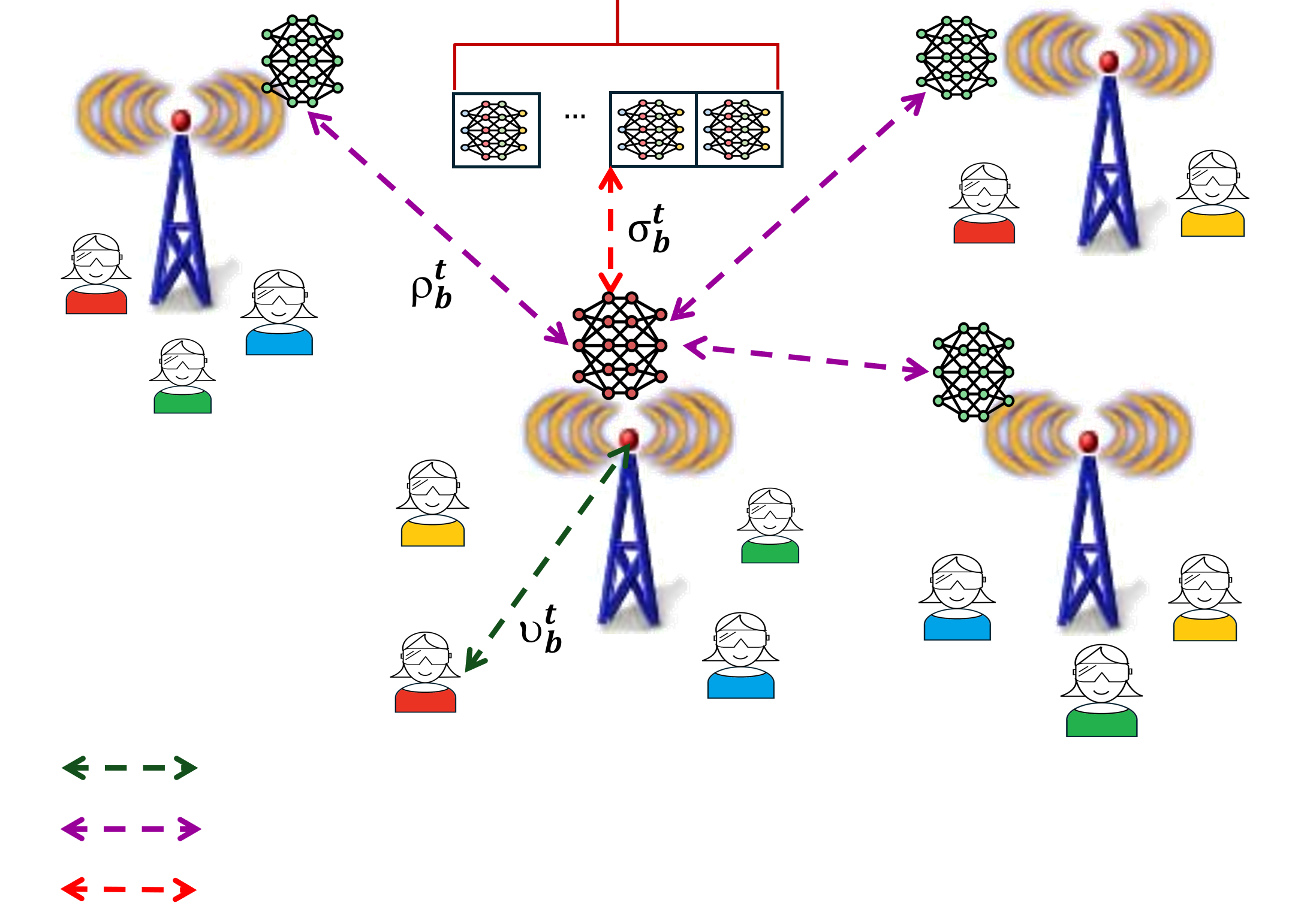}};
			\node [] (C) at (0.65,-2.2) {\small{ $\bm{\upsilon_b^t}$ spatial correlation weights among neighboring users}};
			\node [] (C) at (0.65,-2.6) {\small{$\bm{\rho_b^t}$ spatial correlation weights among neighboring BSs}};
			\node [] (C) at (-.15,-3) {\small{ $\bm{\sigma_b^t}$ temporal correlation weights at the BS}};
			\node [] (C) at (0.25,3.5) { \small{$\tau$ past temporal caching strategies}};
		\end{tikzpicture}
		\caption{Decentralized and personalized federated learning in VR system.}
		\label{fig:sys_model}  
	\end{figure}
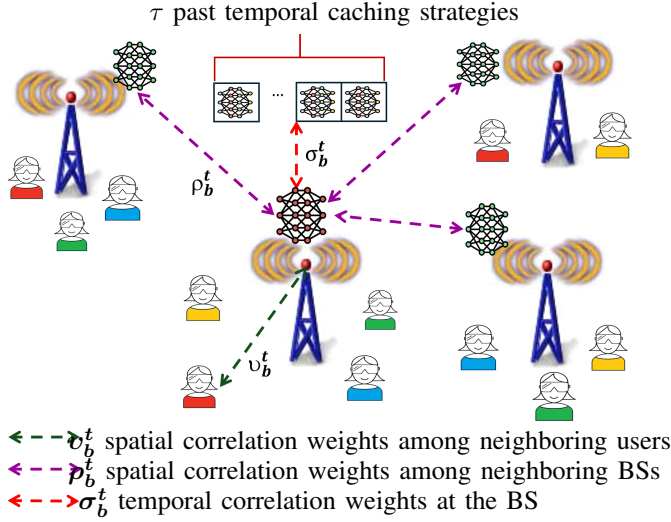

	%\begin{figure}[t!]
	%\centering    
	%\includegraphics[width=1.035\linewidth]{system_model_5.png}
	%\caption{Decentralized and personalized federated learning in VR system} 
	%\label{fig:sys_model} 
	%\end{figure}
	
	As shown in Fig.~\ref{fig:sys_model}, each BS gathers data in a non-iid fashion across the network. Following the sign of the stochastic gradient descent (signSGD) method from~\cite{bernstein18}, one-bit gradient quantization is performed before transmitting the gradients to other BSs. Furthermore, the amount of data on each node can also differ significantly. However, a structure exists between different models, and this is captured in the weighted caching strategies across spatial and temporal domains. The rationale behind employing a linear combination of caching strategies draws inspiration from the online learning literature, particularly from scenarios involving non-i.i.d data~\cite{Anava13}. Thus, by integrating principles from both statistical and adversarial learning, a robust caching strategy is proposed to effectively handle the unpredictable nature of highly non-stationary FoV requests in distributed networks. To make the caching strategy personalized by considering the statistical heterogeneity across the BSs, the caching strategy ($\bm{\bm{\tilde{\phi}}}_{b}^{T+1}$) is assumed to be the weighted average of a sequence of caching strategies $\bm{\phi}_{b}^t$ from time slot $t= T-\tau + 1$ to $T$. Let $\sigma_{b,i,f}$ be the temporal weight associated with the $i$-th user connected to the $b$-th BS at $t$-th time, the weighted caching strategy $\bm{\bm{\tilde{\phi}}}_{b}^{T+1}$ is defined as 
	\begin{equation}\label{eq:phi_time}
		\bm{\bm{\tilde{\phi}}}_{b}^{T+1} := \sum_{t=T-{\tau}+1}^T \sigma_{b,i,f}^t{\phi}_{b,i,f}^t,
	\end{equation}
	where $\sigma_{b,i,f}^t$'s are the non-negative weights that satisfy $\sum_{t = T-\tau + 1}^T \sigma_{b,i,f}^t = 1$. Let  $\bm{\sigma_b^t}: = \times_{i = 1}^U \times_{f = 1}^F \sigma_{b,i,f}^t $ be a matrix of dimension $U \times F$. To take into account the shared FoVs among the users connected to the same BS, the caching strategy $(\bm{\bm{\bar{\phi}}}_{b}^{T+1})$ is assumed to be a weighted linear combination of all the neighboring users caching strategies connected to the same BS. Let $\upsilon_{b,i,f}$ be the spatial weight associated with the $i$-th user connected to the $b$-th BS at $t$-th time. Then, the weighted caching strategy is defined as follows: 
	\begin{equation}\label{eq:phi_fov}
		\bm{\bm{\bar{\phi}}}_{b}^{T+1} := {\upsilon}_{b,i,f}^t{\tilde{{\phi}}}_{b,i,f}^t + \sum_{\substack{i^{'} \in 
				\mathcal{U}_b^t \\ i^{'} \neq i}} {\upsilon}^{t}_{b,i^{'},f} {\tilde{{\phi}}}_{b,i^{'},f}^{t},
	\end{equation}
	where $\upsilon_{b,i,f}^t$'s are the non-negative weights that satisfy $\sum_{i^{'} \in \mathcal{U}_b^t} \upsilon^{t}_{b,i^{'}, f} + \upsilon^{t}_{b,i,f} = 1$ $\forall$ BS $b$, and the set of users associated to the $b$-th BS at $t$-th time slot be denoted by $\mathcal{U}_{b}^{t}$. Let  $\bm{\upsilon_b^t}: = \times_{i = 1}^U \times_{f = 1}^F \upsilon_{b,i,f}^t $ be a matrix of dimension $U \times F$. Further, to take into account spatial heterogeneity and personalize caching strategies in the spatial domain, the paper formulates the caching strategy $(\bm{\phi}_{b}^{T+1})^{(av)}$ as a weighted linear combination of all the neighboring BSs caching strategies, by considering the spatial heterogeneity across the BSs. Let $\rho_{b,i,f}$ be the spatial weight associated with the $b$-th BS for the $i$-th user at $t$-th time. $(\bm{\phi}_{b}^{T+1})^{(av)}$ can then be written as follows
	\begin{equation} \label{eq:phi_weights}
		(\bm{\phi}_{b}^{T+1})^{(av)} := \rho^{T+1}_{b,i,f} {\bar{\phi}}_{b,i,f}^{T+1}  + \sum_{\substack{b^{'} \in \mathcal{N}_b^t \\ b^{'} \neq b}}\rho^{T+1}_{b^{'},i,f} {\bar{\phi}}_{b^{'},i,f}^{T+1},
	\end{equation}
	where the map $\mathcal{N}_b^t$ denotes the set of neighboring BSs to which the BS is connected at $t$-th time. The weights are chosen to be non-negative with the constraint given by $\sum_{b^{'} \in \mathcal{N}_b^t} \rho^{T+1}_{b^{'},i,f} + \rho^{T+1}_{b,i,f} = 1$ $\forall$ BS $b$. Let $\bm{\rho_b^t}: = \times_{i = 1}^U \times_{f = 1}^F\rho_{b,i,f}^t $ be a matrix of dimension $U \times F$.
	Thus, the original optimization problem in~\eqref{eq:the_problem} can be rewritten as follows\footnote{Note that the subscript of $\mathcal{Q}$ has been dropped in some instances for ease of notation.}: 	
	\begin{equation}
		\begin{aligned}
			&\max_{\bm{\phi}, \sigma,\rho, \upsilon} {\sum_{b}\mathbb{E}\{\mathcal{Q}_{\bm{\rho}_b^t,\bm{\sigma}_b^t,\bm{\upsilon}_b^t}
				(\bm{\phi}_{b}^T)^{(av)}|D_{b}^t \}}\\
			& \text{s. t.} \sum_i \sum_f\phi_{b,i,f}^t  \leq C_b. 
			%  &&\sum_f {\phi_{b,f}^t}T_b^{wireless} + (1 - \phi_{b,f}^t)\big(T_b^{wireless} + T_b^{fronthaul}\big) \leq \sum_f T_b^{th}
		\end{aligned}
		\label{eq:the_mod_problem}
	\end{equation}
	The selection of the weights $\rho^{T+1}_{b^{'},i,f}$, as well as $\sigma_{b,i,f}^t$, and $\upsilon_{b,i,f}^t$ now depend on how relevant (i) the past caching decisions are to the current demands, (ii) the caching decisions of neighboring users are, and (iii) the caching decisions of neighboring BSs are to the BS $b$. These aspects are quantified through the concepts of disparity, divergence, variance, and regret, as defined below.
	\begin{defn} (Disparity):
		The disparity $\texttt{S}_{b}^{T+1}({\bm{\rho}}_{\neq b}^{ T})$ between a BS $b$ and its neighbors with weights $\rho^{T+1}_{b^{'}}$, $b^{'} \in \mathcal{N}_b^t$ is given by
		\begin{equation}
			\texttt{S}_{b}^{T+1}({\bm{\rho}}_{\neq b}^{ T}) := \sum_{b^{'}  \in \mathcal{N}_b^t} \rho^{T+1}_{b^{'} } \Delta_{b}^{T+1}(\bm{\upsilon}_{b}^{T+1}, \bm{\upsilon}_{b^{'}}^{T+1}),
		\end{equation}
		where the weight vector $\bm{\rho}_{\neq b}^{ T}: = (\rho^{T+1}_{b^{'}}: b^{'} \in \mathcal{N}_b^t )$, and
		\begin{eqnarray} \nonumber
			\Delta_{b}^{T+1}(\bm{\upsilon}_{b}^{T+1}, \bm{\upsilon}_{b^{'}}^{T+1}) := \mathbb{E}\{\mathcal{Q}({\bm{\bar{\phi}}_{b^{'}}^{T+1}})\left \vert \right. D_{b}^{T}\} 
			- \mathbb{E}\{\mathcal{Q}{(\bm{\bar{\phi}}_{b}^{T+1})\left \vert \right. D_{b}^{T}}\}.
		\end{eqnarray}
	\end{defn}
	\noindent If the disparity is small for a BS $b$, it essentially indicates that the neighboring BSs' strategy is effective for BS $b$. 
	\begin{defn} (Divergence):
		The divergence $\texttt{H}_{b}^{T+1}({\upsilon}_{\neq i}^{ T})$ between a user $i$ at BS $b$ and its neighbor users with weights $\upsilon_{i^{'}}^{T+1}, i^{'} \in \mathcal{U}_b^t$ is given by
		\begin{equation}
			\texttt{H}_{b}^{T+1}({\upsilon}_{\neq i}^{ T}) := \sum_{i^{'}  \in \mathcal{U}_b^t} \upsilon^{T+1}_{i^{'} } \Delta_{i}^{T+1}({\sigma}_{b,i,f}^{T+1}, {\sigma}_{b,i^{'},f}^{T+1}),
		\end{equation}
		where the weight vector ${\upsilon}_{\neq i}^{ T}: = (\upsilon^{T+1}_{b,i^{'},f}: i^{'} \in \mathcal{U}_b^t)$, and
		\begin{eqnarray} \nonumber
			\Delta_{i}^{T+1}({\sigma}_{b,i,f}^{T+1}, {\sigma}_{b,i^{'},f}^{T+1}) := \mathbb{E}\{\mathcal{Q}({{\bar{\phi}}_{b,i^{'},f}^{T+1}})\left \vert \right. D_{b}^{T}\}  - \mathbb{E}\{\mathcal{Q}{({\tilde{\phi}}_{b,i,f}^{T+1})\left \vert \right. D_{b}^{T}}\}.   
		\end{eqnarray}
	\end{defn}
	\noindent Similarly, in assessing the caching strategies' relevance across different time slots to the current one, the primary tool is the variance across time, defined as follows:
	\begin{defn}(Variance):
		Given local information at the BS $b$ with caching strategies $\bm{\phi}_{b}^t$ for $t=T-\tau + 1,\ldots,T$, the variance $\mathbb{V}_{b}^{T}(\bm{\sigma}_{b}^{T})$ at the end of time slot $T$ is defined by 
		\begin{equation} \label{eq:disclg}
			\mathbb{V}_{b}^{T}(\bm{\sigma}_{b}^{T}) : = \sup_{\bm{\phi}_{b}^{t}: t=T-\tau+1,\ldots,T} \bigg|{\sum_{t=T-\tau + 1}^T \bm{\sigma}_{b}^{t}\Delta \bar{\mathcal{Q}}(\bm{\tilde{\phi}}_{b}^t)}\bigg|,
		\end{equation}
		where $\Delta \bar{\mathcal{Q}}(\bm{\tilde{\phi}}_{b}^t):=\mathbb{E}\{\mathcal{Q}(\bm{\tilde{\phi}}_{b}^{T})\left \vert \right. D_{b}^{T}\} - \mathbb{E}\{\mathcal{Q}(\bm{\tilde{\phi}}_{b}^{t})\left \vert \right. D_{b}^{t}\}$.
	\end{defn}
	\noindent Following the usual convention in the online learning literature~\cite{paria21, krishnendu24}, the Regret is defined as the difference in the reward (cache hit) when using the best caching strategy ${\phi_b^t}^*$ and that of the online policy $\phi_b^t$.
	\begin{defn} (Regret):
		The regret $\texttt{Reg}(\bm{\phi}_{b}^{t})$ at the BS $b$ at time $t$ with respect to a sequence of strategies $\bm{\phi}_{b}^{t}$ is defined as 
		\begin{eqnarray}
			\label{eq:regret}
			\texttt{Reg}(\bm{\phi}_{b}^{t}) : = \sup_{{\bm{(\phi}_{b}^{t}})^{*}}\sum_{t=T-\tau + 1}^T \mathcal{Q}({\bm{\phi}_{b}^{t}}^{*}) -\sum_{t=T-\tau + 1}^T \mathcal{Q}(\bm{\phi}_{b}^{t}).
		\end{eqnarray}
	\end{defn}
	\noindent It is important to note that minimizing the regret function eventually aims to optimize against the adversarial FoV request sequences. This assumption is commonly used in the caching algorithm. It is further shown that providing strong performance guarantees for the proposed algorithm helps both theoretically and practically. The defined metrics above act as auxiliary variables that intertwine internally to guarantee the proposed bound and are collectively used to derive the PAC bound of the DP-FL algorithm presented in the next section.
	\section{Theoretical Guarantees \& Proposed Algorithm}
	\label{sec:theo_guarant}
	In this section, a high probability bound on the performance of the proposed DP-FL algorithm is provided. This analysis offers insights into selecting appropriate weights and optimizing the sequence of caching policies over time. Theorem~\ref{thm:mainresult_cachingvr} establishes a high-probability lower bound on the average cache hit rate achieved by the caching strategy defined in \eqref{eq:phi_weights}. Later in this section, the communication cost and convergence analysis of the proposed DP-FL algorithm is also provided. This section further tackles the optimization problem that incorporates the delay constraint associated with transmitting VR tiles so as to account for the practical VR system delay requirements.
	\begin{theorem}
		\label{thm:mainresult_cachingvr}
		(PAC Bound) Given the caching weights and a sequence of caching strategies as in \eqref{eq:phi_weights}, with a probability of at least $1 - \delta$, $\delta>0$, we can establish a lower bound on the conditional expectation of cache hit. i.e. with a high probability the conditional expectation of cache hit is lower bounded by the difference between the true average cache hit and the error term as follows:
		\begin{equation}
			\begin{aligned}
				\mathbb{E}\left[\mathcal{Q}(\bm{\phi_{b}^{T+1}})^{(av)} \left | \right. D_{b}^t\right] \geq \sum_{t=T-\tau}^T\sigma_{b}^t\mathcal{Q}{(\bm{\phi}_{b}^{t}})^{*}  - \mathcal{E}_{\bm{\rho},\bm{\sigma},\upsilon}^T, 
			\end{aligned}
			\label{eq:the_bound}
		\end{equation} where $\mathcal{E}_{\bm{\rho},\bm{\sigma},\upsilon}^T := {\mathcal{C}_{max}} \norm{\bm{\sigma_{b}^T}}_2 \sqrt{\frac{2}{\tau} \log\frac{1}{\delta}} + \mathcal{S}_{b}^{T+1}(\bm{\rho}_{\neq b}^{ T})  + \mathbb{V}_{b}^{T}(\bm{\sigma}_{b}^T)
		+ \texttt{H}_{b}^{T+1}({\upsilon}_{\neq i}^{T}) +  \mathcal{C}_{max}\sum_{t= T -\tau + 1}^{T} \bigg|\bm{\sigma}_{b}^t - \frac{1}{\tau}\bigg| + \frac{2\texttt{Reg}(\bm{\phi}_{b}^{t})}{\tau} $, and 
		$C_{\texttt{max}}$ is the maximum possible cache hit rate.
		% $\mathcal{S}_{b}^{T+1}(\bm{\rho}_{\neq b}^{ T}) = \sum_{b^{'} \in {\color{red}\mathcal{N}_b^t}} \rho^{T+1}_{b^{'} } \Delta_{b}^{T+1}(\bm{\sigma}_{b}^{T+1}, \bm{\sigma}_{b^{'}}^{T+1})$,\\
		%$\mathbb{V}_{b}^{T}(\bm{\sigma_{b}^T}) = \sup_{\bm{\phi}_{b}^t} \abs{\sum_{t=T-\tau + 1}^T \bm{\sigma}_{b}^t\Delta \bar{\mathcal{Q}}(\bm{\phi}_{b}^t)}$,\\ 
		%$\texttt{H}_{b}^{T+1}({\upsilon}_{\neq i}^{T}):=  \sum_{i^{'} \in {\color{red}\mathcal{U}_b^t}} \upsilon^{T+1}_{i^{'} }\Delta_{i}^{T+1}({\sigma}_{b,i,f}^{T+1}, {\sigma}_{b,i^{'},f}^{T+1})$.
	\end{theorem}
	\begin{IEEEproof}\normalfont
		See Appendix \ref{appendix:proof_main_res_cachingvr}.
	\end{IEEEproof} 
	\subsection{Algorithm}
	The PAC bound states that given the caching weights and a sequence of caching strategies, with a probability of at least $1 - \delta$, $\delta>0$, we can establish a lower bound on the conditional expectation of cache hit. i.e. with a high probability the conditional expectation of cache hit is lower bounded by the difference between the true average cache hit and the error term as given in Theorem~\ref{thm:mainresult_cachingvr}. Thus, the aim is to minimize the error term given by $\mathcal{E}_{\bm{\rho},\bm{\sigma},\upsilon}^T$, ensuring that the PAC bound remains tight with a high probability ( i.e., the aim is to maximize the right-hand side of the performance bound in eq.~\eqref{eq:the_bound}). The error term $\mathcal{E}_{\bm{\rho},\bm{\sigma},\upsilon}^T$ includes the terms: disparity, divergence, variance, and regret. Thus maximizing the right-hand side of eq.~\eqref{eq:the_bound} also implies minimizing the error term $\mathcal{E}_{\bm{\rho},\bm{\sigma},\upsilon}^T$, which effectively corresponds to minimizing the disparity, divergence, variance, and regret terms \footnote{Note that these terms appear as negative values in the expression \eqref{eq:the_bound}}. Such a result is one of the major seeds for deriving the steps of our algorithm, i.e., proposed DP-FL method. More specifically, the minimization is achieved through a two-step process: (i) In the first step, the caching strategies are chosen to minimize the regret term in the bound as follows:
	\begin{eqnarray}
		\label{eq:min_regret}
		\min_{\sum {\phi}_{b,i,f}^t \leq C_b} \left[\sup_{(\bm{\phi}_{b}^t)^{*}}\sum_{t=T-\tau + 1}^T \mathcal{Q}({\bm{\phi}_{b}^t}^{*}) -\sum_{t=T-\tau + 1}^T \mathcal{Q}(\bm{\phi}_{b}^t)\right],
	\end{eqnarray}
	to obtain a sequence of caching strategies $(\bm{\phi}_b^t)^c$.
	(ii) In the second step, the right hand side of eq.~\eqref{eq:the_bound} is maximized excluding the regret as follows:
	\begin{eqnarray}
		\label{eq:second_step}
		& &\hspace{-0.6cm}\max_{\bm{\sigma_{b}^t},\bm{\upsilon}_{b}^{t}, \bm{\rho}_{b}^{t} } \sum_{t=T-\tau}^T\sigma_{b}^t\mathcal{Q}{(\bm{\phi}_{b}^{t}})^{c}-{\mathcal{C}_{max}} \norm{\bm{\sigma_{b}^T}}_2 \sqrt{\frac{2}{\tau} \log\frac{1}{\delta}}  - \mathbb{V}_{b}^{T}(\bm{\sigma}_{b}^T)
		\nonumber\\  &&\hspace{1.1cm} - \mathcal{S}_{b}^{T+1}(\bm{\rho}_{\neq b}^{ T})   
		- \texttt{H}_{b}^{T+1}({\upsilon}_{\neq i}^{T}) 
		- \mathcal{C}_{max}\sum_{t= T -\tau + 1}^{T} \bigg|\bm{\sigma}_{b}^t - \frac{1}{\tau}\bigg|
	\end{eqnarray}
	However, this involves the discrepancy, divergence and variance terms which are unknown. Since the terms, i.e., discrepancy, divergence and variance are unknown, their estimates are used instead. Additionally, the BS have access to local data, and hence the terms, i.e., discrepancy, divergence and variance are estimated in a distributed manner. More specifically, a natural approach to solving for the estimates of discrepancy, divergence and variance is to employ a distributed gradient descent method. However, the stochastic gradient descent (SGD) requires the exchange of the gradient during each round, which eventually leads to a large communication overhead. Such a cost can be reduced if the gradients are compressed before sending. One way of compressing the gradients is to use the signSGD during each round. In fact, it is shown that signSGD achieves highly compressed gradients with SGD-convergence rate~\cite{bernstein18}. In the algorithm, computing the gradient of the function $\mathcal{{Q}}_{\bm{\sigma}_b^t, \bm{\rho}_b^t,\bm{\upsilon}_b^t}(\bm{\phi}_b^t)$ involves finding the sign of the gradient with respect to $\bm{\phi}_b^t$ instead of the full gradient. Thus, using OBSGD leads to Algorithm~\ref{alg:depe_fl_algo} shown at the top of the page.
	The described implementation of DP-FL highlights how the proposed approach can be implemented in a distributed fashion across the network, thereby facilitating the reliable computation of the estimates of the individual caching strategies (i.e., on a per BS basis). The next section analyzes the communication cost of Algorithm~\ref{alg:depe_fl_algo}.

	%Theorem \ref{thm:mainresult_cachingvr} establishes a high-probability lower bound on the average cache hit rate achieved by the caching strategy defined in \eqref{eq:phi_weights}. At time slot $T + 1$, the aim is to strategically select individual caching strategies $\phi_{b,i,f}^t$ to construct an optimal aggregated strategy $(\phi_{b}^{T})^{(av)}$, as defined in~\eqref{eq:phi_weights}. This optimal strategy aims to maximize the right-hand side of the performance bound in~\eqref{eq:the_bound}. The maximization is achieved through a two-step process: (i) In the first step, the caching strategies are chosen to minimize the regret term in the bound, as shown below.
	
	\begin{algorithm}[H]
		\captionof{algorithm}{DP-FL Algorithm}%
		\label{alg:depe_fl_algo}
		\begin{algorithmic}[1]
			\Procedure{Proposed DP-FL}{}
			\State \texttt{Initialize} $\bm{\sigma}_b^0$, $\bm{\rho}_b^0$, and $\bm{\upsilon}_b^0$ for $b = 1, \ldots, B$
			\For {$t = 1,2, \ldots, T$}
			\State \text{Run regret minimization as in \eqref{eq:min_regret}}
			\text{\hspace{1cm} to get a sequence of $(\bm{\phi}_b^t)^R$}, $\forall t$
			\For{$\forall$ BS $\forall b = 1, \ldots, B$}
			\State \text{Get ${g}_{b}^t$ $\gets$ stochastic gradient }
			\EndFor       
			\State \text{Call  \textbf{Subroutine} \big(($\bm{\phi}_b^t)^R$, $\bm{\sigma}_b^t$, $\bm{\rho}_b^t$, $\bm{\upsilon}_b^t$, $\forall$  $b$\big)}
			\text{\hspace{1cm}to get $\bm{\phi}_b^{t+1}$.}
			\EndFor
			\EndProcedure
		\end{algorithmic}
		\vspace{0.2cm}	
		\hrule
		\vspace{0.2cm}	
		\textbf{Subroutine \big(($\bm{\phi}_b^t)^R$, $\bm{\sigma}_b^t$, $\bm{\rho}_b^t$, $\bm{\upsilon}_b^t$, $\forall$ $b$\big):}
		\begin{itemize}
			\item Signed Gradient descent step on $\bm{\phi}_b$ for $b = 1,\ldots, B$:
			\begin{eqnarray}
				\bm{\phi} &=& \bm{\phi}_b^{t} - \eta \nabla_{\bm{\phi}_b^t, \rm{sign}}\mathcal{{Q}}_{\bm{\sigma}_b^t, \bm{\rho}_b^t, \bm{\upsilon}_b^t}(\bm{\phi}_b^t) \nonumber\\
				\bm{\phi}_b^{t+1} &=& \argmin_{x \in \Delta_N}|| x - \bm{\phi}||_1
			\end{eqnarray}   
			\item Gradient descent with projection step on $\bm{\sigma}_b^t$ for $b = 1,\ldots, B$:
			
			\begin{eqnarray}
				\sigma &=& \bm{\sigma}_b^{t} - \mu \nabla_{\bm{\sigma}_b^t}\mathcal{{Q}}_{\bm{\sigma}_b^t, \bm{\rho}_b^t,\bm{\upsilon}_b^t}(\bm{\phi}_b^t) \nonumber\\
				\bm{\sigma}_b^{t+1} &=& \argmin_{x \in \Delta_N}|| x - \sigma||_1
			\end{eqnarray}
			\item Gradient descent with projection step on $\bm{\rho}_b^t$ for $b = 1,\ldots, B$:
			\begin{eqnarray}
				\rho &=& \bm{\rho}_b^{t} - \nu \nabla_{\bm{\rho}_b^t}\mathcal{{Q}}_{\bm{\sigma}_b^t, \bm{\rho}_b^t,\bm{\upsilon}_b^t}(\bm{\phi}_b^t) \nonumber\\
				\bm{\rho}_b^{t+1} &=& \argmin_{x \in \Delta_N}|| x - \rho||_1
			\end{eqnarray}
			\item Gradient descent with projection step on $\bm{\upsilon}_b^t$ for $b = 1,\ldots, B$:
			\begin{eqnarray}
				\upsilon &=& \bm{\upsilon}_b^{t} - \iota \nabla_{\upsilon_b^t}\mathcal{{Q}}_{\sigma_b^t, \rho_b^t,\upsilon_b^t}(\bm{\phi}_b^t) \nonumber\\
				\bm{\upsilon}_b^{t+1} &=& \argmin_{x \in \Delta_N}|| x - \upsilon||_1
			\end{eqnarray}	
			\item Broadcast $\bm{\phi}_b^{t+1}$ to all BS $b = 1,2, \ldots B$
		\end{itemize}
		\hrule
	\end{algorithm}
	
	\subsection{Communication Cost}
	We would first like to recall from Algorithm I that maximizing the right hand side of the bound given in Theorem~\ref{thm:mainresult_cachingvr} involves an estimate of the terms namely disparity, divergence, and variance which are unknown. Thus, a natural approach to solving for the estimates of disparity, divergence, and variance is to employ a distributed gradient descent method. This requires the exchange of the gradient terms during each communication round. For the ease of notation let, $\mathcal{F}_b(\bm{\phi}_b^t) := \mathcal{{Q}}_{\bm{\sigma}_b^t, \bm{\rho}_b^t, \bm{\upsilon}_b^t}(\bm{\phi}_b^t)$ be the function whose gradient is calculated in each communication round. In each communication round, say the $t$-th round, each BS computes a local estimate using its local dataset $D_b^t$. Let $g_b^t$ denote the local estimate at the $b$-th BS at the $t$-th time. Thus we have the following:
	\begin{equation}
		g_b^t = \nabla_{\bm{\phi}_b^t} \mathcal{{F}}(\bm{\phi}_b^t), 
	\end{equation}
	where $\nabla_{\bm{\phi}_b^t}$ represents the gradient operator with respect to $\bm{\phi}_b^t$. For large models, this step is likely to be the bottleneck of the algorithm due to multiple factors \cite{konecny16}. Thus, a naive implementation of the FL using SGD would require repeated exchanges of gradients of the losses, which leads to relatively large radio resource requirements. The incurring communication overhead can be overcome by compressing the gradient information before being transmitted. Thus, inspired by the signSGD, we employ one-bit quantization of local gradient estimates by taking the element-wise signs of the local gradient parameters \cite{bernstein18}:
	\begin{equation}
		\hat{g}_b^t = \textrm{sign}(g_b^t),  \hspace{0.7cm}\forall \hspace{0.2cm}b,t.
	\end{equation}
	The above one-bit quantized gradient is then broadcast to the neighboring BSs, and so each BS uses such a quantized one-bit gradient to update its current estimate using the gradient descent method based on the following equation:
	\begin{eqnarray}
		\bm{\phi} &=& \bm{\phi}_b^{t} - \eta \nabla_{\bm{\phi}_b^t, \rm{sign}}\mathcal{{Q}}_{\bm{\sigma}_b^t, \bm{\rho}_b^t, \bm{\upsilon}_b^t}(\bm{\phi}_b^t) \nonumber\\
		\bm{\phi}_b^{t+1} &=& \argmin_{x \in \Delta_N}|| x - \bm{\phi}||_1
	\end{eqnarray}
	Note that the gradient step is followed by the projection step as $\bm{\phi}_b^t$ should satisfy the cache constraint in \eqref{eq:the_mod_problem}. Such a process shows how the the communication cost of the DP-FL algorithm on the same par as to the signSGD. The next section analyzes the convergence of Algorithm~\ref{alg:depe_fl_algo}.
	\subsection{Convergence}
	This subsection presents the convergence analysis of the proposed DP-FL algorithm, i.e., Algorithm~\ref{alg:depe_fl_algo}. In order for the Algorithm to converge it is sufficient to show that the gradient of the function $\mathcal{{Q}}_{\bm{\sigma}_b^t, \bm{\rho}_b^t,\bm{\upsilon}_b^t}(\bm{\phi}_b^t)$ with respect to $\bm{\phi}_b^t$, $\bm{\sigma}_b^t$, $\bm{\rho}_b^t$, and $\bm{\upsilon}_b^t$ converges. To establish one of the main convergence results, the following standard assumptions are made about the regret function, $\texttt{Reg}(\bm{\phi}_{b}^{t})$, similar to~\cite{bernstein18}:
	\begin{assmon} For all $\bm{\phi}_b^t$, $\sigma_b^t$, $\rho_b^t$, and $\upsilon_b^t$, $\texttt{Reg}(\bm{\phi}_{b}^{t})$ $\geq$ $\texttt{Reg}^*$, where $\texttt{Reg}^*$ is the optimal regret.
	\end{assmon}
	\begin{assmon} $\beta$-smoothness: Let $\nabla\texttt{Reg}(\bm{\phi}_b^t)$ denote the gradient of the objective function $\texttt{Reg}(\cdot)$ evaluated at point $\bm{\phi}_b^t$. Then $\forall$ $\bm{\phi}_1^t, \bm{\phi}_2^t$, we require that for some non-negative constant $L:=[L_1,\ldots,L_d]$
		\begin{eqnarray}
			\bigg|\texttt{Reg}(\bm{\phi}_1^t) -[\texttt{Reg}(\bm{\phi}_2^t) + \nabla\texttt{Reg}(\bm{\phi}_{1}^t)^T(\bm{\phi}_1^t-\bm{\phi}_2^t)]\bigg|  
			\leq \frac{1}{2}\sum_iL_i(\bm{\phi}_{1i}^t -\bm{\phi}_{2i}^t)^2 \nonumber
		\end{eqnarray}
	\end{assmon}
	\begin{assmon}: The stochastic gradient gives an independent unbiased estimate $\hat{g}_{uv}$, for $u,v = 1,2,\ldots, N$ that has coordinate bounded variance:
		\begin{equation}
			\mathbb{E}[\hat{g}_{uv}] = g_{uv}, \hspace{1cm} \mathbb{E}[\hat{g}_{uv} - {g}_{uv}] \leq \omega_{uv}^2 \nonumber
		\end{equation}
	\end{assmon}
	\begin{assmon}: The function $\texttt{Reg}(\bm{\phi}_{b}^{t})$ is Lipschitz in $\sigma_b^t$ with Lipschitz constant $\delta$.
	\end{assmon}
	\begin{assmon}: The function $\texttt{Reg}(\bm{\phi}_{b}^{t})$ is Lipschitz in $\rho_b^t$ with Lipschitz constant $\lambda$.
	\end{assmon}
	\begin{assmon}: The function $\texttt{Reg}(\bm{\phi}_{b}^{t})$ is Lipschitz in $\upsilon_b^t$ with Lipschitz constant $\zeta$.
	\end{assmon}
	\begin{defn} (Projected Gradient) Let $f: \mathcal{K} \rightarrow \mathbb{R}$ be a differentiable function on a closed (but not necessarily bounded) convex set $\mathcal{K} \subseteq \mathbb{R}^n$. Define $\nabla_{\mathcal{K}, \pi}f: \mathcal{K} \rightarrow \mathbb{R}^n$, the $(\mathcal{K}, \pi)$-projected gradient of $f$, by
		\begin{equation}
			\nabla_{\mathcal{K}, \pi}f(x) = \pi\big( x - \Pi_{\mathcal{K}}[x - \pi\nabla_f(x)] \big)
		\end{equation}
		where $\pi > 0$ and $\Pi_{\mathcal{K}}$ is the orthogonal projection onto $\mathcal{K}$.
	\end{defn}
	The following results (Lemma~\ref{lm:martingale} and Lemma~\ref{lm:gradient}) from~\cite{chung06} and~\cite{hazan17} prove to be useful in proving the convergence of the proposed algorithm and, hence, are presented in the following lemma for future reference.
	
	\begin{lemma} \label{lm:martingale}
		Let $\bm{Y}^t_b$ be a process of fetching FoV from the BS $b$ in each time slot $t$. The sequence $\bm{Y}_b^t$ is a super martingale with a bounded difference, i.e. $\bm{Y}^t_b - \bm{Y}^{t-1}_b \leq C$ as follows \cite{chung06}:
		\begin{equation}
			\mathbb{P}[\bm{Y}^T_b \geq \lambda]  \leq \exp\bigg({\frac{-\lambda^2}{2TC^2}}\bigg),
		\end{equation}
		which is one generalized version of Azuma’s inequality to supermartingales.
	\end{lemma}
	
	\begin{lemma} \label{lm:gradient}
		Let $\mathcal{K} \in \mathbb{R}^n$ be a closed convex set, and let $\eta > 0$. Suppose $f: \mathcal{K} \rightarrow \mathbb{R}$ is differentiable. Then, the following inequality holds for any $x \in \mathbb{R}$ ~\cite{hazan17}:
		\begin{equation}
			\langle \nabla f(x), \nabla_{\mathcal{K},\eta} f(x)\rangle \geq ||\nabla_{\mathcal{K},\eta} f(x)||^2.
		\end{equation}
	\end{lemma}
	\begin{theorem}\label{thm:guarant}
		After $T$ iterations, choosing the learning rates $\eta_t= \frac{1}{\sqrt{T}}$ , $\mu_t = \frac{1}{\sqrt{T}}$, $\nu_t = \frac{1}{\sqrt{T}}$, $\iota = \frac{1}{\sqrt{T}}$, and the batch size $ \theta^t = T$, the following holds:
		\begin{equation}
			\begin{aligned}
				\mathbb{E}\bigg[\frac{1}{T}\sum_{t=0}^{T-1}\Delta^t_b \bigg]  \leq \frac{1}{\sqrt{T}} \bigg( 2\sum_{m=1}^B|\omega_{bm}||_1 + \frac{||L||_1}{2} + \mathcal{Q}^{*} - \mathcal{Q}_{\bm{\rho}_b^0, \bm{\sigma}_b^0,\bm{\upsilon}_b^0}(\bm{\phi}_{b}^0)   \bigg),  
			\end{aligned}
		\end{equation}
		where $\Delta^t_b = \bigg(\sum_{m=1}^{B}||g_{mb}^t||_1 +
		\big(1 - \frac{\delta}{2\sqrt{T}}\big)||\nabla_{\mathcal{K},\bm{\sigma}_b^t}\mathcal{{Q}}_{ \bm{\rho}_b^{t+1},\bm{\sigma}_b^{t+1},\bm{\upsilon}_b^t}(\bm{\phi}_b^t)  ||_2^2 
		\\+ \big(1 - \frac{\lambda}{2\sqrt{T}}\big)||\nabla_{\mathcal{K}, \bm{\rho}_b^t }\mathcal{{Q}}_{ \bm{\rho}_b^t,\bm{\sigma}_b^t,\bm{\upsilon}_b^t}(\bm{\phi}_b^t) ||_2^2  
		+ \big(1 - \frac{\zeta}{2\sqrt{T}}\big)||\nabla_{\mathcal{K}, \bm{\upsilon}_b^t }\mathcal{{Q}}_{ \bm{\rho}_b^t,\bm{\sigma}_b^t,\bm{\upsilon}_b^t}(\bm{\phi}_b^t)  ||_2^2  
		\bigg) $.
	\end{theorem}
	\begin{IEEEproof}\normalfont
		See Appendix \ref{appendix:proof_convg_cachingvr}.
	\end{IEEEproof}  
	\vspace{0.5cm}
	It is observed that as $T \rightarrow \infty$, the right-hand side goes to zero by appropriately choosing the right-hand side terms to be arbitrarily small, compared to $T$. This shows that the rate of convergence is $\mathcal{O}(1/\sqrt{T})$ similar to~\cite{bernstein18}.
	
	It is important to note that in the optimization problem in~\eqref{eq:the_problem}, no assumption has been made on the channel conditions. This paper initially formulates the optimization problem in~\eqref{eq:the_problem} while assuming that the transmission of VR tile requests is instantaneous, and so there is no delay associated with sending these requests from VR users to the BSs. While these assumptions simplify the initial problem, they are not representative of real-world scenarios. Moreover, in scenarios where the user-to-base station connection is unknown, an outage may occur, preventing the requested FoV from being delivered. To address such a shortcoming, the paper introduces a more realistic model accounting for the communication model between BSs and VR users. Thus, the optimization problem in~\eqref{eq:the_problem} is modified to incorporate the delay constraint associated with transmitting VR tiles. In addition to incorporating channel and delay considerations, the paper further expands its model by exploring the joint streaming of VR tiles from BSs. This approach considers the spatial and content correlations between VR users' requests, allowing for more efficient broadcasting of VR tiles to target user groups.
	\subsection{Delay-aware Caching Optimization}
	This subsection presents the updated caching algorithm, incorporating delay constraints and enabling joint streaming of VR tiles from multiple BSs. Let the maximum VR interaction latency requirement for the $f$-th tile requested by the $i$-th user at the $b$-th BS be denoted by $T_{b,i,f}(th)$. A rendered FoV is then considered successfully delivered to a VR device if the actual interaction latency satisfies $T_{b,i,f} < T_{b,i,f}(th)$. The VR interaction latency consists of two main components: (i) the time taken to render the requested FoV at the BS, and (ii) the time required to transmit the rendered FoV from the BS to the VR user and can be written as %\begin{eqnarray}
	$T_{b,i,f} =  T_{b,i,f}(r) + T_{b,i,f}(c)$, 
	%\nonumber\end{eqnarray}
where $c \in \{w, fl\}$ is the transmit time of the FoV, where $c$ can be either $w$ (multicast/unicast to VR users) or $fl$ (fetched from the server). Denoting the execution capability of the GPU at the $b$-th BS as $F_{b,i,f}$ and the number of cycles needed to process one bit of input data as $f_{b,i,f}$, the rendering time can be expressed as
$T_{b,i,f}(r) = {f_{b,i,f}}/{F_{b,i,f}}$. To calculate $T_{b,i,f}(c)$, the transmission model is introduced first. The multicast transmission rate between the $i$-th VR user in the $j$-th multicast group and the $b$-th BS at the $t$-th time slot can be expressed as
\begin{equation}
	R_{b,i,f}^{t}(m) = \log_2\bigg(1 + \frac{|{(\bm{h}_{b,ij,f}^t)}^H\bm{v}_{b,j}^{t}|^2}{\bm{I}_{ij,b}^{t} + {\sigma^t_{ij}}^2} \bigg),
\end{equation}
where $\bm{I}_{ij,b}^{t} = \sum_{\mathcal{G}^{m}_{b^{'}s} \in \mathcal{G}^{m}_b/\mathcal{G}^{m}_{bs} } |{(\bm{h}_{b^{'},ij,s}^t)}^H\bm{v}_{b^{'},s}^{t}|^2$. 
Similarly, the unicast transmission rate between the $i$-th VR user in the $l$-th unicast group and the $b$-th BS for the $f$-th tile at the $t$-th time slot can be expressed as
\begin{equation}
	R_{b,i,f}^{t}(u) = \log_2\bigg(1 + \frac{|{(\bm{g}_{b,il,f}^t)}^H\bm{v}_{b,l}^{t}|^2}{\bm{I}_{il,b}^{t} + {\sigma^t_{il}}^2} \bigg),
\end{equation}
where $\bm{I}_{il,b}^{t} = \sum_{\mathcal{G}^{u}_{b^{'}s} \in \mathcal{G}^{u}_b/\mathcal{G}^{u}_{bs} } |{(\bm{g}_{b^{'},ij,s}^t)}^H\bm{v}_{b^{'},s}^{t}|^2 $.
Further, to calculate $T_{b, i,f}(w)$, we denote the size of the FoV to be transmitted as $C$. Usually, the FoV has to be compressed before downlink transmission. By assuming the compression ratio as $C^R_{b,i,f}$, the size of the compressed data for downlink transmission can be calculated as $C/C^R_{b,i,f}$~\cite{liu21}.
Thus, the $T_{b,i,f}(w)$ becomes $T_{b,i,f}(w) = C/C^R_{b,i,f} R_{b,i,f}^t(w)$, where $R_{b,i,f}^t(w) \in \{R^{t}_{b,i,f}(m), R^{t}_{b,i,f}(u) \}$.
Whenever the FoV is not cached at the BS, it is fetched from the server, and the delay is given by $T_{b,i,f}(fl) = {C}/{C^R_{b,i,f} R^{t}_{b,i,f}(fl)}$. Thus the total delay $T_{b,i,f}$ is given by 
\begin{equation}
	T_{b,i,f} = \frac{f_{b,i,f}}{F_{b,i,f}} + \frac{C}{C^R_{b,i,f} R^{t}_{b,i,f}(c)},   
\end{equation}
where $c \in \{w, fl\}$ can be either $w$ (multicast/unicast to VR users) or $fl$ (fetched from the server). Thus, the modified optimization problem can be written as follows:
\begin{equation} 
	\begin{aligned}
		&\max_{\bm{\phi}, \sigma,\rho, \upsilon} {\sum_{b}\mathbb{E}\{\mathcal{Q}_{\bm{\rho}_b^t,\bm{\sigma}_b^t,\bm{\upsilon}_b^t}
			(\bm{\phi}_{b}^T)^{avg}|D_{b}^{t-1} \}} \\
		&\text{s. t.} \sum_f \sum_i \phi_{b,i,f}^t  \leq C_b, \\
		&\sum_{i,f}{\phi_{b,i,f}^t}T_{b,i,f}(w) + (1 - \phi_{b,i,f}^t)\bigg(T_{b,i,f}(w) + T_{b,i,f}(fl)\bigg) \leq \sum_{i,f} T_{b,i,f}(th), 
	\end{aligned}
	\label{eq:the_problem_vr}
\end{equation}
where the first constraint enforces the storage capacity constraint at the BS, while the second constraint ensures adherence to the delay requirement.

The steps for solving the optimization problem in \eqref{eq:the_problem_vr} mirror those of Algorithm~\ref{alg:depe_fl_algo} with an additional delay constraint. That is, instead of solving \eqref{eq:regret}, in the first step, the following optimization problem is solved:
\begin{equation}
	\begin{aligned}
		&\min_{\sum {\phi}_{b,i,f}^t \leq C_b} \left[\sup_{(\bm{\phi}_{b}^t)^{*}}\sum_{t=T-\tau + 1}^T \mathcal{Q}({\bm{\phi}_{b}^t}^{*}) -\sum_{t=T-\tau + 1}^T \mathcal{Q}(\bm{\phi}_{b}^t)\right] \\
		&\sum_{i,f}{\phi_{b,i,f}^t}T_{b,i,f}(w) + (1 - \phi_{b,i,f}^t)\big(T_{b,i,f}(w) + T_{b,i,f}(fl)\big) \leq \sum_{i,f} T_{b,i,f}(th),   
	\end{aligned}
	\label{eq:min_regret_vr}
\end{equation}
The remaining steps of the modified Algorithm are provided in Algorithm~\ref{alg:depe_fl_algo_vr} at the top of the page.
\begin{algorithm}[H]
	\captionof{algorithm}{Delay-aware DP-FL Algorithm}%
	\label{alg:depe_fl_algo_vr}
	\begin{algorithmic}[1]
		\Procedure{Proposed Delay-aware DP-FL}{}
		\State \texttt{Initialize} $\bm{\sigma}_b^0$, $\bm{\rho}_b^0$, and $\bm{\upsilon}_b^0$ for $b = 1, \ldots, B$
		\For {$t = 1,2, \ldots, T$}
		\State \text{Run regret minimization as in~\eqref{eq:min_regret_vr}}
		\text{\hspace{1cm} to get a sequence of $(\bm{\phi}_b^t)^R$}, $\forall t$
		\For{ $\forall$ BS $\forall k = 1, \ldots, B$}
		\State \text{$ \hat{g}_{t}$ $\gets$ stochastic gradient}
		\EndFor       
		\State \text{Call \textbf{Subroutine} \big(($\bm{\phi}_b^t)^R$, $\bm{\sigma}_b^t$, $\bm{\rho}_b^t$, $\bm{\upsilon}_b^t$, $\forall$  $b$\big)}
		\text{\hspace{1cm}to get $\bm{\phi}_b^{t+1}$.}
		\EndFor
		\EndProcedure
	\end{algorithmic}
\end{algorithm}

Similar to Theorem~\ref{thm:guarant}, the convergence guarantees for the delay-aware DP-FL algorithm are derived in Theorem~\ref{thm:guarant_delay}.

\begin{theorem}  \label{thm:guarant_delay}
	For the delay-aware caching scenario after $T$ iterations, choosing the learning rates $\eta_t= \frac{1}{\sqrt{T}}$ , $\mu_t = \frac{1}{\sqrt{T}}$, $\nu_t = \frac{1}{\sqrt{T}}$, $\iota_t = \frac{1}{\sqrt{T}}$, and the batch size $ \theta^t = T$, the following holds
	\begin{equation}
		\begin{aligned}
			& \mathbb{E}\bigg[\frac{1}{T}\sum_{t=0}^{T-1}\Delta^t_b \bigg] \leq \frac{D_{\max}}{T}\bigg( 1 - \frac{\lambda^2}{2TC^2}\bigg) \\
			& + \frac{1}{\sqrt{T}} \bigg( 2\sum_{m=1}^B||\omega_{bm}||_1 + \frac{||L||_1}{2} + \mathcal{Q}^{*} -\mathcal{Q}_{\bm{\rho}_b^0, \bm{\sigma}_b^0,\bm{\upsilon}_b^0}(\bm{\phi}_{b}^0)  \bigg),
		\end{aligned}
	\end{equation}
	where $\Delta^t_b = \bigg(\sum_{m=1}^{B}||g_{mb}^t||_1  +
	\big(1 - \frac{\delta}{2\sqrt{T}}\big)||\nabla_{\mathcal{K},\bm{\sigma}_b^t}\mathcal{{Q}}_{\bm{\rho}_b^{t+1},\bm{\sigma}_b^t,\bm{\upsilon}_b^t}(\bm{\phi}_b^t) ||_2^2  
	\\+ \big(1 - \frac{\lambda}{2\sqrt{T}}\big)||\nabla_{\mathcal{K}, \bm{\rho}_b^t }\mathcal{{Q}}_{\bm{\rho}_b^t,\bm{\sigma}_b^t,
		\bm{\upsilon}_b^t}(\bm{\phi}_b^t) ||_2^2 
	+ \big(1 - \frac{\zeta}{2\sqrt{T}}\big)||\nabla_{\mathcal{K}, \bm{\upsilon}_b^t }\mathcal{{Q}}_{\bm{\rho}_b^t,\bm{\sigma}_b^t,\bm{\upsilon}_b^t}(\bm{\phi}_b^t) ||_2^2 \bigg)$ 
	and $D_{max}$ is the maximum bounded value $\sum_{t=0}^{T-1} \Delta_b^T$ can take.
\end{theorem} 

As in the delay-unconstrained case of Theorem \ref{thm:guarant}, it is again observed here that as $T \rightarrow \infty$, the right-hand side goes to zero. Thus, the convergence rate is of the order of $\mathcal{O}(1/{T})$. 
\begin{IEEEproof}\normalfont
	See Appendix \ref{appendix:proof_convg_cachingvr_delay}.
\end{IEEEproof}  
\section{Simulation Results}
\label{sec:sim_res}
\begin{table}[t!]
	\centering
	\caption{Summary of simulation parameters.}
	\begin{tabular}{ | c | c| c | c | c |} 
		\hline
		\multicolumn{5}{|c|}{\textbf{Parameters for Fig.~~\ref{fig:com},~\ref{fig:delay}, ~\ref{fig:comp_bs}, ~\ref{fig:delay_bs}, ~\ref{fig:convg}\&~\ref{fig:tile_size}}} \\ 
		\hline
		$\eta_t= \frac{1}{\sqrt{T}}$  & $\mu_t = \frac{1}{\sqrt{T}}$ & $\nu_t = \frac{1}{\sqrt{T}}$ &
		$\iota =  \frac{1}{\sqrt{T}}$  & $ \theta^t = T$ \\ 
		\hline
	\end{tabular}
	\label{table:1}
\end{table}
\subsection{Data sets:} To validate the performance of the proposed algorithms, we used two different datasets that are widely used in the literature, as follows: 
\begin{itemize}
	\item Dataset 1: The dataset comprises tracked head movements of $50$ users while watching a catalog of $10$ high definition $360^{\circ}$ YouTube videos from~\cite{lo17}. For each video, the dataset includes $1800$ samples for every user, the length of the video is $60$s, and the FoV is $100^{\circ} \times 100^{\circ}$. 
	\item Dataset 2: The dataset contains real head movement patterns of $48$ unique VR users viewing $18$ long-duration videos using an HTC Vive headset~\cite{wu17dataset}.
	%The dataset contains data collected from $59$ users, each watching five $70$ s long $360^\circ$ videos and each user has at least $3000$ samples~\cite{xavier17}.
\end{itemize}
To build the tiled-FoV, the equirectangular projection of each of the video frames is divided into $ N \times P$ tiles. Without loss of generality, it is assumed that mobile users can connect to different BSs at different times and request the FoV. The values for the learning rates are shown in Table~\ref{table:1}. The caching strategy is updated at each time slot $t$. Hence, we assume that the VR video frame prediction horizon is one video frame.
%chosen as $\eta_t= \frac{1}{\sqrt{T}}$ , $\mu_t = \frac{1}{\sqrt{T}}$, $\nu_t = \frac{1}{\sqrt{T}}$, $\iota =  \frac{1}{\sqrt{T}}$ and the batch size $ \theta^t = T$ as .
\begin{figure*}[t]
	\begin{minipage}{0.33\linewidth}\centering
		\resizebox{1.0\columnwidth}{!}{\begin{tikzpicture}[thick,scale=1, every node/.style={scale=1.3},font=\Huge]
\definecolor{mycolor1}{rgb}{0.92900,0.69400,0.12500}%
\definecolor{mycolor2}{rgb}{0.49400,0.18400,0.55600}%
\definecolor{mycolor3}{rgb}{0.46600,0.67400,0.18800}%
\definecolor{mycolor4}{rgb}{0.30100,0.74500,0.93300}%
\definecolor{mycolor5}{rgb}{0.63500,0.07800,0.18400}%
\definecolor{mycolor6}{rgb}{1.00000,0.00000,1.00000}%
\definecolor{mycolor7}{rgb}{0.00000,0.00000,1.00000}%

\begin{axis}[%
width=9.2in,
height=7.9in,
at={(1.06in,0.651in)},
scale only axis,
xmin=10,
xmax=25,
xtick={10, 15,..., 25},
xlabel style={font=\color{white!15!black}},
xlabel={\Huge{Cache size}},
ymin=10,
ymax=50,
ytick={10,15,..., 50},
ylabel style={font=\color{white!15!black}},
ylabel={\Huge{Average cache hit}},
axis background/.style={fill=white},
xmajorgrids,
ymajorgrids,
legend style={legend cell align=left, align=left, draw=white!15!black}
]
\addplot [color=red, line width=2.0pt, mark=asterisk, mark size=8.0pt, mark options={solid, red}]
  table[row sep=crcr]{%
10	34.5953117647059\\
15	42.2070941176471\\
20	46.9924176470588\\
25	49.9736823529412\\
};
\addlegendentry{Algo -1}

\addplot [color=black, dashed, line width=2.0pt, mark=+, mark size=8.0pt, mark options={solid, black}]
  table[row sep=crcr]{%
10	34.2011764705883\\
15	41.5079588235294\\
20	46.2376823529412\\
25	49.5837529411765\\
};
\addlegendentry{Algo -2}

\addplot [color=mycolor1, line width=2.0pt, mark=o, mark size=6.0pt, mark options={solid, mycolor1}]
  table[row sep=crcr]{%
10	21.6412926470588\\
15	28.4252911764706\\
20	34.099275\\
25	38.9459226470588\\
};
\addlegendentry{sgdalgo1}

\addplot [color=mycolor2, dashed, line width=2.0pt, mark=o, mark size=6.0pt, mark options={solid, mycolor2}]
  table[row sep=crcr]{%
10	21.2315764705882\\
15	27.7613\\
20	33.2497344117647\\
25	37.7932738235294\\
};
\addlegendentry{sgdalgo2}

\addplot [color=mycolor7, dashed, line width=2.0pt, mark=diamond, mark size=8.0pt, mark options={solid, mycolor7}]
  table[row sep=crcr]{%
10	16.4130\\
15	22.5234\\
20	29.4918\\
25	34.1413\\
};
\addlegendentry{\text{ FedAvg}}

\addplot [color=mycolor3, line width=2.0pt, mark=triangle, mark size=8.0pt, mark options={solid, mycolor3}]
  table[row sep=crcr]{%
10	15.3207271176471\\
15	21.7884076470588\\
20	27.6632110588235\\
25	33.0374632352941\\
};
\addlegendentry{$\rho\text{ learning}$}

\addplot [color=mycolor4, dashed, line width=2.0pt, mark=triangle, mark size=8.0pt, mark options={solid, mycolor4}]
  table[row sep=crcr]{%
10	14.4925797058824\\
15	20.6106558823529\\
20	26.1679023529412\\
25	31.2516544117647\\
};
\addlegendentry{$\sigma\text{ learning}$}

\addplot [color=mycolor5, line width=2.0pt, dashed, mark=square, mark size=6.0pt, mark options={solid, mycolor5}]
  table[row sep=crcr]{%
10	10.6152097058824\\
15	15.7575476470588\\
20	20.2460464705882\\
25	24.22805\\
};
\addlegendentry{$\rho{}_\text{1}^\text{t}\text{ = }\rho{}_\text{2}^\text{t}\text{ = 0.5}$}

\addplot [color=mycolor6, line width=2.0pt, mark=square, mark size=6.0pt, mark options={solid, mycolor6}]
  table[row sep=crcr]{%
10	11.6157970588235\\
15	14.9814017647059\\
20	17.6722147058824\\
25	19.84872\\
};
\addlegendentry{$\sigma{}_\text{1}^\text{t}\text{= }\sigma{}_\text{2}^\text{t}\text{ = 0.5}$}

\end{axis}
				\node[above,font=\Huge\bfseries] at (current bounding box.north) {Dataset 1};
		\end{tikzpicture}}
		\caption{{Average cache hit versus cache size.}}
		\label{fig:com}
	\end{minipage}
	\begin{minipage}{0.33\linewidth}\centering
		\resizebox{1.0\columnwidth}{!}{\begin{tikzpicture}[thick,scale=1, every node/.style={scale=1.3},font=\Huge]
\definecolor{mycolor1}{rgb}{1.00000,0.00000,1.00000}%
\definecolor{mycolor2}{rgb}{0.63529,0.07843,0.18431}%
\definecolor{mycolor3}{rgb}{0.30196,0.74510,0.93333}%
\definecolor{mycolor4}{rgb}{0.46667,0.67451,0.18824}%
\definecolor{mycolor5}{rgb}{0.92941,0.69412,0.12549}%
\definecolor{mycolor6}{rgb}{0.49412,0.18431,0.55686}%
\definecolor{mycolor7}{rgb}{0.00000,0.00000,1.00000}%

\begin{axis}[%
width=9.2in,
height=7.9in,
at={(1.06in,0.651in)},
scale only axis,
xmin=10,
xmax=25,
xtick={10,15,...,25},
xlabel style={font=\color{white!15!black}},
xlabel={\Huge{Cache Size}},
ymin=0.02,
ymax=0.055,
ytick={0.02,0.025,...,0.055},
ylabel style={font=\color{white!15!black}},
ylabel={\Huge{Average Delay (s)}},
axis background/.style={fill=white},
xmajorgrids,
ymajorgrids,
legend style={legend cell align=left, align=left, draw=white!15!black}
]
\addplot [color=mycolor1, line width=2.0pt, mark=square, mark size=6.0pt, mark options={solid, mycolor1}]
  table[row sep=crcr]{%
10	0.0527\\
15	0.0518\\
20	0.0509\\
25	0.05\\
};
\addlegendentry{$\sigma{}_\text{1}^\text{t}\text{= }\sigma{}_\text{2}^\text{t}\text{ = 0.5}$}

\addplot [color=mycolor2, dashed, line width=2.0pt, mark=square, mark size=6.0pt, mark options={solid, mycolor2}]
  table[row sep=crcr]{%
10	0.0505\\
15	0.049735834\\
20	0.049735834\\
25	0.0493532816470588\\
};
\addlegendentry{$\rho{}_\text{1}^\text{t}\text{ = }\rho{}_\text{2}^\text{t}\text{ = 0.5}$}

\addplot [color=mycolor3, line width=2.0pt, dashed, mark=triangle, mark size=8.0pt, mark options={solid, mycolor3}]
  table[row sep=crcr]{%
10	0.0485288294117647\\
15	0.0478083823529412\\
20	0.0467624294117647\\
25	0.0458322058823529\\
};
\addlegendentry{$\sigma\text{ learning}$}

\addplot [color=mycolor4, line width=2.0pt, mark=triangle, mark size=8.0pt, mark options={solid, mycolor4}]
  table[row sep=crcr]{%
10	0.0476029411764706\\
15	0.0460835176470588\\
20	0.0444914647058824\\
25	0.0426895470588235\\
};
\addlegendentry{$\rho\text{ learning}$}

\addplot [color=mycolor7, dashed, line width=2.0pt, mark=diamond, mark size=8.0pt, mark options={solid, mycolor7}]
  table[row sep=crcr]{%
10	0.0457824\\
15	0.042951\\
20	0.041459\\
25	0.039245\\
};
\addlegendentry{\text{ Fedavg}}

\addplot [color=mycolor5, line width=2.0pt, mark=o, mark size=6.0pt, mark options={solid, mycolor5}]
  table[row sep=crcr]{%
10	0.0403375592941176\\
15	0.0381331217647059\\
20	0.0360903805294118\\
25	0.0339943507647059\\
};
\addlegendentry{sgdalgo1}

\addplot [color=mycolor6, dashed, line width=2.0pt, mark=o, mark size=6.0pt, mark options={solid, mycolor6}]
  table[row sep=crcr]{%
10	0.0395338575882353\\
15	0.0374372845882353\\
20	0.0354661754117647\\
25	0.0335776948823529\\
};
\addlegendentry{sgdalgo2}

\addplot [color=red, line width=2.0pt, mark=asterisk, mark size=8.0pt, mark options={solid, red}]
  table[row sep=crcr]{%
10	0.0324987235294118\\
15	0.0279997\\
20	0.0245325470588235\\
25	0.0216206941176471\\
};
\addlegendentry{Algo -1}

\addplot [color=black, dashed, line width=2.0pt, mark=+, mark size=8.0pt, mark options={solid, black}]
  table[row sep=crcr]{%
10	0.0322032176470588\\
15	0.0274961705882353\\
20	0.0239089470588235\\
25	0.0211540176470588\\
};
\addlegendentry{Algo -2}

\end{axis}
				\node[above,font=\Huge\bfseries] at (current bounding box.north) {Dataset 1};
		\end{tikzpicture}}
		\caption{Average delay versus cache size.}
		\label{fig:delay}
	\end{minipage}
	\begin{minipage}{0.32\linewidth}\centering
		\resizebox{1.0\columnwidth}{!}{\begin{tikzpicture}[thick,scale=1, every node/.style={scale=1.3},font=\Huge]
\definecolor{mycolor1}{rgb}{0.92900,0.69400,0.12500}%
\definecolor{mycolor2}{rgb}{0.49400,0.18400,0.55600}%
\definecolor{mycolor3}{rgb}{0.46600,0.67400,0.18800}%
\definecolor{mycolor4}{rgb}{0.30100,0.74500,0.93300}%
\definecolor{mycolor5}{rgb}{0.63500,0.07800,0.18400}%
\definecolor{mycolor6}{rgb}{1.00000,0.00000,1.00000}%
\definecolor{mycolor7}{rgb}{0.00000,0.00000,1.00000}%

\begin{axis}[%
width=9.2in,
height=7.9in,
at={(1.06in,0.651in)},
scale only axis,
xmin=3,
xmax=13,
xtick={3,4,...,13},
xlabel style={font=\color{white!15!black}},
xlabel={\Huge{no. of BS}},
ymin=0,
ymax=50,
ytick={0,5,...,50},
ylabel style={font=\color{white!15!black}},
ylabel={\Huge{Average cache hit}},
axis background/.style={fill=white},
xmajorgrids,
ymajorgrids,
legend style={legend cell align=left, align=left, draw=white!15!black}
]
\addplot [color=red, line width=2.0pt, mark=asterisk, mark size=8.0pt, mark options={solid, red}]
  table[row sep=crcr]{%
3	27.0790117647059\\
5	34.6528941176471\\
7	40.9555\\
9	45.8024941176471\\
11	49.3083176470588\\
13	49.3083176470588\\
};
\addlegendentry{Algo -1}

\addplot [color=black, dashed, line width=2.0pt, mark=+, mark size=8.0pt, mark options={solid, black}]
  table[row sep=crcr]{%
3	26.4098816470588\\
5	33.9647016470588\\
7	40.0554708235294\\
9	44.8509451764706\\
11	48.1260821176471\\
13	48.1260821176471\\
};
\addlegendentry{Algo -2}

\addplot [color=mycolor1, line width=2.0pt, mark=o, mark size=6.0pt, mark options={solid, mycolor1}]
  table[row sep=crcr]{%
3	12.8647457647059\\
5	17.2784788235294\\
7	21.4682588235294\\
9	25.5994694117647\\
11	29.6571470588235\\
13	29.6572129411765\\
};
\addlegendentry{sgdalgo1}

\addplot [color=mycolor2, dashed, line width=2.0pt, mark=o, mark size=6.0pt, mark options={solid, mycolor2}]
  table[row sep=crcr]{%
3	12.5487190941176\\
5	16.7528996470588\\
7	20.8271629411765\\
9	24.8584757647059\\
11	28.8640637647059\\
13	28.8640718823529\\
};
\addlegendentry{sgdalgo2}

\addplot [color=mycolor7, dashed, line width=2.0pt, mark=diamond, mark size=8.0pt, mark options={solid, mycolor7}]
  table[row sep=crcr]{%
3	11.5205235\\
5	13.529582\\
7	16.205235\\
9	19.931044\\
11  22.41903\\
13  22.51304\\
};
\addlegendentry{\text{ Fedavg}}

\addplot [color=mycolor3, line width=2.0pt, mark=triangle, mark size=8.0pt,, mark options={solid, mycolor3}]
  table[row sep=crcr]{%
3	8.93178352941177\\
5	11.8739823529412\\
7	14.7899411764706\\
9	17.6914705882353\\
11	20.5649470588235\\
13	20.947058823554\\
};
\addlegendentry{$\rho\text{ learning}$}

\addplot [color=mycolor4, dashed, line width=2.0pt, mark=triangle, mark size=8.0pt, mark options={solid, mycolor4}]
  table[row sep=crcr]{%
3	5.70967588235294\\
5	9.40338647058823\\
7	12.9902529411765\\
9	16.3891058823529\\
11	21.0342411764706\\
13	20.6647941176471\\
};
\addlegendentry{$\sigma\text{ learning}$}

\addplot [color=mycolor5, line width=2.0pt, dashed, mark=square, mark size=6.0pt, mark options={solid, mycolor5}]
  table[row sep=crcr]{%
3	1.53117411764706\\
5	7.61011882352941\\
7	9.50044470588235\\
9	11.3866235294118\\
11	13.2694352941176\\
13	13.3872352941176\\
};
\addlegendentry{$\rho{}_\text{1}^\text{t}\text{ = }\rho{}_\text{2}^\text{t}\text{ = 0.5}$}

\addplot [color=mycolor6, line width=2.0pt, mark=square, mark size=6.0pt, mark options={solid, mycolor6}]
  table[row sep=crcr]{%
3	3.90435088235294\\
5	5.151895\\
7	6.36549117647059\\
9	7.53271441176471\\
11	8.655955\\
13	8.655955\\
};
\addlegendentry{$\sigma{}_\text{1}^\text{t}\text{ = }\sigma{}_\text{2}^\text{t}\text{ = 0.5}$}

\end{axis}
				\node[above,font=\Huge\bfseries] at (current bounding box.north) {Dataset 1};
		\end{tikzpicture}}
		\caption{Average cache hit versus no. of BS.}
		\label{fig:comp_bs}
	\end{minipage}
\end{figure*}

\begin{figure*}[t]
	\begin{minipage}{0.33\linewidth}\centering
		\resizebox{1.0\columnwidth}{!}{\begin{tikzpicture}[thick,scale=1, every node/.style={scale=1.3},font=\Huge]
\definecolor{mycolor1}{rgb}{0.92900,0.69400,0.12500}%
\definecolor{mycolor2}{rgb}{0.49400,0.18400,0.55600}%
\definecolor{mycolor3}{rgb}{0.46600,0.67400,0.18800}%
\definecolor{mycolor4}{rgb}{0.30100,0.74500,0.93300}%
\definecolor{mycolor5}{rgb}{0.63500,0.07800,0.18400}%
\definecolor{mycolor6}{rgb}{1.00000,0.00000,1.00000}%
\definecolor{mycolor7}{rgb}{0.00000,0.00000,1.00000}%

\begin{axis}[%
width=9.2in,
height=7.9in,
at={(1.06in,0.651in)},
scale only axis,
xmin=10,
xmax=25,
xtick={10, 15,..., 25},
xlabel style={font=\color{white!15!black}},
xlabel={\Huge{Cache size}},
ymin=0,
ymax=45,
ytick={0,5,10,15,..., 45},
ylabel style={font=\color{white!15!black}},
ylabel={\Huge{Average cache hit}},
axis background/.style={fill=white},
xmajorgrids,
ymajorgrids,
legend style={legend cell align=left, align=left, draw=white!15!black}
]
\addplot [color=red, line width=2.0pt, mark=asterisk, mark size=8.0pt, mark options={solid, red}]
  table[row sep=crcr]{%
10	20.9088\\
15	28.3245\\
20	36.0935\\
25	43.7846\\
};
\addlegendentry{Algo -1}

\addplot [color=black, dashed, line width=2.0pt, mark=+, mark size=8.0pt, mark options={solid, black}]
  table[row sep=crcr]{%
10	19.9131\\
15	26.8805\\
20	33.2319\\
25	40.6044\\
};
\addlegendentry{Algo -2}

\addplot [color=mycolor1, line width=2.0pt, mark=o, mark size=6.0pt, mark options={solid, mycolor1}]
  table[row sep=crcr]{%
10	8.79626\\
15	13.1714\\
20	17.5059\\
25	21.7718\\
};
\addlegendentry{sgdalgo1}

\addplot [color=mycolor2, dashed, line width=2.0pt, mark=o, mark size=6.0pt, mark options={solid, mycolor2}]
  table[row sep=crcr]{%
10	8.56732\\
15	12.8221\\
20	17.039\\
25	21.1916\\
};
\addlegendentry{sgdalgo2}

\addplot [color=mycolor3, line width=2.0pt, mark=triangle, mark size=8.0pt, mark options={solid, mycolor3}]
  table[row sep=crcr]{%
10	7.32729\\
15	10.9879\\
20	14.6353\\
25	18.264\\
};
\addlegendentry{$\rho\text{ learning}$}

\addplot [color=mycolor7, dashed, line width=2.0pt, mark=diamond, mark size=8.0pt, mark options={solid, mycolor7}]
  table[row sep=crcr]{%
10	5.836\\
15	11.2459\\
20	16.6192\\
25	18.6937\\
};
\addlegendentry{\text{ FedAvg}}

\addplot [color=mycolor4, dashed, line width=2.0pt, mark=triangle, mark size=8.0pt, mark options={solid, mycolor4}]
  table[row sep=crcr]{%
10	6.93122\\
15	10.394\\
20	13.8442\\
25	17.2768\\
};
\addlegendentry{$\sigma\text{ learning}$}

\addplot [color=mycolor5, line width=2.0pt, dashed, mark=square, mark size=6.0pt, mark options={solid, mycolor5}]
  table[row sep=crcr]{%
10	5.04289\\
15	7.57032\\
20	10.0876\\
25	12.6211\\
};
\addlegendentry{$\rho{}_\text{1}^\text{t}\text{ = }\rho{}_\text{2}^\text{t}\text{ = 0.5}$}

\addplot [color=mycolor6, line width=2.0pt, mark=square, mark size=6.0pt, mark options={solid, mycolor6}]
  table[row sep=crcr]{%
10	4.5504\\
15	6.76188\\
20	8.89734\\
25	10.9278\\
};
\addlegendentry{$\sigma{}_\text{1}^\text{t}\text{= }\sigma{}_\text{2}^\text{t}\text{ = 0.5}$}

\end{axis}
				\node[above,font=\Huge\bfseries] at (current bounding box.north) {Dataset 2};
		\end{tikzpicture}}
		\caption{{Average cache hit versus cache size.}}
		\label{fig:com2}
	\end{minipage}
	\begin{minipage}{0.33\linewidth}\centering
		\resizebox{1.0\columnwidth}{!}{\begin{tikzpicture}[thick,scale=1, every node/.style={scale=1.3},font=\Huge]
\definecolor{mycolor1}{rgb}{1.00000,0.00000,1.00000}%
\definecolor{mycolor2}{rgb}{0.63529,0.07843,0.18431}%
\definecolor{mycolor3}{rgb}{0.30196,0.74510,0.93333}%
\definecolor{mycolor4}{rgb}{0.46667,0.67451,0.18824}%
\definecolor{mycolor5}{rgb}{0.92941,0.69412,0.12549}%
\definecolor{mycolor6}{rgb}{0.49412,0.18431,0.55686}%
\definecolor{mycolor7}{rgb}{0.00000,0.00000,1.00000}%

\begin{axis}[%
width=9.2in,
height=7.9in,
at={(1.06in,0.651in)},
scale only axis,
xmin=10,
xmax=25,
xtick={10,15,...,25},
xlabel style={font=\color{white!15!black}},
xlabel={\Huge{Cache Size}},
ymin=0.025,
ymax=0.065,
ytick={0.025,0.03,...,0.06, 0.065},
ylabel style={font=\color{white!15!black}},
ylabel={\Huge{Average Delay (s)}},
axis background/.style={fill=white},
xmajorgrids,
ymajorgrids,
legend style={legend cell align=left, align=left, draw=white!15!black}
]
\addplot [color=mycolor1, line width=2.0pt, mark=square, mark size=6.0pt, mark options={solid, mycolor1}]
  table[row sep=crcr]{%
10	0.06324\\
15	0.06216\\
20	0.06108\\
25	0.05990\\
};
\addlegendentry{$\sigma{}_\text{1}^\text{t}\text{= }\sigma{}_\text{2}^\text{t}\text{ = 0.5}$}

\addplot [color=mycolor2, dashed, line width=2.0pt, mark=square, mark size=6.0pt, mark options={solid, mycolor2}]
  table[row sep=crcr]{%
10	0.0598949\\
15	0.0587643\\
20	0.0572105\\
25	0.0569328\\
};
\addlegendentry{$\rho{}_\text{1}^\text{t}\text{ = }\rho{}_\text{2}^\text{t}\text{ = 0.5}$}

\addplot [color=mycolor3, line width=2.0pt, dashed, mark=triangle, mark size=8.0pt, mark options={solid, mycolor3}]
  table[row sep=crcr]{%
10	0.0587406\\
15	0.0580425\\
20	0.0575806\\
25	0.0565079\\
};
\addlegendentry{$\sigma\text{ learning}$}

\addplot [color=mycolor7, dashed, line width=2.0pt, mark=diamond, mark size=8.0pt, mark options={solid, mycolor7}]
  table[row sep=crcr]{%
10	0.0535037\\
15	0.0517683\\
20	0.0499474\\
25	0.0475195\\
};
\addlegendentry{\text{ Fedavg}}

\addplot [color=mycolor4, line width=2.0pt, mark=triangle, mark size=8.0pt, mark options={solid, mycolor4}]
  table[row sep=crcr]{%
10	0.0523168\\
15	0.0502386\\
20	0.0482836\\
25	0.0466237\\
};
\addlegendentry{$\rho\text{ learning}$}

\addplot [color=mycolor5, line width=2.0pt, mark=o, mark size=6.0pt, mark options={solid, mycolor5}]
  table[row sep=crcr]{%
10	0.0519342\\
15	0.0491124\\
20	0.0465215\\
25	0.0445579\\
};
\addlegendentry{sgdalgo1}

\addplot [color=mycolor6, dashed, line width=2.0pt, mark=o, mark size=6.0pt, mark options={solid, mycolor6}]
  table[row sep=crcr]{%
10	0.0496849\\
15	0.0470722\\
20	0.0446947\\
25	0.0414032\\
};
\addlegendentry{sgdalgo2}

\addplot [color=red, line width=2.0pt, mark=asterisk, mark size=8.0pt, mark options={solid, red}]
  table[row sep=crcr]{%
10	0.037182\\
15	0.034508\\
20	0.031003\\
25	0.028314\\
};
\addlegendentry{Algo -1}

\addplot [color=black, dashed, line width=2.0pt, mark=+, mark size=8.0pt, mark options={solid, black}]
  table[row sep=crcr]{%
10	0.0347894\\
15	0.0335801\\
20	0.0302194\\
25	0.0273032\\
};
\addlegendentry{Algo -2}

\end{axis}
				\node[above,font=\Huge\bfseries] at (current bounding box.north) {Dataset 2};
		\end{tikzpicture}}
		\caption{Average delay versus cache size.}
		\label{fig:delay2}
	\end{minipage}
	\begin{minipage}{0.32\linewidth}\centering
		\resizebox{1.0\columnwidth}{!}{\begin{tikzpicture}[thick,scale=1, every node/.style={scale=1.3},font=\Huge]
\definecolor{mycolor1}{rgb}{0.92900,0.69400,0.12500}%
\definecolor{mycolor2}{rgb}{0.49400,0.18400,0.55600}%
\definecolor{mycolor3}{rgb}{0.46600,0.67400,0.18800}%
\definecolor{mycolor4}{rgb}{0.30100,0.74500,0.93300}%
\definecolor{mycolor5}{rgb}{0.63500,0.07800,0.18400}%
\definecolor{mycolor6}{rgb}{1.00000,0.00000,1.00000}%
\definecolor{mycolor7}{rgb}{0.00000,0.00000,1.00000}%

\begin{axis}[%
width=9.2in,
height=7.9in,
at={(1.06in,0.651in)},
scale only axis,
xmin=3,
xmax=13,
xtick={3,4,...,13},
xlabel style={font=\color{white!15!black}},
xlabel={\Huge{no. of BS}},
ymin=0,
ymax=45,
ytick={0,5,...,45},
ylabel style={font=\color{white!15!black}},
ylabel={\Huge{Average cache hit}},
axis background/.style={fill=white},
xmajorgrids,
ymajorgrids,
legend style={legend cell align=left, align=left, draw=white!15!black}
]
\addplot [color=red, line width=2.0pt, mark=asterisk, mark size=8.0pt, mark options={solid, red}]
  table[row sep=crcr]{%
3	19.41984\\
5	25.41983 \\
7	30.14309\\
9	35.50954\\
11	39.41842\\
13	43.13284\\
};
\addlegendentry{Algo -1}

\addplot [color=black, dashed, line width=2.0pt, mark=+, mark size=8.0pt, mark options={solid, black}]
  table[row sep=crcr]{%
3	17.2835\\
5	23.6309\\
7	29.5353 \\
9	33.5256 \\
11	38.14985 \\
13	42.24253 \\
};
\addlegendentry{Algo -2}

\addplot [color=mycolor1, line width=2.0pt, mark=o, mark size=6.0pt, mark options={solid, mycolor1}]
  table[row sep=crcr]{%
3	10.4295\\
5	12.5253\\
7	15.2498\\
9	18.1438\\
11	20.9853\\
13	23.4183\\
};
\addlegendentry{sgdalgo1}

\addplot [color=mycolor2, dashed, line width=2.0pt, mark=o, mark size=6.0pt, mark options={solid, mycolor2}]
  table[row sep=crcr]{%
3	9.4246\\
5	10.2732\\
7	13.5231\\
9	16.1319\\
11	19.3120\\
13	22.5195\\
};
\addlegendentry{sgdalgo2}

\addplot [color=mycolor3, line width=2.0pt, mark=triangle, mark size=8.0pt,, mark options={solid, mycolor3}]
  table[row sep=crcr]{%
3	8.52872\\
5	9.14395\\
7	12.5290\\
9	14.5285\\
11	18.2831\\
13	20.41934\\
};
\addlegendentry{$\rho\text{ learning}$}

\addplot [color=mycolor7, dashed, line width=2.0pt, mark=diamond, mark size=8.0pt, mark options={solid, mycolor7}]
  table[row sep=crcr]{%
3	7.1493\\
5	9.2194\\
7	12.459\\
9	15.349\\
11  18.934\\
13  21.4198\\
};
\addlegendentry{\text{ Fedavg}}

\addplot [color=mycolor4, dashed, line width=2.0pt, mark=triangle, mark size=8.0pt, mark options={solid, mycolor4}]
  table[row sep=crcr]{%
3	5.98248\\
5	7.89452\\
7	9.914930\\
9	12.92583\\
11	15.2394\\
13	18.4219\\
};
\addlegendentry{$\sigma\text{ learning}$}

\addplot [color=mycolor5, line width=2.0pt, dashed, mark=square, mark size=6.0pt, mark options={solid, mycolor5}]
  table[row sep=crcr]{%
3	4.1278\\
5	6.6023\\
7	8.1489\\
9	10.5387\\
11	12.5312\\
13	14.23894\\
};
\addlegendentry{$\rho{}_\text{1}^\text{t}\text{ = }\rho{}_\text{2}^\text{t}\text{ = 0.5}$}

\addplot [color=mycolor6, line width=2.0pt, mark=square, mark size=6.0pt, mark options={solid, mycolor6}]
  table[row sep=crcr]{%
3	3.874\\
5	5.9128\\
7	6.9283\\
9	8.4924\\
11	10.4293\\
13	12.5329\\
};
\addlegendentry{$\sigma{}_\text{1}^\text{t}\text{ = }\sigma{}_\text{2}^\text{t}\text{ = 0.5}$}

\end{axis}
				\node[above,font=\Huge\bfseries] at (current bounding box.north) {Dataset 2};
		\end{tikzpicture}}
		\caption{Average cache hit versus no. of BS.}
		\label{fig:comp_bs2}
	\end{minipage}
\end{figure*}

\begin{figure*}[t!]
	\begin{minipage}{0.33\linewidth}\centering
		\resizebox{1\columnwidth}{!}{\begin{tikzpicture}[thick,scale=1, every node/.style={scale=1.3},font=\Huge]
\definecolor{mycolor1}{rgb}{1.00000,0.00000,1.00000}%
\definecolor{mycolor2}{rgb}{0.63529,0.07843,0.18431}%
\definecolor{mycolor3}{rgb}{0.30196,0.74510,0.93333}%
\definecolor{mycolor4}{rgb}{0.46667,0.67451,0.18824}%
\definecolor{mycolor5}{rgb}{0.92941,0.69412,0.12549}%
\definecolor{mycolor6}{rgb}{0.49412,0.18431,0.55686}%
\definecolor{mycolor7}{rgb}{0.00000,0.00000,1.00000}%

\begin{axis}[%
width=9.2in,
height=7.9in,
at={(1.06in,0.651in)},
scale only axis,
xmin=3,
xmax=13,
xtick={3,4,...,13},
xlabel style={font=\color{white!15!black}},
xlabel={\Huge{no. of BS}},
ymin=0,
ymax=0.07,
ytick={0,0.01,...,0.07},
ylabel style={font=\color{white!15!black}},
ylabel={\Huge{Average delay (s)}},
axis background/.style={fill=white},
xmajorgrids,
ymajorgrids,
legend style={legend cell align=left, align=left, draw=white!15!black}
]
\addplot [color=mycolor1, line width=2.0pt, mark=square, mark size=6.0pt, mark options={solid, mycolor1}]
  table[row sep=crcr]{%
3	0.0605310042352941\\
5	0.0538349477647059\\
7	0.0478703887058824\\
9	0.0423984\\
11	0.0373682597647059\\
13	0.0373682597647059\\
};
\addlegendentry{$\sigma{}_\text{1}^\text{t}\text{= }\sigma{}_\text{2}^\text{t}\text{ = 0.5}$}

\addplot [color=mycolor2, line width=2.0pt, dashed, mark=square, mark size=6.0pt, mark options={solid, mycolor2}]
  table[row sep=crcr]{%
3	0.0430063674705882\\
5	0.0418442121176471\\
7	0.0408039250588235\\
9	0.0398383607647059\\
11	0.0367754385882353\\
13	0.0367754385882353\\
};
\addlegendentry{$\rho{}_\text{1}^\text{t}\text{ = }\rho{}_\text{2}^\text{t}\text{ = 0.5}$}

\addplot [color=mycolor3, line width=2.0pt, dashed, mark=triangle, mark size=8.0pt, mark options={solid, mycolor3}]
  table[row sep=crcr]{%
3	0.0400630294117647\\
5	0.0377878294117647\\
7	0.0354945823529412\\
9	0.0334608470588235\\
11	0.0300322058823529\\
13	0.0300322058823529\\
};
\addlegendentry{$\sigma\text{ learning}$}

\addplot [color=mycolor4, line width=2.0pt, mark=triangle, mark size=8.0pt, mark options={solid, mycolor4}]
  table[row sep=crcr]{%
3	0.0386002647058824\\
5	0.0361335176470588\\
7	0.0337105470588235\\
9	0.0314098705882353\\
11	0.0289671\\
13	0.0289671\\
};
\addlegendentry{$\rho\text{ learning}$}

\addplot [color=mycolor7, dashed, line width=2.0pt, mark=diamond, mark size=8.0pt, mark options={solid, mycolor7}]
  table[row sep=crcr]{%
3	0.0359248\\
5	0.033529\\
7	0.030581\\
9	0.029417\\
11  0.026925\\
13  0.026506\\
};
\addlegendentry{\text{ Fedavg}}

\addplot [color=mycolor5, line width=2.0pt, mark size=6.0pt, mark=o, mark options={solid, mycolor5}]
  table[row sep=crcr]{%
3	0.0342734025882353\\
5	0.0318423444705882\\
7	0.0294850138235294\\
9	0.0272253190588235\\
11	0.0249006216176471\\
13	0.0249006216176471\\
};
\addlegendentry{sgdalgo1}

\addplot [color=mycolor6, dashed, line width=2.0pt, mark=o, mark size=6.0pt, mark options={solid, mycolor6}]
  table[row sep=crcr]{%
3	0.0317186682352941\\
5	0.0298527152941176\\
7	0.0280245976470588\\
9	0.0263228564705882\\
11	0.0244818023529412\\
13	0.0244818023529412\\
};
\addlegendentry{sgdalgo2}

\addplot [color=red, line width=2.0pt, mark=asterisk, mark size=8.0pt, mark options={solid, red}]
  table[row sep=crcr]{%
3	0.0227706638823529\\
5	0.0169253851764706\\
7	0.0125394873882353\\
9	0.00928340767058823\\
11	0.00661207330588235\\
13	0.00661207330588235\\
};
\addlegendentry{Algo -1}

\addplot [color=black, dashed, line width=2.0pt, mark=+, mark size=8.0pt, mark options={solid, black}]
  table[row sep=crcr]{%
3	0.0195661117647059\\
5	0.0145813941176471\\
7	0.0107361611764706\\
9	0.00778680529411765\\
11	0.00554808\\
13	0.00554808\\
};
\addlegendentry{Algo -2}

\end{axis}
				\node[above,font=\Huge\bfseries] at (current bounding box.north) {Dataset 1};
		\end{tikzpicture}}
		\caption{Average delay versus no. of BS.}
		\label{fig:delay_bs}
	\end{minipage}
	\begin{minipage}{0.33\linewidth}\centering
		\resizebox{1.0\columnwidth}{!}{\begin{tikzpicture}[thick,scale=1, every node/.style={scale=1.3},font=\Huge]
\definecolor{mycolor1}{rgb}{0.85000,0.32500,0.09800}%

\begin{axis}[%
width=9.2in,
height=7.9in,
at={(1.063in,0.721in)},
scale only axis,
xmin=18,
xmax=200,
xtick={0,20,...,200},
xlabel style={font=\color{white!15!black}},
xlabel={\Huge{Time}},
ymin=0,
ymax=250,
ytick={0,50,...,250},
ylabel style={font=\color{white!15!black}},
ylabel={\Huge{Loss}},
axis background/.style={fill=white},
xmajorgrids,
ymajorgrids,
legend style={legend cell align=left, align=left, draw=white!15!black}
]
\addplot [color=black, line width=1.0pt]
  table[row sep=crcr]{%
1	210\\
2	210\\
3	210\\
4	210\\
5	210\\
6	210\\
7	210\\
8	210\\
9	210\\
10	210\\
11	210\\
12	210\\
13	210\\
14	210\\
15	210\\
16	210\\
17	210\\
18	210\\
19	210\\
20	52.843137254902\\
21	51.3255813953488\\
22	48.5625\\
23	44.5154639175258\\
24	20.4347826086957\\
25	67.4927536231884\\
26	12.0487804878049\\
27	17.0333333333333\\
28	57.7741935483871\\
29	56.1590909090909\\
30	35\\
31	27.1509433962264\\
32	42.1190476190476\\
33	40.7397260273973\\
34	44.1666666666667\\
35	21.296875\\
36	53.8181818181818\\
37	26.5522388059701\\
38	46.7627118644068\\
39	47.9\\
40	31.2\\
41	17.1875\\
42	12.5128205128205\\
43	54.8333333333333\\
44	35.5806451612903\\
45	30\\
46	13.4054054054054\\
47	35.4406779661017\\
48	38.746835443038\\
49	54.2837837837838\\
50	39\\
51	30.2641509433962\\
52	49.8061224489796\\
53	55.36\\
54	40.0961538461538\\
55	48\\
56	36.4705882352941\\
57	36.32\\
58	53.7413793103448\\
59	28.71\\
60	41.4408602150538\\
61	18.4666666666667\\
62	42.9583333333333\\
63	62.1269841269841\\
64	68.7407407407407\\
65	48.4736842105263\\
66	38.1428571428571\\
67	44.5714285714286\\
68	48.3207547169811\\
69	48.5714285714286\\
70	54.2325581395349\\
71	43.5294117647059\\
72	25.5102040816327\\
73	54.325\\
74	27.8157894736842\\
75	47.7439024390244\\
76	42.8131868131868\\
77	39.8\\
78	45.9666666666667\\
79	28.6271186440678\\
80	61.4146341463415\\
81	48.1428571428571\\
82	36.125\\
83	40.2361111111111\\
84	55.4597701149425\\
85	43.5975609756098\\
86	19.25\\
87	41\\
88	29.551724137931\\
89	33.4361702127659\\
90	51.05\\
91	48.2881355932203\\
92	18.15625\\
93	52.7407407407407\\
94	45.1566265060241\\
95	50.54\\
96	60.5081967213115\\
97	51.8333333333333\\
98	59.2222222222222\\
99	51.9574468085106\\
100	50.5471698113208\\
101	53.4047619047619\\
102	53.5\\
103	38.4366197183099\\
104	43.8181818181818\\
105	30.1777777777778\\
106	47.8604651162791\\
107	58.3333333333333\\
108	42.219512195122\\
109	19.741935483871\\
110	24.3181818181818\\
111	49.7605633802817\\
112	30.0714285714286\\
113	42.3023255813953\\
114	44.6046511627907\\
115	43.525641025641\\
116	57.2619047619048\\
117	52.9390243902439\\
118	30.2428571428571\\
119	28.3333333333333\\
120	43.8229166666667\\
121	43.135593220339\\
122	53.3382352941177\\
123	69.1475409836066\\
124	36.8636363636364\\
125	22.6626506024097\\
126	58.4\\
127	48.7173913043478\\
128	34.7594936708861\\
129	12.5483870967742\\
130	56.8225806451613\\
131	22.8854166666667\\
132	35.5454545454546\\
133	22.2407407407407\\
134	45.948717948718\\
135	29.1971830985916\\
136	46.98\\
137	53.0449438202247\\
138	28.640625\\
139	47.4226804123711\\
140	78.948717948718\\
141	47.7916666666667\\
142	42.7592592592593\\
143	14.6521739130435\\
144	57.953488372093\\
145	60.1785714285714\\
146	42.1076923076923\\
147	51.8840579710145\\
148	48.2209302325581\\
149	48.6478873239437\\
150	50.2340425531915\\
151	44.0555555555555\\
152	58.5365853658537\\
153	19.6315789473684\\
154	91.3763440860215\\
155	29.7\\
156	47.9090909090909\\
157	47\\
158	34.6190476190476\\
159	52.5576923076923\\
160	41.6736842105263\\
161	15.7272727272727\\
162	51.3617021276596\\
163	23.2439024390244\\
164	27.2394366197183\\
165	55.2168674698795\\
166	23.1363636363636\\
167	55.8522727272727\\
168	17.1777777777778\\
169	14.1666666666667\\
170	31.134328358209\\
171	21.421875\\
172	30.3809523809524\\
173	25.8846153846154\\
174	39.7179487179487\\
175	30.5\\
176	78.1170212765958\\
177	54.9125\\
178	23.8648648648649\\
179	43.4651162790698\\
180	57.7948717948718\\
181	31.1777777777778\\
182	39.078125\\
183	47.5\\
184	52.390243902439\\
185	48.609756097561\\
186	49.8510638297872\\
187	45.6885245901639\\
188	44.047619047619\\
189	31.05\\
190	8.63333333333333\\
191	39.3387096774194\\
192	37.0422535211268\\
193	2.8125\\
194	12.4179104477612\\
195	51.3384615384615\\
196	62.4545454545454\\
197	25.0694444444445\\
198	16.6944444444445\\
199	52.5588235294118\\
};
\addlegendentry{SGD}

\addplot [color=mycolor1, line width=1.0pt]
  table[row sep=crcr]{%
1	220\\
2	220\\
3	220\\
4	220\\
5	220\\
6	220\\
7	220\\
8	220\\
9	220\\
10	220\\
11	220\\
12	220\\
13	220\\
14	220\\
15	220\\
16	220\\
17	220\\
18	220\\
19	220\\
20	62\\
21	61\\
22	58\\
23	54\\
24	30\\
25	77\\
26	22\\
27	27\\
28	67\\
29	66\\
30	45\\
31	37\\
32	52\\
33	50\\
34	54\\
35	31\\
36	63\\
37	36\\
38	56\\
39	57\\
40	41\\
41	27\\
42	22\\
43	64\\
44	45\\
45	40\\
46	23\\
47	45\\
48	48\\
49	64\\
50	49\\
51	40\\
52	59\\
53	65\\
54	50\\
55	58\\
56	46\\
57	46\\
58	63\\
59	38\\
60	51\\
61	28\\
62	52\\
63	72\\
64	78\\
65	58\\
66	48\\
67	54\\
68	58\\
69	58\\
70	64\\
71	53\\
72	35\\
73	64\\
74	37\\
75	57\\
76	52\\
77	49\\
78	55\\
79	38\\
80	71\\
81	58\\
82	46\\
83	50\\
84	65\\
85	53\\
86	29\\
87	51\\
88	39\\
89	43\\
90	61\\
91	58\\
92	28\\
93	62\\
94	55\\
95	60\\
96	70\\
97	61\\
98	69\\
99	61\\
100	60\\
101	63\\
102	63\\
103	48\\
104	53\\
105	40\\
106	57\\
107	68\\
108	52\\
109	29\\
110	34\\
111	59\\
112	40\\
113	52\\
114	54\\
115	53\\
116	67\\
117	62\\
118	40\\
119	38\\
120	53\\
121	53\\
122	63\\
123	79\\
124	46\\
125	32\\
126	68\\
127	58\\
128	44\\
129	22\\
130	66\\
131	32\\
132	45\\
133	32\\
134	55\\
135	39\\
136	56\\
137	63\\
138	38\\
139	57\\
140	88\\
141	57\\
142	52\\
143	24\\
144	67\\
145	70\\
146	52\\
147	61\\
148	58\\
149	58\\
150	60\\
151	54\\
152	68\\
153	29\\
154	101\\
155	39\\
156	57\\
157	57\\
158	44\\
159	62\\
160	51\\
161	25\\
162	61\\
163	33\\
164	37\\
165	65\\
166	33\\
167	65\\
168	27\\
169	24\\
170	41\\
171	31\\
172	40\\
173	35\\
174	49\\
175	40\\
176	88\\
177	64\\
178	33\\
179	53\\
180	67\\
181	41\\
182	49\\
183	57\\
184	62\\
185	58\\
186	59\\
187	55\\
188	54\\
189	41\\
190	18\\
191	49\\
192	47\\
193	12\\
194	22\\
195	61\\
196	72\\
197	35\\
198	26\\
199	62\\
};
\addlegendentry{signSGD}

\end{axis}
				\node[above,font=\Huge\bfseries] at (current bounding box.north) {Dataset 1};
		\end{tikzpicture}}
		\caption{Average regret versus iterations.}
		\label{fig:convg}
	\end{minipage}
	\begin{minipage}{0.32\linewidth}\centering
		\resizebox{1.0\columnwidth}{!}{\begin{tikzpicture}[thick,scale=1, every node/.style={scale=1.3},font=\Huge]
\definecolor{mycolor1}{rgb}{0.63529,0.07843,0.18431}%
\definecolor{mycolor2}{rgb}{0.46667,0.67451,0.18824}%

\begin{axis}[%
width=9.2in,
height=7.9in,
at={(1.065in,0.72in)},
scale only axis,
bar shift auto,
symbolic x coords={0.05,0.07,0.24,0.95},
enlarge x limits=0.2,
xtick=data,
xlabel style={font=\color{white!15!black}},
xlabel={\Huge{Tile size ratio}},
ymin=0,
ymax=40,
ytick={0,5,...,40},
ylabel style={font=\color{white!15!black}},
ylabel={\Huge{Average cache hit}},
axis background/.style={fill=white},
legend style={legend cell align=left, align=left, draw=white!15!black}
]
\addplot[ybar, bar width = 25.0pt, fill=mycolor1, draw=black, area legend] table[row sep=crcr] {%
0.95	35\\
0.24	35.5\\
0.07	36.4\\
0.05	35.6\\
};
\addplot[forget plot, color=white!15!black] table[row sep=crcr] {%
0.05	0\\
0.95	0\\
};
\addlegendentry{Algo -1}

\addplot[ybar, bar width = 25.0pt, fill=mycolor2, draw=black, area legend] table[row sep=crcr] {%
0.95	34\\
0.24	34.2\\
0.07	34.5\\
0.05	34.1\\
};
\addplot[forget plot, color=white!15!black] table[row sep=crcr] {%
0.05	0\\
0.95	0\\
};
\addlegendentry{Algo -2}

\end{axis}
				\node[above,font=\Huge\bfseries] at (current bounding box.north) {Dataset 1};
		\end{tikzpicture}}
		\caption{Average cache hit versus tile size ratio.}
		\label{fig:tile_size}
	\end{minipage}
\end{figure*}

%\begin{figure*}[t]
%	\centering
%	\begin{minipage}{.5\columnwidth}
	%		\centering
	%		\includegraphics[width=0.8\textwidth]{comp_new.eps}
	%		\caption{\scriptsize{Average cache hit versus cache size.}}
	%		\label{fig:com}
	%	\end{minipage}%
%	\begin{minipage}{.48\columnwidth}
	%		\centering
	%		\includegraphics[width=0.8\textwidth]{delay_new1.eps}
	%		\caption{\scriptsize{Average delay versus cache size.}}
	%		\label{fig:delay}
	%	\end{minipage}
%	\begin{minipage}{.5\columnwidth}
	%		\centering
	%		\includegraphics[width=0.8\textwidth]{comp_bs.eps}
	%		\caption{\scriptsize{Average cache hit versus no. of BS.}}
	%		\label{fig:comp_bs}
	%	\end{minipage}%
%	\begin{minipage}{.5\columnwidth}
	%		\centering
	%		\includegraphics[width=0.8\textwidth]{delay_bs.eps}
	%		\caption{\scriptsize{Average delay versus no. of BS.}}
	%		\label{fig:delay_bs}
	%	\end{minipage}
%\end{figure*}

In the following figures, Algorithm~\ref{alg:depe_fl_algo} corresponds to the solution of the optimization problem in~\eqref{eq:the_problem} and Algorithm~\ref{alg:depe_fl_algo_vr} corresponds to the solution of the optimization problem in~\eqref{eq:the_problem_vr}. To understand the importance of past demands and the neighboring BSs and users' demands, it is important to compare the proposed scheme under various conditions. In particular, the proposed DP FL-based algorithm is compared with the following benchmark methods:
\begin{itemize}
	\item sgdalgo1 - using a stochastic gradient descent for Algorithm~\ref{alg:depe_fl_algo}.
	\item sgdalgo2 - using a stochastic gradient descent for Algorithm~\ref{alg:depe_fl_algo_vr}.
	\item $\rho$ learning - when the $\sigma$ and $\upsilon$ are kept constant, i.e. $\sigma = 1/T$, and $\upsilon = 1/U$ and only $\rho$ is learned.
	\item $\sigma$ learning - when $\rho$ and $\upsilon$ is kept constant, i.e. $\rho = 1/B$, and $\upsilon = 1/U$ and only $\sigma$ is learned.
	\item when only one nearest BS is connected, i.e. $\rho_1^t = \rho_2^t = 0.5$.
	\item when only the last time slot's caching strategy is considered, i.e. $\sigma_1^t = \sigma_2^t = 0.5$.
	\item Conventional FedAvg: The central node gathers the gradients from each BS's local loss to train a global model that minimizes the overall loss across all BSs.
\end{itemize}

%(a) sgdalgo1 - using stochastic gradient descent for Algorithm~\ref{alg:depe_fl_algo}, (b) sgdalgo2 - using a stochastic gradient for Algorithm~\ref{alg:depe_fl_algo_vr}, (c) $\rho$ learning - when the $\sigma$ and $\upsilon$ are kept constant, i.e. $\sigma = 1/T$, and $\upsilon = 1/U$ and only $\rho$ is learned, (d) $\sigma$ learning - when $\rho$ and $\upsilon$ is kept constant, i.e. $\rho = 1/B$, and $\upsilon = 1/U$ and only $\sigma$ is learned, (e) when only one nearest BS is connected, i.e. $\rho_1^t = \rho_2^t = 0.5$, (f) when only the last time slot's caching strategy is considered, i.e. $\sigma_1^t = \sigma_2^t = 0.5$, and {\color{blue}(g) Conventional FedAvg: The central node gathers the gradients from each BS's local loss to train a global model that minimizes the overall loss across all BSs.} 

Fig.~\ref{fig:com} illustrates the impact of the cache size of the BS on the average cache hit. As depicted in the figure, a clear trend emerges, indicating that as the cache size of the BS increases, the average cache hit naturally increases. It is clear from Fig.~\ref{fig:com} that the proposed algorithms (both algorithm 1 and 2) perform better than sgdalgo1, sgdalgo2, FedAvg, $\rho$ learning, $\sigma$ learning when $\rho_1^t = \rho_2^t = 0.5$, and as well as when $\sigma_1^t = \sigma_2^t = 0.5$, demonstrating the benefit of using the proposed scheme. For instance, the average cache hit for both algorithms is at least $38$ \% higher as compared to the benchmark algorithms. This is attributed to the proposed algorithms' quick adaptation to the dynamic FoV request pattern.

Fig.~\ref{fig:delay} illustrates the inverse relationship between the BS cache size and average delay. This decrease in delay can be attributed to the concurrent reduction in the fronthaul load, as the larger cache size enables the BS to store and serve a greater amount of data locally, reducing the need for frequent data transfers over the fronthaul.  Furthermore, we can observe that Algorithm~$2$ performs better when compared with Algorithm~$1$ since the FoVs are grouped into multicast groups during transmission in Algorithm~$2$. This improvement is particularly noteworthy compared to benchmark algorithms, where the proposed algorithm consistently achieves the lowest average delay across all cache sizes.

In Fig.~\ref{fig:comp_bs}, we observe the trade-off between the average cache hit and the number of BSs in the network. Initially, when the number of BSs is low, the average cache hit remains at a low level. However, as we gradually increase the number of BSs, the average cache hit increases for all the algorithms. Notably, the performance gains from increasing the number of BSs plateau beyond a certain point for all algorithms. This suggests that simply adding more BSs, and consequently more data, does not necessarily translate to improved cache hit rates. 

These findings provide valuable insights into how the performance of different algorithms varies with an increase in the number of BSs, and they emphasize the advantages of the proposed algorithm in effectively managing larger datasets even under increased network demands. As such, the proposed algorithm proves to be a compelling solution that holds the potential to enhance the overall user experience and bolster the efficiency of the communication system in scenarios with varying data.

Fig.~\ref{fig:com2} presents the simulation results using Dataset $2$. From Fig.~\ref{fig:com2}, we can infer that as the cache size of the BS increases, the average cache hit naturally increases. Fig.~\ref{fig:com2} evidently illustrates that the proposed algorithms (both algorithm $1$
and 2) achieve superior performance compared to sgdalgo1, sgdalgo2, $\rho$ learning, $\sigma$ learning when $\rho_1^t = \rho_2^t = 0.5$, $\sigma_1^t = \sigma_2^t = 0.5$, and as well as FedAvg, demonstrating the advantages of the proposed approach. Furthermore, Fig.~\ref{fig:com2} indicates that the results align with those obtained using Dataset $1$ and thus highlighting that the proposed algorithms generalizes well across different data attributes. This observation accentuates the promising potential and benefits offered by the proposed algorithm.

Fig.~\ref{fig:delay2} displays the simulation results obtained with Dataset $2$. More specifically, in Fig.~\ref{fig:delay2}, similar to Fig.~\ref{fig:delay}, shows comparable trends regarding the average cache hit with respect to average delay. Fig.~\ref{fig:delay2} also illustrates how the proposed Algorithm $1$ and Algorithm $2$ perform better when compared with the benchmark algorithms. One can, therefore, infer that the results of Fig.~\ref{fig:delay2} are contingent with the results using Dataset $1$, thus reinforcing our proposed algorithms applicability to various datasets and distinct attributes.    

In Fig.~\ref{fig:comp_bs2} we observe the relationship between the BS cache size and number of BSs using Dataset $2$. Similar to Fig.~\ref{fig:comp_bs}, Fig.~\ref{fig:comp_bs2} also shows an increasing average cache hit when the number of BSs increases. Further the proposed Algorithm $1$ and $2$ shows significant improvement in terms of average cache hit when compared with the benchmark algorithms. These results further highlight the advantages of the proposed algorithms. 

In Fig.~\ref{fig:delay_bs}, the average delay is plotted against the cache size of the BS. Similar to the findings in Fig.~\ref{fig:delay}, Fig.~\ref{fig:delay_bs} also reveals a consistent pattern of decreasing average delay as the number of BSs increases. The decline in the average delay is attributed to the reduction in fronthaul load, which is made possible by the increased number of BS's enhanced caching capabilities. Moreover, just as observed in Fig.~\ref{fig:delay}, Fig.~\ref{fig:delay_bs} reaffirms the superior performance of the proposed algorithm when compared to other existing benchmark algorithms. Once again, the proposed algorithm stands out by consistently exhibiting the least average delay, outperforming its counterparts. These results strongly suggest that the proposed algorithm can significantly enhance communication systems, particularly in future deployments with numerous BSs, by enabling low-latency, high-quality user experiences.

Fig.~\ref{fig:convg} demonstrates that signSGD achieves a convergence rate comparable to traditional SGD, highlighting the effectiveness of one-bit gradient quantization. It is observed that both algorithms have the same convergence rate, thus highlighting the efficiency of using one-bit quantization of the stochastic gradient. By transmitting only the sign of each minibatch stochastic gradient, signSGD significantly reduces communication overhead without sacrificing the convergence rate. This allows for efficient distributed training with compressed gradients while maintaining performance comparable to standard SGD. Hence, it can get the best of both worlds: compressed gradients and SGD-level convergence rate.

Fig.~\ref{fig:tile_size} shows the variation of the average cache hit with respect to the tile size for the two proposed algorithms. To capture the variation of tile
sizes, a parameter $ \chi = S_T /S_V$ is defined, where $S_T$ denotes the tile size and $S_V$ denotes the FoV size. In the simulation, the VR video was segmented into $6 \times 4$ (that means the full frame will be segmented into $6$ tiles horizontally and $4$ tiles vertically), $8 \times 6$, $10 \times 8$ and $12 \times 10$ tiles. Fig.~\ref{fig:tile_size} shows the average cache hit rate changes for different tile size ratios. It can be seen
from Fig.~\ref{fig:tile_size} that the best tile size ratio is approximately $0.07$ (corresponding to the $10 \times 8$ tile partition for the VR video). This trade-off ratio arises from striking a balance between individual tile data size and flexibility in accessing the entire FoV. Smaller tiles offer more flexibility but increase overhead, while larger tiles reduce flexibility.
\section{Conclusion}
\label{sec:concl}
Virtual reality systems are bound to radically change the interactions between devices and their supportive communications landscape. This paper addresses the challenge of VR content caching in highly non-stationary environments within a FL framework. A novel algorithm named DP-FL is proposed so as to leverage both statistical and adversarial learning principles to create a robust caching strategy. DP-FL algorithm, supported by theoretical guarantees from Theorems~\ref{thm:mainresult_cachingvr}, and~\ref{thm:guarant}, utilizes a regret minimization in conjunction with disparity, divergence, and variance measures to optimize cache content selection in distributed networks. Further, to include the wireless characteristics of the channel, the FoVs are grouped together into multicast or unicast groups based on the number of requesting VR users. Through various simulation results, it is shown that the proposed algorithm performs better than the existing baseline methods. Finally, as possible research direction, we relegate accounting for online, data-driven user association strategy in the context of our work for future investigation.
%This underscores the numerical potential of the proposed algorithm within the context of future future VR systems.
\appendices
\section{Proof of Theorem~\ref{thm:mainresult_cachingvr}} \label{appendix:proof_main_res_cachingvr}
Assume that each BS $b$ employs the caching strategy in \eqref{eq:phi_weights} based on the local data $D_{b}^T$. Then, the corresponding conditional average of the hit rate is given by
\begin{equation}
	\begin{aligned}
		\mathbb{E}\left[\mathcal{Q}(\bm{\phi_{b}^{T+1}})^{(av)} \left | \right. D_{b}^T\right] 
		&\stackrel{(a)}{=} \bm{\rho}_{b}^{T+1} \mathbb{E}\left[\mathcal{Q}(\bar{\bm{\phi}}_{b}^{T+1}) \left | \right. D_{b}^T\right] 
		+ \sum_{b^{'} \in \mathcal{N}_b^t} \bm{\rho}_{b^{'}}^{T+1} \mathbb{E}\left[\mathcal{Q}(\bar{\bm{\phi}}_{b^{'}}^{T+1}) \left | \right. D_{b}^T\right] \\
		& \stackrel{(b)}{=}  \mathbb{E}\left[\mathcal{Q}(\bar{\bm{\phi}}_{b}^{T+1}) \left | \right. D_{b}^T\right] - \texttt{S}_{b}^{T+1}(\bm{\rho}_{\neq b}^{T}), \\
		&\stackrel{(c)}{=}  \upsilon_{b,i,f}^{T+1} \sum_{t=T-\tau}^T \bm{\sigma}_{b}^{t} \mathbb{E}\left[\mathcal{Q}(\tilde{\phi}_{b,i,f}^{t}) \left | \right. D_{b}^T\right]  + \sum_{i^{'} \in \mathcal{U}_b^t} \upsilon_{b,i^{'},f}^{T+1} \sum_{t=T-\tau}^T \bm{\sigma}_{b^{'}}^{t} \mathbb{E}\left[\mathcal{Q}(\tilde{\phi}_{b^{'},i,f}^{t}) \left | \right. D_{b}^T\right] \\ 
		&\stackrel{(d)}{=} \sum_{t=T-\tau}^T \bm{\sigma}_{b}^{t} \mathbb{E}\left[\mathcal{Q}(\tilde{\phi}_{b,i,f}^t) \left | \right. D_{b}^T\right]  -  \texttt{S}_{b}^{T+1}(\bm{\rho}_{\neq b}^{T}) - \texttt{H}_{b}^{T+1}({\upsilon}_{\neq i}^{T}), \\
		& \stackrel{(e)}{\geq} \sum_{t=T-\tau}^T \bm{\sigma}_{b}^{t}\mathbb{E}\left[\mathcal{Q}(\tilde{\phi}_{b,i,f}^{t}) \left | \right. D_{b}^{t-1}\right] - \texttt{S}_{b}^{T+1}(\bm{\rho}_{\neq b}^{T}) - \texttt{H}_{b}^{T+1}({\upsilon}_{\neq i}^{T}) - \mathbb{V}_{b}^{T}(\bm{\sigma}_{b}^{T}), 
	\end{aligned}
	\label{eq:first_equality_coded_caching}
\end{equation}
where $(a)$ follows from substituting for $\bm{\phi_{b}^{T+1}}$ from~\eqref{eq:phi_weights}, $(b)$ follows from (i) adding and subtracting the term $\sum_{b^{'} \in \mathcal{N}_b^t} \bm{\rho}_{b^{'}}^{T+1}  \mathbb{E}\left[\mathcal{Q}(\bar{\bm{\phi}}_{b^{'}}^{t}) \left | \right. D_{b}^T\right]$, and using the definition of $\texttt{S}_{b}^{T+1}(\bm{\rho}_{\neq b}^{T})$, and (ii) using the fact that $\bm{\rho}_{b}^{T+1} + \sum_{b^{'} \in \mathcal{N}_b^t} \bm{\rho}_{b^{'}}^{T+1} = 1$ $\forall$ $b \in \mathcal{N}_b^t$.
%\begin{equation} 
%\begin{aligned}
%&\mathbb{E}\left[\mathcal{Q}(\bm{\phi_{b}^{T+1}})^{(av)}  \left | \right. D_{b}^T\right]  \stackrel{(c)}{=}  \upsilon_{b,i,f}^{T+1} \sum_{t=T-\tau}^T \bm{\sigma}_{b}^{t} \mathbb{E}\left[\mathcal{Q}(\tilde{\phi}_{b,i,f}^{t}) \left | \right. D_{b}^T\right] \\
%& + \sum_{i^{'} \in \mathcal{U}_b^t} \upsilon_{b,i^{'},f}^{T+1} \sum_{t=T-\tau}^T \bm{\sigma}_{b^{'}}^{t} \mathbb{E}\left[\mathcal{Q}(\tilde{\phi}_{b^{'},i,f}^{t}) \left | \right. D_{b}^T\right] \\ 
%&\stackrel{(d)}{=}\!\!\! \sum_{t=T-\tau}^T \!\!\!\bm{\sigma}_{b}^{t} \mathbb{E}\left[\mathcal{Q}_{b}^{T+1}(\tilde{\phi}_{b,i,f}^t) \left | \right. D_{b}^T\right]  - \texttt{S}_{b}^{T+1}(\bm{\rho}_{\neq b}^{T}) - \texttt{H}_{b}^{T+1}({\upsilon}_{\neq i}^{T}),
%\end{aligned}
% \label{eq:lowerbound_deterministic}
%\end{equation} 
$(c)$ follows simply from substituting for $\bm{\bar{\phi}}_{b}^{T+1}$ from \eqref{eq:phi_fov}. $(d)$ is obtained from (i) adding and subtracting the term $\sum_{i^{'} \in \mathcal{U}_b^t} \upsilon_{b,i^{'},f}^{T+1} \sum_{t=T-\tau}^T \bm{\sigma}_{b}^{t} \mathbb{E}\left[\mathcal{Q}(\tilde{\phi}_{b,i,f}^{t}) \left | \right. D_{b}^T\right]$, and using the definition of $\texttt{H}_{b}^{T+1}({\upsilon}_{\neq i}^{T})$, and (ii) using the fact that $\upsilon_{b,i,f}^{T+1} + \sum_{i^{'} \in \mathcal{U}_b^t} \upsilon_{b,i^{'},f}^{T+1} = 1$ $\forall$ $i \in \mathcal{U}_b^t$. 
%\begin{equation}
%\begin{aligned}
%& \mathbb{E}\left[\mathcal{Q}(\bm{\phi_{b}^{T+1}})^{(av)}  \left | \right. D_{b}^T\right] \stackrel{(e)}{\geq} \sum_{t=T-\tau}^T \bm{\sigma}_{b}^{t}\mathbb{E}\left[\mathcal{Q}(\tilde{\phi}_{b,i,f}^{t}) \left | \right. D_{b}^{t-1}\right] \nonumber \\
%&- \texttt{S}_{b}^{T+1}(\bm{\rho}_{\neq b}^{T})  - \texttt{H}_{b}^{T+1}({\upsilon}_{\neq i}^{T}) - \mathbb{V}_{b}^{T}(\bm{\sigma}_{b}^{T}), 
%\end{aligned} 
%\label{eq:lowerbound_deterministic}
%\end{equation} 
Finally, $(e)$ follows by adding and subtracting $\sum_{t=T-\tau}^T\bm{\sigma}_{b}^{t} \mathbb{E}\left[\mathcal{Q}(\tilde{\phi}_{b,i,f}^{t}) \left | \right. D_{b}^{t-1} \right] $, and using the definition of $\mathbb{V}_{b}^{T}(\bm{\sigma}_{b}^{T})$  in \eqref{eq:disclg}, the above equation can be lower bounded. 

Similarly, an upper bound can also be obtained as follows
\begin{equation}
	\begin{aligned}
		\mathbb{E}\left[\mathcal{Q}(\bm{\phi_{b}^{T+1}})^{(av)} \left | \right. D_{b}^T\right]  &\leq  \sum_{t=T-\tau}^T \bm{\sigma}_{b}^{t} \mathbb{E}\left[\mathcal{Q}(\tilde{\phi}_{b,i,f}^{t}) \left | \right. D_{b}^{t-1}\right] \\
		&+ \texttt{S}_{b}^{T+1}(\bm{\rho}_{\neq b}^{T})  + \texttt{H}_{b}^{T+1}({\upsilon}_{\neq i}^{ T})+  \mathbb{V}_{b}^{T}(\bm{\sigma_{b}^{T}}),
	\end{aligned}
	\label{eq:upperbound_deterministic}
\end{equation} 
where the above upper bound follows by adding the disparity term instead of subtraction. Note that the term $$M_t := \bm{\sigma}_{b}^{t} \mathcal{Q}(\tilde{\phi}_{b,i,f}^{t})  - \bm{\sigma}_{b}^{t} \mathbb{E}\left[\mathcal{Q}(\tilde{\phi}_{b,i,f}^{t}) \left | \right. D_{b}^{t}\right]$$ is a Martingale difference, i.e., $\mathbb{E}\left\{M_t \left | \right. D_{b}^{t}\right\} = 0$. Thus, the following event occurs with a probability of at least $1-\delta$, which follows from the Azuma's inequality \cite{chung06}
\begin{equation} 
	\sum_{t=T-\tau}^T M_t \leq C_{\texttt{max}} \norm{\bm{\sigma_{b}^{T}}}_2 \sqrt{\frac{2}{\tau} \log\frac{1}{\delta}},
\end{equation}
where $C_{\texttt{max}}$ is the maximum possible cache hit rate. The above implies that
\begin{equation}
	\begin{aligned} 
		\sum_{t=T-\tau}^T \bm{\sigma}_{b}^{t} \mathbb{E}\left[\mathcal{Q}(\tilde{\phi}_{i,b,f}^{t}) \left | \right. D_{b}^{t-1}\right] 
		\geq \sum_{t=T-\tau}^T  \bm{\sigma}_{b,t} \mathcal{Q}(\tilde{\phi}_{i,b,f}^{t})
		- C_{\texttt{max}} \norm{\bm{\sigma_{b}^{T}}}_2 \sqrt{\frac{2}{\tau} \log\frac{1}{\delta}} 
	\end{aligned}
	\label{eq:azuma_At_firstequation}
\end{equation}
Since $-M_t$ is also a Martingale difference, using Azuma's inequality, the following holds good with a probability of at least $1-\delta$
\begin{equation}
	\begin{aligned}
		\sum_{t=T-\tau}^T \bm{\sigma}_{b}^{t} \mathcal{Q}(\tilde{\phi}_{i,b,f}^{t}) \geq  \sum_{t=T-\tau}^T  \bm{\sigma}_{b}^{t} \mathbb{E}\left[\mathcal{Q}(\tilde{\phi}_{i,b,f}^{t}) \left | \right. D_{b}^{t-1}\right] -  C_{\texttt{max}} \norm{\bm{\sigma_{b}^{T}}}_2 \sqrt{\frac{2}{\tau} \log\frac{1}{\delta}}
	\end{aligned}
	\label{eq:azuma_At_secondequation}
\end{equation}
Using \eqref{eq:azuma_At_firstequation} in \eqref{eq:first_equality_coded_caching}, the following holds good with a probability of at least $1-\delta$
\begin{equation}
	\begin{aligned}
		\mathbb{E}\left[\mathcal{Q}(\bm{\phi_{b}^{T+1}})^{(av)} \left | \right. D_{b}^T\right]  &\geq \sum_{t=T-\tau}^T\bm{\sigma}_{b}^{t} \mathcal{Q}(\tilde{\phi}_{i,b,f}^{t})  - C_{\texttt{max}} \norm{\bm{\sigma_{b}^{T}}}_2 \sqrt{\frac{2}{\tau} \log\frac{1}{\delta}} - \mathcal{S}_{b}^{T+1}(\bm{\rho}_{\neq b}^{T}) - \mathbb{V}_{b}^{T}(\bm{\sigma}_{b}^T)\\
		&-  \texttt{H}_{b}^{T+1}({\upsilon}_{\neq i}^{T}) . 
	\end{aligned}
	\label{eq:azuma_At_thirdequation}
\end{equation} 
Similar to the above equation, using \eqref{eq:azuma_At_secondequation} in \eqref{eq:upperbound_deterministic}, the following holds good with a probability of at least $1 - \delta$
\begin{equation}
	\begin{aligned}
		\mathbb{E}\left[\mathcal{Q}(\bm{\phi_{b}^{T+1}})^{(av)} \left | \right. D_{b}^T\right]  &\leq \sum_{t=T-\tau}^T\bm{\sigma}_{b}^{t} \mathcal{Q}(\tilde{\phi}_{i,b,f}^{t})  +  C_{\texttt{max}} \norm{\bm{\sigma_{b}^{T}}}_2 \sqrt{\frac{2}{\tau} \log\frac{1}{\delta}} \\ 
		&+ \mathcal{S}_{b}^{T+1}(\bm{\rho}_{\neq b}^{T}) + \mathbb{V}_{b}^{T}(\bm{\sigma}_{b}^T) \nonumber + \texttt{H}_{b}^{T+1}({\upsilon}_{\neq i}^{T}) .  
	\end{aligned}
	\label{eq:azuma_At_fourthequation}
\end{equation}
Let ${(\bm{\phi}_{b}^{t}})^{*}$ be the the optimal caching strategy used. Let's now consider the following term:
\begin{equation}
	\begin{aligned}
		\sum_{t=T-\tau}^T \bm{\sigma}_{b}^{t} \mathcal{Q}{(\bm{\phi}_{b}^{t}})^{*} - \sum_{t=T-\tau}^T \bm{\sigma}_{b}^{t} \mathcal{Q}(\tilde{\phi}_{b,i,f}^{t}) &\leq \sum_{t=T-\tau}^T \big(\bm{\sigma}_{b}^{t} - \frac{1}{\tau} \big)\bigg(\mathcal{Q}{(\bm{\phi}_{b}^{t}})^{*}- \mathcal{Q}(\tilde{\phi}_{b,i,f}^{t}) \bigg)\\
		& + \frac{1}{\tau}\sum_{t=T-\tau}^T\big(\mathcal{Q}{(\tilde{\phi}_{b,i,f}^{t}})^{*} - \mathcal{Q}(\bm{\phi}_{b}^{t})\big) \\
		&\stackrel{(a)}{\leq} \mathcal{C}_{max}\sum_{t= T -\tau + 1}^{T} \bigg|\bm{\sigma}_{b}^t - \frac{1}{\tau}\bigg| + \frac{2\texttt{Reg}(\bm{\phi}_{b}^{t})}{\tau}, 
	\end{aligned}
\end{equation}
where $(a)$ follows from the definition of Regret from \eqref{eq:regret}. Consequently, we have
\begin{equation}
	\begin{aligned}
		\sum_{t=T-\tau}^T \bm{\sigma}_{b}^{t} \mathcal{Q}(\tilde{\phi}_{b,i,f}^{t}) \geq \sum_{t=T-\tau}^T \bm{\sigma}_{b}^{t} \mathcal{Q}{(\bm{\phi}_{b}^{t}})^{*} - \frac{2\texttt{Reg}(\bm{\phi}_{b}^{t})}{\tau} -\mathcal{C}_{max}\sum_{t= T -\tau + 1}^{T} \bigg|\bm{\sigma}_{b}^t - \frac{1}{\tau}\bigg|    
	\end{aligned} 
\end{equation}
From \eqref{eq:azuma_At_thirdequation}, we have 
\begin{equation}
	\begin{aligned}
		\mathbb{E}\left[\mathcal{Q}(\bm{\phi_{b}^{T+1}})^{(av)}  \left | \right. D_{b}^T\right]  &\stackrel{(e)}{\geq} \sum_{t=T-\tau}^T \bm{\sigma}_{b}^{t} \mathcal{Q}{(\bm{\phi}_{b}^{t}})^{*}  - C_{\texttt{max}} \norm{\bm{\sigma_{b}^{T}}}_2 \sqrt{\frac{2}{\tau} \log\frac{1}{\delta}} - \texttt{S}_{b}^{T+1}(\bm{\rho}_{\neq b}^{T})  - \texttt{H}_{b}^{T+1}({\upsilon}_{\neq i}^{T}) \\
		&- \mathbb{V}_{b}^{T}(\bm{\sigma}_{b}^{T}) - \frac{2\texttt{Reg}(\bm{\phi}_{b}^{t})}{\tau} -\mathcal{C}_{max}\sum_{t= T -\tau + 1}^{T} \bigg|\bm{\sigma}_{b}^t - \frac{1}{\tau}\bigg|. \\
		\text{This proves the theorem.}
	\end{aligned}  
\end{equation}
\section{Proof of Theorem \ref{thm:guarant}} \label{appendix:proof_convg_cachingvr}
In order to prove convergence of the Algorithm, it suffices to prove that $\nabla_{\bm{\phi_b}}\mathcal{{Q}}_{\bm{\sigma}_b^t, \bm{\rho}_b^t,\bm{\upsilon}_b^t}(\bm{\phi}_b^t)$, $\nabla_{\bm{\sigma}_b^t}\mathcal{{Q}}_{\bm{\sigma}_b^t, \bm{\rho}_b^t,\bm{\upsilon}_b^t}(\bm{\phi}_b^t)$, $\nabla_{\bm{\rho_b^t}}\mathcal{{Q}}_{\bm{\sigma}_b^t, \bm{\rho}_b^t,\bm{\upsilon}_b^t}(\bm{\phi}_b^t)$, and $\nabla_{\bm{\upsilon}_b^t}\mathcal{{Q}}_{\bm{\sigma}_b^t, \bm{\rho}_b^t,\bm{\upsilon}_b^t}(\bm{\phi}_b^t)$ converges $\forall$ $b = 1,2,\ldots,B$. 

\emph{Proof:} Consider the $\beta$ smoothness of $\mathcal{Q}_{\bm{\sigma}_b^t,\bm{\rho}_b^t,\bm{\upsilon}_b^t}(\bm{\phi}_b^t)$ w.r.t. $\phi_b^t$ $\forall$ $ b = 1,2,\ldots, B$.
\begin{eqnarray}
	\mathcal{Q}_{\bm{\sigma}_b^{t+1}, \bm{\rho}_b^{t+1},\bm{\upsilon}_b^{t+1}}(\bm{\phi}_b^{t}) - \mathcal{Q}_{\bm{\sigma}_b^{t+1}, \bm{\rho}_b^{t+1},\bm{\upsilon}_b^{t+1}}(\bm{\phi}_b^{t+1}) &&\leq \langle \nabla_{\bm{\phi}_{b}^t} \mathcal{Q}_{\bm{\sigma}_b^{t+1}, \bm{\rho}^{t+1},\bm{\upsilon}_b^{t+1}}(\bm{\phi}_b^{t}), \bm{\phi}_b^{t} - \bm{\phi}_b^{t+1} \rangle 
	\nonumber\\
	&& + \sum_i\frac{L_i}{2} (\bm{\phi}_b^{t} - \bm{\phi}_b^{t+1})^2_i \nonumber
\end{eqnarray}
Let $\Delta_{\bm{\phi}_b^t} \mathcal{Q}_{\bm{\sigma}_b^{t+1}, \bm{\rho}_b^{t+1},\bm{\upsilon}_b^{t+1}}(\bm{\phi}_b^{t+1}) $ := $\mathcal{Q}_{\bm{\sigma}_b^{t+1}, \bm{\rho}_b^{t+1},\bm{\upsilon}_b^{t+1}}(\bm{\phi}_b^{t}) - \mathcal{Q}_{\bm{\sigma}_b^{t+1}, \bm{\rho}_b^{t+1},\upsilon^{t+1}}(\bm{\phi}_b^{t+1})$. From the Algorithm, we have $\bm{\phi}_b^{t} - \bm{\phi}_b^{t+1} = -\eta^t \nabla_{\bm{\phi}_b^t, \rm{sign}} \mathcal{Q}_{\bm{\sigma}_b^{t+1}, \bm{\rho}_b^{t+1},\bm{\upsilon}_b^{t+1}}(\bm{\phi}_b^{t}) $. Thus we have 
\begin{eqnarray}
	\Delta_{\bm{\phi}_b^t}\mathcal{Q}_{\bm{\sigma}_b^{t+1}, \bm{\rho}_b^{t+1},\bm{\upsilon}_b^{t+1}}(\bm{\phi}_b^{t+1})    && \leq  - \eta^t \nabla_{\bm{\phi}_b^t, \rm{sign}} \mathcal{Q}_{\bm{\sigma}_b^{t+1}, \bm{\rho}_b^{t+1},\bm{\upsilon}_b^{t+1}}(\bm{\phi}_b^{t})\nabla_{\bm{\phi}_b^t} \mathcal{Q}_{\bm{\sigma}_b^{t+1}, \bm{\rho}_b^{t+1},\bm{\upsilon}_b^{t+1}}(\bm{\phi}_b^{t}) \nonumber \\
	&&	+ \sum_i^d \frac{L_i}{2}\big(-\eta^t \nabla_{\bm{\phi}_b^t} \mathcal{Q}_{\bm{\sigma}_b^{t+1}, \bm{\rho}_b^{t+1},\bm{\upsilon}_b^{t+1}}(\bm{\phi}_b^{t})\big)_i^2, 
	\label{eq:prob}
\end{eqnarray}
The true gradient is given by $\nabla_{\bm{\phi}_b^t} \mathcal{Q}_{\bm{\sigma}_b^{t+1}, \bm{\rho}_b^{t+1},\bm{\upsilon}_b^{t+1}}(\bm{\phi}_b^{t}) = \sum_m^{B}\sigma_{mb}^{t+1}\rho_{mb}^{t+1}g_{mk}^t$ and the signed gradient is given by $\nabla_{\bm{\phi}_b^t, \rm{sign}} \mathcal{Q}_{\bm{\sigma}_b^{t+1}, \bm{\rho}_b^{t+1},\bm{\upsilon}_b^{t+1}}(\bm{\phi}_b^{t}) = \sum_m^{B}\sigma_{mb}^{t+1}\rho_{mb}^{t+1}{\rm sign} (\tilde{g}^t_{mb})$, thus the above \eqref{eq:prob} becomes:
\begin{eqnarray}
	\Delta_{\bm{\phi}_b^t}\mathcal{Q}_{\bm{\sigma}_b^{t+1}, \bm{\rho}_b^{t+1},\bm{\upsilon}_b^{t+1}}(\bm{\phi}_b^{t+1}) &&\leq - \eta^t\bigg( \sum_m^B \sigma_{mb}^{t+1}\rho_{mb}^{t+1}||g_{mb}^t||_1\bigg) + \frac{(\eta^t)^2}{2}\sum_i^dL_i\sum_m^B\bigg( \sigma_{mb}^t\rho_{mb}^t {\rm sign} (\tilde{g}_{mb}^t) \bigg)^2 \nonumber \\
	&& + 2\eta^t\bigg(\sum_i^d\sum_m^B \sigma_{mb}^t\rho_{mb}^t |g_{mb,i}^t| \mathbbm{1} [{\rm sign }(\tilde{g}_{mb,i}^t) \neq {\rm sign}(g_{mb,i}^t)]\bigg) \nonumber
\end{eqnarray}
Next, we take the expected improvement at time $t+1$ conditioned on the previous iterate,
\begin{equation}
	\begin{aligned}
		\mathbb{E}\bigg[\Delta_{\bm{\phi}_b^t}\mathcal{Q}_{\bm{\sigma}_b^{t+1}, \bm{\rho}_b^{t+1},\bm{\upsilon}_b^{t+1}}(\bm{\phi}_b^{t+1})\bigg|\bm{\phi}_{b}^t\bigg] \leq &- \eta^t\bigg( \sum_m^B \sigma_{mb}^{t+1}\rho_{mb}^{t+1}||g_{mb}^t||_1\bigg)  \\
		&+ \frac{(\eta^t)^2}{2} \sum_i^dL_i\sum_m^B\bigg( \sigma_{mb}^t\rho_{mb}^t {\rm sign} (\tilde{g}_{mb}^t) \bigg)^2 \\
		&+ 2\eta^t\bigg(\sum_i^d\sum_m^B \sigma_{mb}^t\rho_{mb}^t |g_{mb,i}^t| \mathbb{P} [{\rm sign }(\tilde{g}_{mb,i}^t) \neq {\rm sign}(g_{mb,i}^t)]\bigg)
		\label{eq:expect}
	\end{aligned}
\end{equation}
So the expected improvement crucially depends on the probability that each component of the sign vector is correct, which is intuitively controlled by the relative scale of the gradient to the noise.
\begin{eqnarray}
	\mathbb{P} [{\rm sign }(\tilde{g}_{mb,i}^t) \neq {\rm sign}(g_{mb,i}^t)] &\leq& \mathbb{P}[|\tilde{g}_{mb,i}^t -g_{mb,i}^t| \geq |g_{mb,i}^t| ] \nonumber \\
	&\stackrel{(a)}\leq& \frac{\mathbb{E}[|\tilde{g}_{mb,i}^t -g_{mb,i}^t|]}{|g_{mb,i}^t|} \nonumber \\
	&\stackrel{(b)}\leq& \frac{\sqrt{\mathbb{E}[(\tilde{g}_{mb,i}^t -g_{mb,i}^t)^2]}}{|g_{mb,i}^t|} \nonumber \\
	&\stackrel{(c)}\leq& \frac{\omega_{mb,i}}{|g_{mb,i}^t|} = \frac{\omega_{mb,i}}{\sqrt{\theta^t}},
\end{eqnarray}
where $(a)$ follows from the Markov’s inequality, $(b)$ is obtained from the Jensen’s inequality, and $(c)$ follows since $\tilde{g}_{mb,i}^t$ is an unbiased estimate of ${g}_{mb,i}^t$ and using the definition of variance.

Substituting these values in the equation \eqref{eq:expect}, and since $\sum_{m}\sigma_{mb} \leq 1$, $\sum_{m}\rho_{mb} \leq 1$, and ${\rm sign}(\hat{g}_{mb,i}) \leq 1$ we have:
\begin{equation}
	\begin{aligned}
		\mathbb{E}\bigg[\Delta_{\phi_b}\mathcal{Q}_{\bm{\sigma}_b^{t+1}, \bm{\rho}_b^{t+1},\bm{\upsilon}_b^{t+1}}(\bm{\phi}_b^{t+1})\bigg|\bm{\phi}_{b}^t\bigg] 
		&\leq \mathbb{E}\bigg[- \eta^t \bigg(\sum_m^B||g_{mb}^t||_1 \bigg) + \frac{(\eta^t)^2}{2}||L||_1 \\
		&+ 2\eta^t\bigg(\sum_m^B \frac{||\sigma_{mb}||_1}{\sqrt{\theta^t}} \bigg) \bigg]
	\end{aligned}
	\label{eq:phi_bound}
\end{equation}
Now, bounding when $\bm{\sigma}_b^{t+1}$ is variable:
\begin{equation}
	\begin{aligned}
		\mathcal{Q}_{\bm{\sigma}_b^{t}, \bm{\rho}_b^{t+1},\bm{\upsilon}_b^{t+1}}(\bm{\phi}_b^{t}) - \mathcal{Q}_{\bm{\sigma}_b^{t+1}, \bm{\rho}_b^{t+1},\bm{\upsilon}_b^{t+1}}(\bm{\phi}_b^{t}) &\leq
		\langle \nabla_{\bm{\sigma}_b^t} \mathcal{Q}_{\bm{\sigma}_b^{t}, \bm{\rho}_b^{t+1}}(\bm{\phi}_b^{t}), \bm{\sigma}_b^{t} - \bm{\sigma}_b^{t + 1} \rangle \nonumber\\
		&+ \frac{\delta}{2} ||\bm{\sigma}_b^{t} - \bm{\sigma}_b^{t +1}||^2_2  
	\end{aligned} 
\end{equation}
From Algorithm, we have $\bm{\sigma}_{b}^{t} - \bm{\sigma}_{b}^{t+1} = - \mu \nabla_{\mathcal{K}, \bm{\sigma}_b^t}\mathcal{Q}_{\bm{\sigma}_b^{t}, \bm{\rho}_b^{t+1},\bm{\upsilon}_b^{t+1}}(\phi_b^{t}) $. Recall from Lemma \ref{lm:gradient}, we have $\langle \nabla f(x), \nabla_{\mathcal{K}, \eta} f(x) \rangle \geq ||\nabla_{\mathcal{K}, \eta}f(x)||^2$.
Thus, we have 
\begin{eqnarray}
	\Delta_{\bm{\sigma}_b^t}\mathcal{Q}_{\bm{\sigma}_b^{t}, \bm{\rho}_b^{t+1},\bm{\upsilon}_b^{t+1}}(\bm{\phi}_b^{t}) \leq \bigg(-\mu^t + \frac{\delta}{2}(\mu^t)^2\bigg) ||\nabla_{\mathcal{K},\bm{\sigma}_b^t}\mathcal{Q}_{\bm{\sigma}_b^{t}, \bm{\rho}_b^{t+1},\bm{\upsilon}_b^{t+1}}(\bm{\phi}_b^{t})||^2_2 
	\label{eq:alpha_bound}
\end{eqnarray}
where $\Delta_{\bm{\sigma}_b^t}\mathcal{Q}_{\bm{\sigma}_b^{t}, \bm{\rho}_b^{t+1},\bm{\upsilon}_b^{t+1}}(\bm{\phi}_b^{t}) = \mathcal{Q}_{\bm{\sigma}_b^{t}, \bm{\rho}_b^{t+1},\bm{\upsilon}_b^{t+1}}(\bm{\phi}_b^{t}) - \mathcal{Q}_{\bm{\sigma}_b^{t + 1}, \bm{\rho}_b^{t+1},\bm{\upsilon}_b^{t+1}}(\bm{\phi}_b^{t})$.
Similarly, bounding when $\bm{\rho}_b^{t+1}$ is variable:
\begin{eqnarray}
	\mathcal{Q}_{\bm{\sigma}_b^{t}, \bm{\rho}_b^{t},\bm{\upsilon}_b^{t}}(\bm{\phi}_b^{t}) - \mathcal{Q}_{\bm{\sigma}_b^{t}, \bm{\rho}_b^{t + 1},\bm{\upsilon}_b^{t+1}}(\bm{\phi}_b^{t}) \leq  \langle \nabla_{\bm{\rho}_b^t} \mathcal{Q}_{\bm{\sigma}_b^{t}, \bm{\rho}_b^{t+1},\bm{\upsilon}_b^{t+1}}(\bm{\phi}_b^{t}), \bm{\rho}_b^{t} - \bm{\rho}_b^{t + 1} \rangle + \frac{\lambda}{2} ||\bm{\rho}_b^{t} - \bm{\rho}_b^{t+1}||^2_2 \nonumber
\end{eqnarray}
From Algorithm, we have $\bm{\rho}_b^{t} - \bm{\rho}_b^{t+1} = - \nu \nabla_{\mathcal{K}, \bm{\rho}_b^t}\mathcal{Q}_{\bm{\sigma}_b^{t}, \bm{\rho}_b^{t},\bm{\upsilon}_b^t}(\bm{\phi}_b^{t}) $ and using the lemma \ref{lm:gradient}, we obtain:
\begin{equation}
	\Delta_{\bm{\rho}_b^t}\mathcal{Q}_{\bm{\sigma}_b^{t}, \bm{\rho}_b^{t},\bm{\upsilon}_b^{t}}(\bm{\phi}_b^{t}) \leq \bigg(- \nu^t + \frac{\lambda}{2}(\nu^t)^2\bigg) ||\nabla_{\mathcal{K},\bm{\rho}_b^t}\mathcal{Q}_{\bm{\sigma}_b^{t}, \bm{\rho}_b^{t},\bm{\upsilon}_b^{t}}(\bm{\phi}_b^{t})||^2_2 
	\label{eq:w_bound}
\end{equation}
where $\Delta_{\bm{\rho}_b^t}\mathcal{Q}_{\bm{\sigma}_b^{t}, \bm{\rho}_b^{t},\bm{\upsilon}_b^{t}}(\bm{\phi}_b^{t}) = \mathcal{Q}_{\bm{\sigma}_b^{t}, \bm{\rho}_b^{t},\bm{\upsilon}_b^{t}}(\bm{\phi}_b^{t}) - \mathcal{Q}_{\bm{\sigma}_b^{t}, \bm{\rho}_b^{t + 1},\bm{\upsilon}_b^{t+1}}(\bm{\phi}_b^{t})$.

Similarly, bounding when $\bm{\upsilon}_b^{t+1}$ is variable:
\begin{eqnarray}
	\mathcal{Q}_{\bm{\sigma}_b^{t}, \bm{\rho}_b^{t},\bm{\upsilon}_b^{t}}(\bm{\phi}_b^{t}) - \mathcal{Q}_{\bm{\sigma}_b^{t}, \bm{\rho}_b^{t },\bm{\upsilon}_b^{t+1}}(\bm{\phi}_b^{t}) \leq  \langle \nabla_{\bm{\upsilon}_b^t}\mathcal{Q}_{\bm{\sigma}_b^{t}, \bm{\rho}_b^{t},\bm{\upsilon}_b^{t+1}}(\bm{\phi}_b^{t}), \bm{\upsilon}_b^{t} - \bm{\upsilon}_b^{t + 1} \rangle + \frac{\zeta}{2} ||\bm{\upsilon}_b^{t} - \bm{\upsilon}_b^{t+1}||^2_2 \nonumber
\end{eqnarray}
From Algorithm, we have $\bm{\upsilon}_b^{t} - \bm{\upsilon}_b^{t+1} = - \iota \nabla_{\mathcal{K}, \bm{\upsilon}_{b}^t}\mathcal{Q}_{\bm{\sigma}_b^{t}, \bm{\rho}_b^{t},\bm{\upsilon}_b^t}(\bm{\phi}_b^{t}) $ and using the lemma \ref{lm:gradient}, the following is obtained:
\begin{equation} \Delta_{\bm{\upsilon}_b^t}\mathcal{Q}_{\bm{\sigma}_b^{t}, \bm{\rho}_b^{t},\bm{\upsilon}_b^{t}}(\bm{\phi}_b^{t})
	\leq \bigg(- \zeta^t + \frac{\lambda}{2}(\zeta^t)^2\bigg) ||\nabla_{\mathcal{K},\upsilon}\mathcal{Q}_{\sigma^{t}, \rho^{t},\upsilon^{t}}(\bm{\phi}_b^{t})||^2_2 
	\label{eq:w_bound}
\end{equation}
where $\Delta_{\upsilon^t}\mathcal{Q}_{\sigma^{t}, \rho^{t},\upsilon^{t}}(\bm{\phi}_b^{t}) = \mathcal{Q}_{\sigma^{t}, \rho^{t},\upsilon^{t}}(\bm{\phi}_b^{t}) - \mathcal{Q}_{\sigma^{t}, \rho^{t},\upsilon^{t+1}}(\bm{\phi}_b^{t})$.

Let $\mathcal{Q}^*_b = \max_{\bm{\phi}_b,\sigma,\rho,\upsilon} \mathcal{Q}_{\sigma^{t}, \rho^{t},\upsilon^{t+1}}(\bm{\phi}_b^{t})$. Consider 
\begin{eqnarray}
	\mathcal{Q}^*_b - \mathcal{Q}_{\sigma^{0}, \rho^{0},\upsilon^{0}}(\bm{\phi}_b^{0})  &\geq& \mathbb{E}\big[\mathcal{Q}_{\sigma^{t+1}, \rho^{t+1},\upsilon^{t+1}}(\bm{\phi}_b^{t+1}) -\mathcal{Q}_{\sigma^{0}, \rho^{0},\upsilon^{0}}(\bm{\phi}_b^{0})\big] ]  \nonumber \\
	&\stackrel{(a)}{=}& \mathbb{E}\bigg[\sum_{t=0}^{T-1} \mathcal{Q}_{\sigma^{t + 1}, \rho^{t + 1},\upsilon^{t+1}}(\bm{\phi}_b^{t + 1}) - \mathcal{Q}_{\sigma^{t}, \rho^{t},\upsilon^{t+1}}(\bm{\phi}_b^{t})\bigg] \nonumber \\
	&\stackrel{(b)}{=}& \hspace{-0.35cm}\mathbb{E}\bigg[\sum_{t=0}^{T-1} \bigg(\mathcal{Q}_{\sigma^{t+1}, \rho^{t+1},\upsilon^{t+1}}(\bm{\phi}_b^{t +1}) - \mathcal{Q}_{\sigma^{t+1}, \rho^{t+1},\upsilon^{t+1}}(\bm{\phi}_b^{t})\bigg) \nonumber \\
	&&+ \bigg(\mathcal{Q}_{\sigma^{t+1}, \rho^{t+1},\upsilon^{t+1}}(\bm{\phi}_b^{t}) - \mathcal{Q}_{\sigma^{t}, \rho^{t},\upsilon^{t}}(\bm{\phi}_b^{t})\bigg)\bigg] \nonumber \\
	&\stackrel{(c)}{=}&\hspace{-0.35cm} \mathbb{E}\bigg[\sum_{t=0}^{T-1} \bigg(\mathcal{Q}_{\sigma^{t +1}, \rho^{t+1},\upsilon^{t+1}}(\bm{\phi}_b^{t+1}) - \mathcal{Q}_{\sigma^{t +1}, \rho^{t+1},\upsilon^{t+1}}(\bm{\phi}_b^{t})\bigg) \nonumber \\
	&&+  \bigg(\mathcal{Q}_{\sigma^{t +1}, \rho^{t+1},\upsilon^{t+1}}(\bm{\phi}_b^{t}) - \mathcal{Q}_{\sigma^{t}, \rho^{t+1},\upsilon^{t+1}}(\phi_b^{t})\bigg) \nonumber \\
	&& + \bigg(\mathcal{Q}_{\sigma^{t}, \rho^{t+1},\upsilon^{t+1}}(\bm{\phi}_b^{t}) - \mathcal{Q}_{\sigma^{t}, \rho^{t+1},\upsilon^{t+1}}(\bm{\phi}_b^{t})\bigg)\bigg] 
	\label{eq:tele}
\end{eqnarray}
where $(a)$ follows from the telescoping sum, $(b)$ follows from adding and subtracting $\mathcal{Q}_{\sigma^{t+1}, \rho^{t+1},\upsilon^{t+1}}(\bm{\phi}_b^{t})$, and $(c)$ follows from adding and subtracting $\mathcal{Q}_{\sigma^{t},\rho^{t+1},\upsilon^{t+1}}(\bm{\phi}_b^{t})$.
This yields, 
\begin{equation}
	\begin{aligned}
		\mathcal{Q}^*_b - \mathcal{Q}_{\sigma^{0}, \rho^{0},\upsilon^{0}}(\bm{\phi}_b^{0})    &\geq  \mathbb{E}\bigg[\sum_{t=0}^{T-1} \Delta_{\bm{\phi}_b}\mathcal{Q}_{\sigma^{t+1}, \rho^{t+1},\upsilon^{t+1}}(\bm{\phi}_b^{t+1}) \bigg]+\mathbb{E}\bigg[\sum_{t=0}^{T-1} \Delta_{\sigma}\mathcal{Q}_{\sigma^{t}, \rho^{t+1},\upsilon^{t+1}}(\bm{\phi}_b^{t}) \bigg] \\
		&+\mathbb{E}\bigg[\sum_{t=0}^{T-1} \Delta_{\rho}\mathcal{Q}_{\sigma^{t}, \rho^{t},\upsilon^{t}}(\bm{\phi}_b^{t}) \bigg]    
	\end{aligned}
\end{equation}

Substituting the values from \eqref{eq:phi_bound}, \eqref{eq:alpha_bound}, and \eqref{eq:w_bound} in the above equation, the following is obtained:
\begin{eqnarray}
	\mathcal{Q}^*_b - \mathcal{Q}_{\sigma^{0}, \rho^{0},\upsilon^{0}}(\bm{\phi}_b^{0})    &&\geq \mathbb{E}\bigg[\sum_{t=0}^{T-1}\eta^t \bigg(\sum_m^B||g_{mb}^t||_1 \bigg) - \frac{(\eta^t)^2}{2}||L||_1  - 2\eta^t\bigg(\sum_m^B \frac{||\omega_{mb}||_1}{\sqrt{\theta^t}} \bigg) \nonumber \\
	&&  +\bigg(\mu^t - \frac{\delta}{2}(\mu^t)^2\bigg) ||\nabla_{\mathcal{K},\sigma }\mathcal{Q}_{\sigma^{t}, \rho^{t+1},\upsilon^{t+1}}(\bm{\phi}_b^{t})
	||^2_2 \nonumber \\
	&& + \bigg(\nu^t - \frac{\lambda}{2}(\nu^t)^2\bigg) ||\nabla_{\mathcal{K},\rho}\mathcal{Q}_{\sigma^{t}, \rho^{t},\upsilon^t}(\bm{\phi}_b^{t})||^2_2 \bigg] 
\end{eqnarray}
Rearranging the terms simplifies to the following: 
\begin{equation}
	\begin{aligned}
		\mathcal{Q}^*_b - \mathcal{Q}_{\sigma^{0}, \rho^{0},\upsilon^{0}}(\bm{\phi}_b^{0}) + \frac{(\eta^t)^2}{2}||L||_1 + 2\eta^t\bigg(\sum_{m=1}^B \frac{||\omega_{mb}||_1}{\sqrt{\theta^t}}\bigg) 
		& \geq \mathbb{E}\bigg[ \sum_{t=0}^{T-1}\eta^t \bigg(\sum_m^B||g_{mb}^t||_1 \bigg)\\
		& + \bigg(\mu^t - \frac{\delta}{2}(\mu^t)^2\bigg) ||\nabla_{\mathcal{K},\sigma}\mathcal{Q}_{\sigma^{t}, \rho^{t+1},\upsilon^{t+1}}(\bm{\phi}_b^{t})||^2_2 \\
		& + \bigg(\nu^t - \frac{\lambda}{2}(\nu^t)^2\bigg) ||\nabla_{\mathcal{K},\rho_{b}}\mathcal{Q}_{\sigma^{t}, \rho^{t},\upsilon^{t}}(\bm{\phi}_b^{t})||^2_2\\
		& + \big(\iota^t - \frac{\zeta}{2}(\iota^t)^2\big)\big[||\nabla_{\mathcal{K}, \upsilon_b }\mathcal{{Q}}_{\bm{\phi}^t, \rho^t,\sigma^t,\upsilon^t} (\bm{\phi}_b^{t})||_2^2  
		\bigg]
	\end{aligned}
\end{equation}
Choosing the learning rates $\eta_t= \frac{1}{\sqrt{T}}$ , $\mu_t = \frac{1}{\sqrt{T}}$, $\iota_t = \frac{1}{\sqrt{T}}$,  $\nu_t = \frac{1}{\sqrt{T}}$ and the batch size $ \theta^t = T$, proves the theorem.
\section{Proof of Theorem \ref{thm:guarant_delay}} \label{appendix:proof_convg_cachingvr_delay}
From the total law of expectation, we write the following:
\begin{equation}
	\mathbb{E}\bigg[\sum_{t=0}^{T-1}\Delta^t_b \bigg] \!\!=\!\! \mathbb{E}\bigg[\sum_{t=0}^{T-1}\Delta^t_b\big|Y^t_b \bigg]\mathbb{P}[Y^t_b] + \mathbb{E}\bigg[\sum_{t=0}^{T-1}\Delta^t_b\big|(Y^t_b)^c \bigg]\mathbb{P}[(Y^t_b)^c], \nonumber
\end{equation}
where the event $Y^t_b$ is the event of fetching the FoV from the BS. Now, from Lemma $\eqref{lm:martingale}$, and $\mathbb{P}[Y^t_b] \leq 1$, we have:
\begin{eqnarray}
	\mathbb{E}\bigg[\sum_{t=0}^{T-1}\Delta^t_b \bigg] \leq \mathbb{E}\bigg[\sum_{t=0}^{T-1}\Delta^t_b\big|Y^t_b \bigg] \exp\bigg({\frac{-\lambda^2}{2TC^2}}\bigg) + \mathbb{E}\bigg[\sum_{t=0}^{T-1}\Delta^t_b\big|(Y^t_b)^c \bigg] 
\end{eqnarray}
Assuming, $\sum_{t=0}^{T-1} \Delta_b^T$ to be bounded and having a maximum value of $D_{\max}$, we have the following bound on the convergence:
\begin{equation}
	\begin{aligned}
		\mathbb{E}\bigg[\frac{1}{T}\sum_{t=0}^{T-1}\Delta^t_b \bigg] &\leq \frac{D_{\max}}{T}\exp\bigg({\frac{-\lambda^2}{2TC^2}}\bigg) + \frac{1}{\sqrt{T}} \bigg( 2\sum_{m=1}^B||\omega_{bm}||_1 + \frac{||L||_1}{2} + \mathcal{Q}_{b}^{*} -\mathcal{Q}_{\sigma^0,\rho^0, \upsilon^0}(\bm{\phi}_{b}^0)  \bigg) \nonumber \\
		&\leq  \frac{D_{\max}}{T}\bigg( 1 - \frac{\lambda^2}{2TC^2}\bigg) + \frac{1}{\sqrt{T}} \bigg( 2\sum_{m=1}^B||\omega_{bm}||_1 + \frac{||L||_1}{2}  + \mathcal{Q}_{b}^{*} -\mathcal{Q}_{\sigma^0,\rho^0, \upsilon^0}(\bm{\phi}_{b}^0)  \bigg). 
	\end{aligned}
\end{equation}
This proves the theorem.

\bibliographystyle{IEEEtran}
\bibliography{strings}

\begin{thebibliography}{10}
\providecommand{\url}[1]{#1}
\csname url@samestyle\endcsname
\providecommand{\newblock}{\relax}
\providecommand{\bibinfo}[2]{#2}
\providecommand{\BIBentrySTDinterwordspacing}{\spaceskip=0pt\relax}
\providecommand{\BIBentryALTinterwordstretchfactor}{4}
\providecommand{\BIBentryALTinterwordspacing}{\spaceskip=\fontdimen2\font plus
\BIBentryALTinterwordstretchfactor\fontdimen3\font minus
  \fontdimen4\font\relax}
\providecommand{\BIBforeignlanguage}[2]{{%
\expandafter\ifx\csname l@#1\endcsname\relax
\typeout{** WARNING: IEEEtran.bst: No hyphenation pattern has been}%
\typeout{** loaded for the language `#1'. Using the pattern for}%
\typeout{** the default language instead.}%
\else
\language=\csname l@#1\endcsname
\fi
#2}}
\providecommand{\BIBdecl}{\relax}
\BIBdecl

\bibitem{grandview23}
\BIBentryALTinterwordspacing
G.~V. Research, ``Virtual reality {(VR)} market size and share report, 2030,''
  July 2023. [Online]. Available:
  \url{https://www.grandviewresearch.com/industry-analysis/virtual-reality-vr-market}
\BIBentrySTDinterwordspacing

\bibitem{guo24}
Y.~Guo, Z.~Qin, X.~Tao, and G.~Y. Li, ``Federated multi-view synthesizing for
  metaverse,'' \emph{IEEE Journal on Selected Areas in Communications},
  vol.~42, no.~4, pp. 867--879, 2024.

\bibitem{fenghe20}
F.~Hu, Y.~Deng, W.~Saad, M.~Bennis, and A.~H. Aghvami, ``Cellular-connected
  wireless virtual reality: Requirements, challenges, and solutions,''
  \emph{IEEE Communications Magazine}, vol.~58, no.~5, pp. 105--111, 2020.

\bibitem{tharakan19}
K.~S. Tharakan, B.~N. Bharath, and V.~Bhatia, ``Cache enabled cellular network:
  Algorithm for cache placement and guarantees,'' \emph{IEEE Wireless
  Communications Letters}, vol.~8, no.~6, pp. 1550--1554, 2019.

\bibitem{liu21}
X.~Liu and Y.~Deng, ``Learning-based prediction, rendering and association
  optimization for {MEC}-enabled wireless virtual reality {(VR)} networks,''
  \emph{IEEE Transactions on Wireless Communications}, vol.~20, no.~10, pp.
  6356--6370, 2021.

\bibitem{chen18}
M.~Chen, W.~Saad, and C.~Yin, ``Virtual reality over wireless networks:
  Quality-of-service model and learning-based resource management,'' \emph{IEEE
  Transactions on Communications}, vol.~66, no.~11, pp. 5621--5635, 2018.

\bibitem{Zink2019}
M.~Zink, R.~K. Sitaraman, and K.~Nahrstedt, ``Scalable 360° video stream
  delivery: Challenges, solutions, and opportunities,'' \emph{Proceedings of
  the IEEE}, vol. 107, pp. 639--650, 2019.

\bibitem{lungaro18}
P.~Lungaro, R.~Sjöberg, A.~J.~F. Valero, A.~Mittal, and K.~Tollmar,
  ``Gaze-aware streaming solutions for the next generation of mobile {VR}
  experiences,'' \emph{IEEE Transactions on Visualization and Computer
  Graphics}, vol.~24, no.~4, pp. 1535--1544, 2018.

\bibitem{ozfatura22}
E.~Ozfatura and D.~Gündüz, ``Uncoded caching and cross-level coded delivery
  for non-uniform file popularity,'' \emph{IEEE Transactions on Information
  Theory}, vol.~68, no.~10, pp. 6842--6859, 2022.

\bibitem{bharath18}
B.~N. Bharath, K.~G. Nagananda, D.~Gündüz, and H.~V. Poor, ``Caching with
  time-varying popularity profiles: A learning-theoretic perspective,''
  \emph{IEEE Transactions on Communications}, vol.~66, no.~9, pp. 3837--3847,
  2018.

\bibitem{paria21}
D.~Paria and A.~Sinha, ``Leadcache : Regret-optimal caching in networks,'' in
  \emph{Advances in Neural Information Processing Systems}, vol.~34, 2021, pp.
  4435 -- 4447.

\bibitem{chen22}
Z.~Chen, H.~Zhu, L.~Song, D.~He, and B.~Xia, ``Wireless multiplayer interactive
  virtual reality game systems with edge computing: Modeling and
  optimization,'' \emph{IEEE Transactions on Wireless Communications}, vol.~21,
  no.~11, pp. 9684--9699, 2022.

\bibitem{gupta23}
S.~Gupta, J.~Chakareski, and P.~Popovski, ``{mmWave} networking and edge
  computing for scalable 360° video multi-user virtual reality,'' \emph{IEEE
  Transactions on Image Processing}, vol.~32, pp. 377--391, 2023.

\bibitem{dang23}
T.~Dang, C.~Liu, and M.~Peng, ``Low-latency mobile virtual reality content
  delivery for unmanned aerial vehicle-enabled wireless networks with energy
  constraints,'' \emph{IEEE Transactions on Vehicular Technology}, vol.~72,
  no.~2, pp. 2189--2201, 2023.

\bibitem{jakob23}
J.~Struye, F.~Lemic, and J.~Famaey, ``Covrage: Millimeter-wave beamforming for
  mobile interactive virtual reality,'' \emph{IEEE Transactions on Wireless
  Communications}, vol.~22, no.~7, pp. 4828--4842, 2023.

\bibitem{zhang22}
R.~Zhang, J.~Liu, F.~Liu, T.~Huang, Q.~Tang, S.~Wang, and F.~R. Yu,
  ``Buffer-aware virtual reality video streaming with personalized and private
  viewport prediction,'' \emph{IEEE Journal on Selected Areas in
  Communications}, vol.~40, no.~2, pp. 694--709, 2022.

\bibitem{li23}
M.~Li, J.~Gao, C.~Zhou, X.~Shen, and W.~Zhuang, ``User dynamics-aware edge
  caching and computing for mobile virtual reality,'' \emph{IEEE Journal of
  Selected Topics in Signal Processing}, vol.~17, no.~5, pp. 1131--1146, 2023.

\bibitem{bakambekova2024interplay}
A.~Bakambekova, N.~Kouzayha, and T.~Al-Naffouri, ``On the interplay of
  artificial intelligence and space-air-ground integrated networks: A survey,''
  \emph{IEEE Open Journal of the Communications Society}, vol.~5, pp.
  4613--4673, 2024.

\bibitem{Zhagypar2023}
R.~Zhagypar, N.~Kouzayha, H.~ElSawy, H.~Dahrouj, and T.~Y. Al-Naffouri,
  ``Characterization of the global bias problem in aerial federated learning,''
  \emph{IEEE Wireless Communications Letters}, vol.~12, no.~8, pp. 1339--1343,
  2023.

\bibitem{amiri20}
M.~Mohammadi~Amiri and D.~Gündüz, ``Machine learning at the wireless edge:
  Distributed stochastic gradient descent over-the-air,'' \emph{IEEE
  Transactions on Signal Processing}, vol.~68, pp. 2155--2169, 2020.

\bibitem{krishnendu22}
K.~S. Tharakan, B.~N. Bharath, N.~Garg, V.~Bhatia, and T.~Ratnarajah,
  ``Learning to cache: Federated caching in a cellular network with correlated
  demands,'' \emph{IEEE Transactions on Communications}, vol.~70, no.~3, pp.
  1653--1665, 2022.

\bibitem{pmlr-mcmahan17a}
B.~McMahan, E.~Moore, D.~Ramage, S.~Hampson, and B.~A.~y. Arcas,
  ``{Communication-Efficient Learning of Deep Networks from Decentralized
  Data},'' in \emph{Proceedings of the 20th International Conference on
  Artificial Intelligence and Statistics}, vol.~54, Apr 2017, pp. 1273--1282.

\bibitem{pmlr-karimireddy20a}
S.~P. Karimireddy, S.~Kale, M.~Mohri, S.~Reddi, S.~Stich, and A.~T. Suresh,
  ``{SCAFFOLD}: Stochastic controlled averaging for federated learning,'' in
  \emph{Proceedings of the 37th International Conference on Machine Learning},
  vol. 119, Jul 2020, pp. 5132--5143.

\bibitem{tian19}
T.~Li, A.~K. Sahu, A.~Talwalkar, and V.~Smith, ``Federated learning:
  Challenges, methods, and future directions,'' \emph{IEEE Signal Processing
  Magazine}, vol.~37, pp. 50--60, 2019.

\bibitem{smith17}
V.~Smith, C.-K. Chiang, M.~Sanjabi, and A.~S. Talwalkar, ``Federated multitask
  learning,'' in \emph{Advances in Neural Information Processing Systems},
  vol.~30, 2017, pp. 4424--4434.

\bibitem{lipsterMATH89}
R.~S. {Liptser} and A.~N. {Shiryayev},
  \emph{\BIBforeignlanguage{English}{{Theory of martingales. Transl. from the
  Russian by K. Dzjaparidze.}}}\hskip 1em plus 0.5em minus 0.4em\relax
  Dordrecht etc.: Kluwer Academic Publishers, 1989.

\bibitem{rzhang24}
R.~Zhang, F.~Liu, J.~Liu, M.~Chen, Q.~Tang, T.~Huang, and F.~R. Yu,
  ``Cpper-{FL}: Clustered parallel training for efficient personalized
  federated learning,'' \emph{IEEE Transactions on Mobile Computing}, vol.~23,
  no.~10, pp. 9424--9436, 2024.

\bibitem{Mangiante2017V}
S.~Mangiante, G.~Klas, A.~Navon, G.~Zhuang, R.~Ju, and M.~D. Silva, ``{VR} is
  on the edge: How to deliver $360^{\circ}$ videos in mobile networks,''
  \emph{Proceedings of the Workshop on Virtual Reality and Augmented Reality
  Network}, 2017.

\bibitem{bernstein18}
J.~Bernstein, Y.~Wang, K.~Azizzadenesheli, and A.~Anandkumar, ``sign{SGD}:
  Compressed optimisation for non-convex problems,'' in \emph{Proceedings of
  the 35th International Conference on Machine Learning}, 2018, pp. 560 -- 569.

\bibitem{Anava13}
O.~Anava, E.~Hazan, S.~Mannor, and O.~Shamir, ``Online learning for time series
  prediction,'' in \emph{Proceedings of the 26th Annual Conference on Learning
  Theory}, vol.~30.\hskip 1em plus 0.5em minus 0.4em\relax PMLR, 12--14 Jun
  2013, pp. 172--184.

\bibitem{krishnendu24}
K.~S. Tharakan, B.~N. Bharath, and V.~Bhatia, ``Online learning to cache and
  recommend in the next generation cellular networks,'' \emph{IEEE Transactions
  on Machine Learning in Communications and Networking}, vol.~2, pp. 511--525,
  2024.

\bibitem{konecny16}
\BIBentryALTinterwordspacing
J.~Kone{\v{c}}n{\'y}, H.~B. McMahan, F.~X. Yu, P.~Richt{\'{a}}rik, A.~T.
  Suresh, and D.~Bacon, ``Federated learning: Strategies for improving
  communication efficiency,'' \emph{CoRR}, vol. abs/1610.05492, 2016. [Online].
  Available: \url{http://arxiv.org/abs/1610.05492}
\BIBentrySTDinterwordspacing

\bibitem{chung06}
F.~Chung and L.~Lu, \emph{Complex graphs and networks}.\hskip 1em plus 0.5em
  minus 0.4em\relax CBMS Regional Conference Series in Mathematics, American
  Mathematical Society, 2006.

\bibitem{hazan17}
E.~Hazan, K.~Singh, and C.~Zhang, ``Efficient regret minimization in non-convex
  games,'' in \emph{Proceedings of the 35th International Conference on Machine
  Learning}, 2017, pp. 1433--1441.

\bibitem{lo17}
W.-C. Lo, C.-L. Fan, J.~Lee, C.-Y. Huang, K.-T. Chen, and C.-H. Hsu,
  ``$360^{\circ}$ video viewing dataset in head-mounted virtual reality,'' in
  \emph{Proceedings of the 8th ACM on Multimedia Systems Conference}, 2017, pp.
  211--216.

\bibitem{wu17dataset}
C.~Wu, Z.~Tan, Z.~Wang, and S.~Yang, ``A dataset for exploring user behaviors
  in {VR} spherical video streaming,'' in \emph{Proceedings of the 8th ACM on
  Multimedia Systems Conference}, ser. MMSys'17, 2017, p. 193–198.

\end{thebibliography}
\end{document}